\documentclass[prd,amsmath,aps,floats,amssymb, floatfix, superscriptaddress,nofootinbib,twocolumn,preprintnumbers]{revtex4-2}
\usepackage{tabularx}
\usepackage{amsmath,amssymb,natbib,latexsym,times}

\usepackage[]{lineno}
\usepackage[T1]{fontenc}
\usepackage{aecompl}

\usepackage{multirow}
\usepackage{rotating}
\usepackage{booktabs}
\usepackage{graphicx}
\usepackage{xspace}

\usepackage{hyperref}
\hypersetup{colorlinks=true, linkcolor=blue, citecolor=blue, filecolor=blue, urlcolor=blue}
\usepackage{ulem}

\newcommand{\blockfont}[1]{{\textsc{#1}}\xspace}

\defcitealias{y6-gold}{\textsc{Y6Gold}}
\defcitealias{y6-piff}{\textsc{Y6PSF}}
\defcitealias{y6-mask}{\textsc{Y6Mask}}
\defcitealias{y6-balrog}{\textsc{Y6Balrog}}
\defcitealias{y6-metadetect}{\textsc{Y6Shear}}
\defcitealias{y6-maglim}{\textsc{Y6Pos}}
\defcitealias{y6-sourcepz}{\textsc{Y6ShearPZ}}
\defcitealias{y6-lenspz}{\textsc{Y6PosPZ}}
\defcitealias{y6-wz}{\textsc{Y6WZ}}
\defcitealias{y6-nzmodes}{\textsc{Y6Modes}}
\defcitealias{y6-magnification}{\textsc{Y6Mag}}
\defcitealias{y6-gglens}{\textsc{Y6GGL}}
\defcitealias{y6-methods}{\textsc{Y6Model}}
\defcitealias{y6-ppd}{\textsc{Y6PPD}}
\defcitealias{y6-imagesims}{\textsc{Y6Imsim}}
\defcitealias{y6-cardinal}{\textsc{Y6Mock}}
\defcitealias{y6-1x2pt}{\textsc{Y6-1x2}pt}
\defcitealias{y6-2x2pt}{\textsc{Y6-2x2}pt}
\defcitealias{y6-3x2pt}{\textsc{Y6-3x2}pt}
\defcitealias{y6-extensions}{\textsc{Y6Extend}}

\newcommand{\omegam}{\ensuremath{\Omega_\mathrm{m}}}

\DeclareRobustCommand{\VAN}[3]{#2}
\let\VANthebibliography\thebibliography
\def\thebibliography{\DeclareRobustCommand{\VAN}[3]{##3}\VANthebibliography}

\usepackage{verbatim}
\usepackage[usenames,dvipsnames]{xcolor}
\usepackage{tabulary}
\usepackage{tabularx}
\usepackage{todonotes}
\usepackage{hyperref}
\usepackage{ulem}
\usepackage{multirow}

\newcommand\lcdm{$\Lambda$CDM}

\newcommand\mcal{{\textsc{Metacalibration}}}
\newcommand\mdet{{\textsc{Metadetection}}}

\newcommand\balrog{{\textsc{Balrog}}}

\defcitealias{y6-methods}{\textsc{Y6Model}}

\newcommand{\LCDM}{$ \Lambda $CDM~}

\newcommand\be{\begin{equation}}
\newcommand\ee{\end{equation}}
\def\bea{\begin{eqnarray}}
\def\eea{\end{eqnarray}}

\allowdisplaybreaks
\usepackage{lineno}

\begin{document}
\title{Dark Energy Survey Year 6 Results: Cosmological Constraints from Cosmic Shear}
% Author list file generated with: mkauthlist 1.3.0 
% mkauthlist --sort /Users/hattifattener/Downloads/DES-2025-0926_author_list.csv 

\author{T.~M.~C.~Abbott}
\affiliation{Cerro Tololo Inter-American Observatory, NSF's National Optical-Infrared Astronomy Research Laboratory, Casilla 603, La Serena, Chile}
\author{M.~Aguena}
\affiliation{INAF-Osservatorio Astronomico di Trieste, via G. B. Tiepolo 11, I-34143 Trieste, Italy}
\affiliation{Laborat\'orio Interinstitucional de e-Astronomia - LIneA, Av. Pastor Martin Luther King Jr, 126 Del Castilho, Nova Am\'erica Offices, Torre 3000/sala 817 CEP: 20765-000, Brazil}
\author{A.~Alarcon}
\affiliation{Institute of Space Sciences (ICE, CSIC),  Campus UAB, Carrer de Can Magrans, s/n,  08193 Barcelona, Spain}
\author{O.~Alves}
\affiliation{Department of Physics, University of Michigan, Ann Arbor, MI 48109, USA}
\author{A.~Amon}
\affiliation{Department of Astrophysical Sciences, Princeton University, Peyton Hall, Princeton, NJ 08544, USA}
\author{D.~Anbajagane}
\affiliation{Kavli Institute for Cosmological Physics, University of Chicago, Chicago, IL 60637, USA}
\author{F.~Andrade-Oliveira}
\affiliation{Physik-Institut, University of Zürich, Winterthurerstrasse 190, CH-8057 Zürich, Switzerland}
\author{W.~d'Assignies}
\affiliation{Institut de F\'{\i}sica d'Altes Energies (IFAE), The Barcelona Institute of Science and Technology, Campus UAB, 08193 Bellaterra (Barcelona) Spain}
\author{S.~Avila}
\affiliation{Centro de Investigaciones Energ\'eticas, Medioambientales y Tecnol\'ogicas (CIEMAT), Madrid, Spain}
\author{D.~Bacon}
\affiliation{Institute of Cosmology and Gravitation, University of Portsmouth, Portsmouth, PO1 3FX, UK}
% ...

\collaboration{DES Collaboration}

\date{\today}
\label{firstpage}

\begin{abstract}
We present legacy cosmic shear measurements and cosmological constraints using six years of Dark Energy Survey (DES) imaging data. From these data, we study $\sim$140 million galaxies (8.29 galaxies/arcmin$^2$) that are 50\% complete at $i=24.0$ mag and extend beyond $z=1.2$. We divide the galaxies into four redshift bins, and obtain cosmic shear measurement with a signal-to-noise of 60, a factor of 1.5 higher than the Year 3 analysis.  We model the uncertainties due to shear and redshift calibrations, and discard measurements on small angular scales to mitigate baryon feedback and other small-scale uncertainties. We consider two fiducial models to account for the intrinsic alignment of the galaxies. We conduct a blind analysis in the context of the \lcdm~model and find $S_8 \equiv \sigma_8(\Omega_{\rm m}/0.3)^{0.5} = 0.798^{+0.014}_{-0.015}$ (marginalized mean with 68$\%$ confidence limits) when using the non-linear alignment model (NLA) and $S_{8} = 0.783^{+0.019}_{-0.015}$ with the tidal alignment and tidal torque model (TATT), providing 1.8\% and 2.5\% uncertainty on $S_8$. Compared to constraints from the cosmic microwave background from {\it Planck} 2018, ACT DR6 and SPT-3G DR1, we find consistency in the full parameter space at 1.1$\sigma$  (1.7$\sigma$) and in $S_8$ at 2.0$\sigma$ (2.3$\sigma$) for NLA (TATT), respectively. The result using the NLA model is preferred according to the Bayesian evidence. We find that the model choice for intrinsic alignment and baryon feedback can impact the value of our $S_8$ constraint up to $1\sigma$. For our fiducial model choices, the resultant uncertainties in $S_8$ are primarily degraded by the removal of scales, as well as the marginalization over the intrinsic alignment parameters. We demonstrate that our result is internally consistent and robust to different choices in calibrating the data, owing to methodological improvements in shear and redshift measurement, which lay the foundation for next-generation cosmic shear programs. 
\end{abstract}

\preprint{DES-2025-0926}
\preprint{FERMILAB-PUB-26-0027-PPD}
\maketitle

\section{Introduction}\label{sec:intro}

The flat $\Lambda$CDM cosmological model --- cold dark matter (CDM) plus a cosmological constant ($\Lambda$) in a spatially flat Universe --- has been remarkably successful in explaining a large number of astrophysical observables that are sensitive to geometry and structure formation of the Universe across cosmic times. Currently, percent-level cosmological constraints are provided both by early-time probes, such as the primary and secondary anisotropies of the cosmic microwave background (CMB) signal \citep{planck2018, ACT_DR6_extended_models, spt_data}, and by late-time probes, such as baryon acoustic oscillations (BAO; e.g., \cite{desidr2, desi_dr2_cosmo, desy6_bao}), type Ia supernovae (SNIa; e.g., \cite{pantheon_plus22, DESY5SN2024, Riess2025, Popovic2025b}) and weak gravitational lensing. Of all the late-time probes, weak gravitational lensing of distant galaxies by large-scale structure, or cosmic shear, is unique in that it directly probes dark matter and is simultaneously sensitive to structure growth and geometry.

Cosmic shear was first detected just 26 years ago \citep{bacon00,kaiser00,vanWaerbeke00,wittman00}. Compared with other probes this first observation came relatively late (e.g., galaxy clustering was detected in the 1930s, the CMB in the 1960s, and strong lensing in 1979), since weak lensing causes only percent-level distortions, meaning it has to be measured statistically over a large population of galaxies. To reach its theoretical promise as a cosmological probe (predicted early on by e.g. \cite{MiraldaEscude91}), the community commissioned larger, more cosmology-oriented, galaxy imaging surveys such as the Dark Energy Survey (DES; \cite{des_proposal}), the Kilo Degree Survey (KiDS; \cite{deJong13}), and the Hyper-Suprime Camera Subaru Strategic Program (HSC-SSP; \cite{aihara22}). To date, the state-of-the-art in terms of cosmic shear cosmology is represented by KiDS-Legacy \citep{kids-legacy}, HSC Year-3 (Y3) \citep{dalal23,li23}, DES Y3 \citep{y3-cosmicshear1, y3-cosmicshear2}, as well as follow-on studies that have combined lensing data sets \citep{decade, des-kids}. This paper presents the first cosmic shear analysis from the full six-year (Y6) dataset from DES, an important milestone for these so-called Stage-III surveys.

In contemporary work, an end-to-end cosmic shear analysis includes many critical building blocks, each of which has seen significant methodological advances in the past 10 years. Roughly, there are three large categories: shear measurement and calibration, redshift estimation and calibration, and modeling of astrophysical effects (in particular intrinsic alignment and baryonic feedback). For shear estimation, in addition to traditional methods that rely solely on image simulations for calibration \citep{miller13, bernstein_bfd1, bernstein_bfd2}, the community has developed new techniques that use the data itself (self-calibration) \citep{mcal1, mcal2, sheldon20} and/or analytical prescriptions \citep{anacal_1} for the first-order bias corrections and image simulations for residual effects \citep{y3-imagesims, li23_kids}. For redshift estimation, the main advance involves the development of more statistically rigorous Bayesian methods \citep{buchs19, y3-sompz} as well as the combination of multiple independent methods that rely on different information in the datasets in addition to photometry (e.g. spatial clustering \citep{y3-sourcewz}, lensing ratios \citep{y3-shearratio}). An additional challenge for current precision is how redshift and shear calibration can be coupled via the effect of blending \citep{y3-imagesims}. Finally, astrophysical systematic effects have come into the spotlight in recent years as the Stage-III analyses come to adopt more advanced models for intrinsic alignment \citep{hirata04,blazek15,blazek19,vlah20,fortuna21,maion24,bakx23} and attempt to model the small-scale matter power spectrum, which is impacted by baryonic feedback \citep{mead21,arico21,giri21,schneider25,schaller25}.

This paper presents constraints placed by DES Y6 cosmic shear on a baseline six-parameter flat $\Lambda$CDM model. The analysis employs state-of-the-art methodologies for shear estimation/calibration described in \citep{y6-piff, y6-metadetect, y6-imagesims}, redshift estimation/calibration described in \citep{y6-sourcepz, y6-wz} and modeling described in \citep{y6-methods}. These results feed into the larger analysis combining three two-point functions to derive cosmological constraints: cosmic shear, galaxy-galaxy lensing and galaxy clustering. This combination is commonly referred to as the ``$3 \times 2$pt'' probes and the results are presented in \citep[][hereafter \citetalias{y6-3x2pt}]{y6-3x2pt}. Cosmic shear is the only two-point function that does not involve galaxy density. The combination of the two other two-point functions is commonly referred to as the ``2$\times$2pt'' probes and we will present constraints from them in a forthcoming paper \citep[][hereafter \citetalias{y6-2x2pt}]{y6-2x2pt}.  

Looking ahead, several new programs are on the horizon to push to even larger datasets: the \textit{Vera C. Rubin} Observatory's Legacy Survey of Space and Time (LSST; \cite{lsst, ivezic_lsst}), the \textit{Euclid} \citep{euclid} mission, and the \textit{Nancy G. Roman} Space Telescope's \citep{spergel_roman} High-Latitude Imaging Survey (HLIS). Many of the advances made in the DES Y6 cosmic shear analysis will form the foundation of these next-generation surveys. 

The paper is structured as follows. Section \ref{sec:data} summarises the DES Y6 data, including construction of associated galaxy catalog. The estimation of cosmic shear two-point functions is then described in Section \ref{sec:cosmicshear}. In Section \ref{sec:method} we outline how we model these two-point data, and set up the inference problem that will be our main result. A selection of external data sets, which we will compare with are discussed in Section \ref{sec:extdata}. Our core cosmological findings are presented in Section \ref{sec:results}, with Section \ref{sec:robustness} exploring the robustness to various systematics and assumptions. Section \ref{sec:insights} considers the results and analyses beyond the direct cosmology constraints. We conclude in Section \ref{sec:con}. 

\section{The DES Year 6 Shear Catalog}\label{sec:data}

The Dark Energy Survey (DES; \citep[][]{des_proposal}) is an imaging survey that took data with the Blanco Telescope on Cerro-Tololo Inter-American Observatory between 2013 and 2019, covering 1/8th of the sky. One of the main products from the full DES dataset is a weak lensing shear catalog comprised of $\sim140$M galaxies, split into 4 tomographic bins. Details of the shape estimation and calibration are described in \cite*[][hereafter \citetalias{y6-metadetect}]{y6-metadetect}; details of the redshift estimation and calibration are described in \cite[][hereafter \citetalias{y6-sourcepz}]{y6-sourcepz} and \cite[][hereafter \citetalias{y6-wz}]{y6-wz}, and details of the shear and redshift calibrations with simulations are described in \cite*[][hereafter \citetalias{y6-imagesims}]{y6-imagesims}. Below we briefly summarize the aspects of the shape measurement (Section~\ref{subsec:shapes}), the redshift estimates (Section~\ref{subsec:redshifts}) and the shear/redshift joint-calibration (Section~\ref{subsec:imagesims}) that are important for this paper. 

\subsection{Shape measurements}\label{subsec:shapes}

In the weak lensing limit, the observed galaxy shapes $e^{\rm obs}$ are connected to the gravitational shear $\gamma$ by 
\begin{equation}\label{eqn:shearbias}
    e^{\rm obs}_i = (1+m_i)(e^{\rm int}_i + \gamma_i) + c_i,
\end{equation}
where $e^{\rm int}$ is the intrinsic galaxy ellipticity, $m$ and $c$ quantify the shear-related systematic biases (characterized as multiplicative and additive shear bias, respectively) with the index $i$ indicating one of the components of the spin-2 shear field. 
The goal of the shape measurement stage is to construct a shape catalog where measured galaxy shapes have as little shear biases as possible. The galaxy shapes are not measured perfectly, however, and are subject to a number of biases, which directly impact inferred $S_8$. Thus, one should calibrate the shear biases from the data itself or from image simulations with known galaxy shear, or both in the case of residual biases.

The Y6 weak lensing galaxy shape catalog \citepalias{y6-metadetect} is the first application of the \mdet\ algorithm \citep{sheldon20} to real images. \mdet\ is a new shear calibration technique that includes detection in the calibration loop to achieve sub-percent level shear calibration biases. In practice, the survey was coadded into small ``cell-based coadds'' \citep{armstrong24}, which are designed to have a continuous PSF within small ``cells'' of 200 $\times$ 200 pixels.  Then the artificial shearing operations in \mdet\ are applied to the entire cell-based coadd. For each artificially sheared image, objects are detected on a multi-band $riz$ coadd. Finally, shape and object measurement algorithms are applied to each detection, resulting in one catalog of objects in the survey for each artificial shear. Per \cite{sheldon20}, shear calibration is achieved by using catalog-average quantities to compute the average response of a selection of objects to shear.
We note that the main difference between \mdet\ and the \mcal\ algorithm used in the DES Y3 analysis \citep{y3-shapecatalog} is that detection is included in the calculation of the response. As such, the shear calibration automatically accounts for any same-redshift blending-induced selection effects at the detection stage (e.g., preferentially detecting objects that are more sheared as they tend to overlap with neighboring objects and result in higher signal-to-noise). This allows us to more reliably calibrate shear in deeper data where detection-related systematic effects become more important.

Within the shape catalog, the quantities of interest for this paper are the galaxy ellipticity, the statistical inverse variance shear weights, and the shear responses over each tomographic bin. The pre-seeing ellipticity for each galaxy is measured as a Gaussian model fit. 
The statistical weight for each galaxy is computed as an inverse variance weight in the grid of S/N and size ratio (see Equation 9 and Figure 5 of \citepalias{y6-metadetect} for the full detail). 
Finally, the shear response for self-calibration of the raw ellipticities can be calculated as
\begin{equation}\label{eqn:response}
    \langle R_{ij} \rangle = \frac{\langle e^{+}_i \rangle - \langle e^{-}_i \rangle}{\Delta\gamma_j},
\end{equation}
using the ellipticities measured in the artificially sheared images $e^{+/-}_i$ with the applied shear of $|\gamma_j|=0.01$ and hence $\Delta\gamma_j = 0.02$.

The \mdet\ catalog passes a series of null tests ensuring that the measurements are not subject to systematic errors due to detector, observation, and PSF-related effects. Through these tests, we found that the galaxy shapes measured on $g$-band images were still affected by PSF mis-modeling, even with the color-dependent modeling implemented in \citep{y6-piff}. However, we are able to measure the $g$-band fluxes by performing forced photometry on $g$-band images at the $riz$ detection, and we use $g$-band flux information for photometric redshifts described in Section~\ref{subsec:redshifts}. The catalog was then tested for robustness at the two-point correlation function and cosmology level. Specifically, we test the impact of the PSF modeling errors (Section~\ref{subsec:psf_sys}), B-modes (Section~\ref{subsec:bmode_sys}) and additive bias correction (Section~\ref{subsec:additive_sys}).

The final catalog consists of 139,662,173 galaxies\footnote{The number of objects used for the final cosmological analysis is slightly different from what is presented in \cite{y6-metadetect} due to additional masks to facilitate combination with galaxy-galaxy lensing and galaxy clustering signals \citep{y6-mask,y6-3x2pt}.} where the overall effective number density is $n_{\rm eff}$=8.29 galaxies per arcmin$^2$ and shape noise\footnote{See Equation 13 and 14 of \cite*{y6-metadetect} for a formal definition of effective number density and shape noise.} of 0.29. Table~\ref{tab:shape_stats} lists the summary statistics of our source data (number of objects, effective number density, shape noise, shear response, mean redshift and mean shear) and shear calibration parameters, compared with previous DES catalogs namely Science Verification (SV) \cite{jarvis16}, Y1 \cite{zuntz18} and Y3 \cite*{y3-shapecatalog, y3-imagesims}. 

\begin{table*}
    \centering
    \caption{\label{tab:shape_stats} The number of objects, survey area, effective number density, shape noise, shear response, mean redshift, weighted residual mean shear, the shear multiplicative bias from image simulations ($m$) with 1$\sigma$ uncertainty.
    }
     \begin{tabular}{ccccccccccc}
\hline\hline
      Release & Bin & num. of objects & area [deg$^2$] & $n_{\rm eff}$ [gals/arcmin$^2$] & $\sigma_{\rm e}$ & $\langle R \rangle$ & $\langle z \rangle $ & $\langle e_1 \rangle \times 10^{-4}$ & $\langle e_2 \rangle \times 10^{-4}$ & $m \times 10^{-3}$ \\ \hline
      \multirow{1}{*}{SV}
        & All & 3,446,533 & 139 & 6.76 & 0.278 & -- & -- & $-0.4$ & $+10.2$  & $-33 \pm 1.5$  \\ 
      \cline{1-11}
      \multirow{1}{*}{Y1}
        & All & 34,839,418 & 1500 & 6.38 & 0.277 & 0.653 & -- & $+3.5$ & $+2.8$ & $12 \pm 13$ \\ 
      \cline{1-11}
      \multirow{1}{*}{Y3} 
        & All & 100,204,026 & 4143 & 5.59 & 0.261 & 0.718 & -- & $+3.5$ & $+0.59$ & $21 \pm 17$  \\  
      \cline{1-11}
      \multirow{5}{*}{Y6}
        & 1 & 33,707,071 & -- & 2.05 & 0.265 & 0.856 & 0.41 & $+1.6$ & $+0.62$ & $-3.4 \pm 5.8$ \\
        & 2 & 34,580,888 & -- & 2.10 & 0.287 & 0.869 & 0.54 & $+0.45$ & $-0.58$ & $6.5 \pm 6.6$ \\
        & 3 & 34,646,416 & -- & 2.14 & 0.282 & 0.819 & 0.85 & $+2.4$ & $-1.7$ &  $15.9 \pm 5.9$ \\
        & 4 & 36,727,798 & -- & 2.32 & 0.347 & 0.692 & 1.16 & $+3.1$ & $+1.1$ & $1.7 \pm 12.2$ \\
        \cline{2-11}
        & All & 139,662,173 & 4031 & 8.29 & 0.289 & 0.819 & -- & $+1.4$ & $-0.15$ &  $3.4 \pm 2.0$ \\
\hline\hline
    \end{tabular}
\end{table*}

\subsection{Photometric redshifts}\label{subsec:redshifts}

To measure the redshift distribution $n(z)$ for DES Y6 source galaxies, we use the Self-Organizing Map photo-$z$ framework (SOMPZ) \citep{buchs19, y3-sompz}. The SOMPZ framework is built on two SOMs: a ``wide SOM'' trained on the wide-field images and $griz$ photometry, and a ``deep SOM'' trained on the DES deep-fields \citep{y3-deepfields} with $ugrizJH K_s$ photometry. For each deep-field phenotype (i.e., cell in the deep SOM), the redshift distribution is obtained from objects in the calibration sample that overlap the deep fields on the sky and are matched in phenotype. The redshift calibration sample was built from spectroscopy, where available, supplemented by high-quality multi-band photometric redshifts from \textsc{PAUS} \citep{paus, paus-cosmos} and \textsc{COSMOS2020} \citep{cosmos2020}. 

To connect deep-field phenotypes to wide-field detections under the same observing conditions, we use the \balrog\ synthetic source injection catalog \citep{y6-balrog}. Deep-field galaxies are injected into DES wide-field images and processed through the same detection, measurement, and selection pipeline, providing a calibrated transfer function that captures selection effects. This transfer function is applied when propagating the deep SOM per-cell $p(z)$ to the wide-field $n(z)$.

To estimate the uncertainty in the $n(z)$, we generate an ensemble of possible redshift distributions, i.e., redshift realizations, to represent the uncertainty in the $n(z)$. The sources of uncertainty and bias include: i) bias and uncertainty from using photometric redshift PAUS and COSMOS; ii) field-to-field variations in DES deep-field photometric calibration; iii) the limited number of galaxies in the deep-field and redshift samples; and iv) the limited area of the deep and redshift field \citep{y6-sourcepz}. We model and propagate each source of uncertainty through the SOM. We found that our redshift sample uncertainty was the dominant source of uncertainty in higher-redshift bins, primarily due to the lack of faint-magnitude and high-redshift spectra. 

The redshift realizations are further constrained by spatially correlating our source galaxies with overlapping spectroscopic galaxies (BOSS-eBOSS; \cite{boss_color,eboss_dawson}). By theoretically predicting the small-scale angular correlation, we select a subset of the redshift realizations that best match the observations \citep{y6-wz}. This use of independent information between methods ensures extra robustness.

\begin{figure}
    \begin{center}
    \includegraphics[width=\columnwidth]{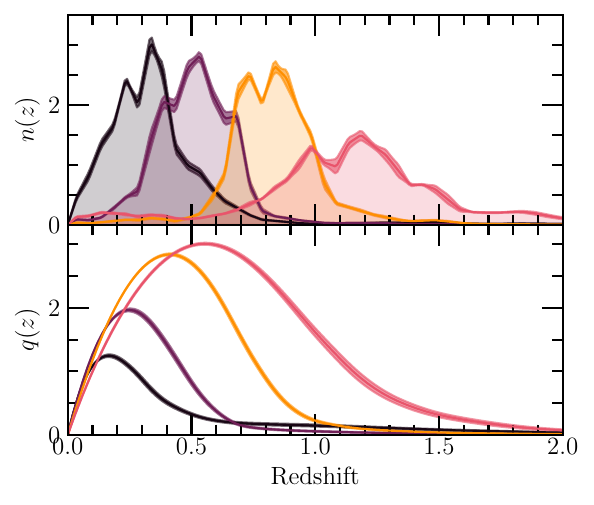}
    \end{center}
    \caption{Estimated Y6 source redshift distributions, divided into four tomographic bins. The shaded curves and band show the mean and standard deviation of 100 $n(z)$ realisations drawn from the posterior from our Y6 TATT analysis. In the lower panel we show the corresponding lensing kernels (see Equation \ref{eq:gz}).
    }
    \label{fig:nz}
\end{figure}

To efficiently sample redshift realizations for cosmological inference, we apply a mode-projection approach following \citep{y6-nzmodes}. This modified principal component analysis (PCA) decomposes the redshift perturbations to modes that capture the largest cosmological information, i.e., we analyze the Fisher matrix on $n(z)$ perturbations and retain those with the most significant effect on theoretical $\xi_{\pm}$ data vector. We keep seven leading modes here, which together encapsulate high-order information on the shape of the redshift distributions. The final redshift distributions for each tomographic bin with realizations reconstructed by sampling over the mode amplitudes are shown in the top panel of Figure~\ref{fig:nz}.

\subsection{Shear \& redshift calibration}\label{subsec:imagesims}

Image simulations are used to evaluate the systematic biases in the measured galaxy shapes and $n(z)$ relative to the truth, and therefore calibrate any residual bias. One major source of biases in the measured shear and $n(z)$ is the blending of source galaxies. Blending has two coupled effects. First, pairs of galaxies with positive tangential shear (i.e., sheared parallel to their separation) are more likely to be blended than those with negative tangential shear (i.e., sheared perpendicular to their separation), introducing shear-dependent blending. Second, in a blended pair, objects at different redshifts respond to shear differently. Since our $n(z)$ is weighted by the shear response and statistical shear weights, the blending of objects at different redshifts couples shear and redshift biases \citep{y3-imagesims}. While the shear-dependent detection bias of objects blended at the same redshift can be mitigated by \mdet\ \citep{sheldon20}, we need to model and correct for the coupled shear and redshift biases using image simulations. 
By simultaneously deriving the shear and redshift distribution calibrations in image simulations, it is also possible to account for covariances between them.

As described in detail in \citepalias{y6-imagesims}, the image simulation process results in an updated effective $n(z)$, redshift uncertainty parameters $u$, as well as shear calibration uncertainty parameters $m$ -- all three together now account for the blending impact on shear and redshift jointly. This effective $n(z)$ shown in Figure~\ref{fig:nz} is incorporated into the data vector from which we infer cosmology. 
We make a final joint adjustment to the priors on $m^i$ and $u$ to ensure that the effects of blending and other systematics in the data can be calibrated as measured in the image simulations. The correction, described in  \citepalias{y6-imagesims} and \citepalias{y6-sourcepz}, results in a prior with small non-zero covariance between the $m^i$ and $u$ parameters. This covariance properly accounts for the effects of blending that entangle the shear and redshift biases, propagated into the parameter inference through these covariant priors.
The measured multiplicative bias in each tomographic bin is $m_1=-0.0034 \pm 0.0058, \, m_2=0.0065 \pm 0.0066, \, m_3=0.0159 \pm 0.0059, \, m_4= 0.0017 \pm 0.0122$ with $1\sigma$ uncertainty measured as the diagonal of the covariance matrix between $m$ and $u$. In terms of the shift in mean redshift, we find $\Delta z = $[$-0.013$, $-0.021$, $-0.002$, $-0.034$]. This blending effect is relatively large compared to DES Y3 results likely due to a combination of more blending, fainter samples and weight assignments of those galaxies in the redshift distribution calibration.

\section{Cosmic Shear Measurement}\label{sec:cosmicshear}

The shear-shear two-point correlation function is calculated via the estimator 
\begin{equation}\label{eq:xipm_estimator}
    \xi^{ij}_{\pm}(\theta_k) = \frac{\sum_{ab} w_a w_b (e^a_{\rm t} e^b_{\rm t} \pm e^a_{\times} e^b_{\times})}{R^i R^j \sum_{ab} w_a w_b } \, ,
\end{equation}
where we sum over galaxy pairs, $(a,b)$, in redshift bins $(i,j)$, that are separated by an angular range, $|\theta_{ab}-\theta_k| < \Delta\theta $. $e_{\rm t}$ and $e_{\times}$ refers to the tangential and radial components of the \mdet\ ellipticities \citep{schneider02b}, $w$ is the galaxy weight and $R$ is the per redshift-bin response. 

\begin{figure*}
    \begin{center}
    \includegraphics[width=0.95\textwidth]{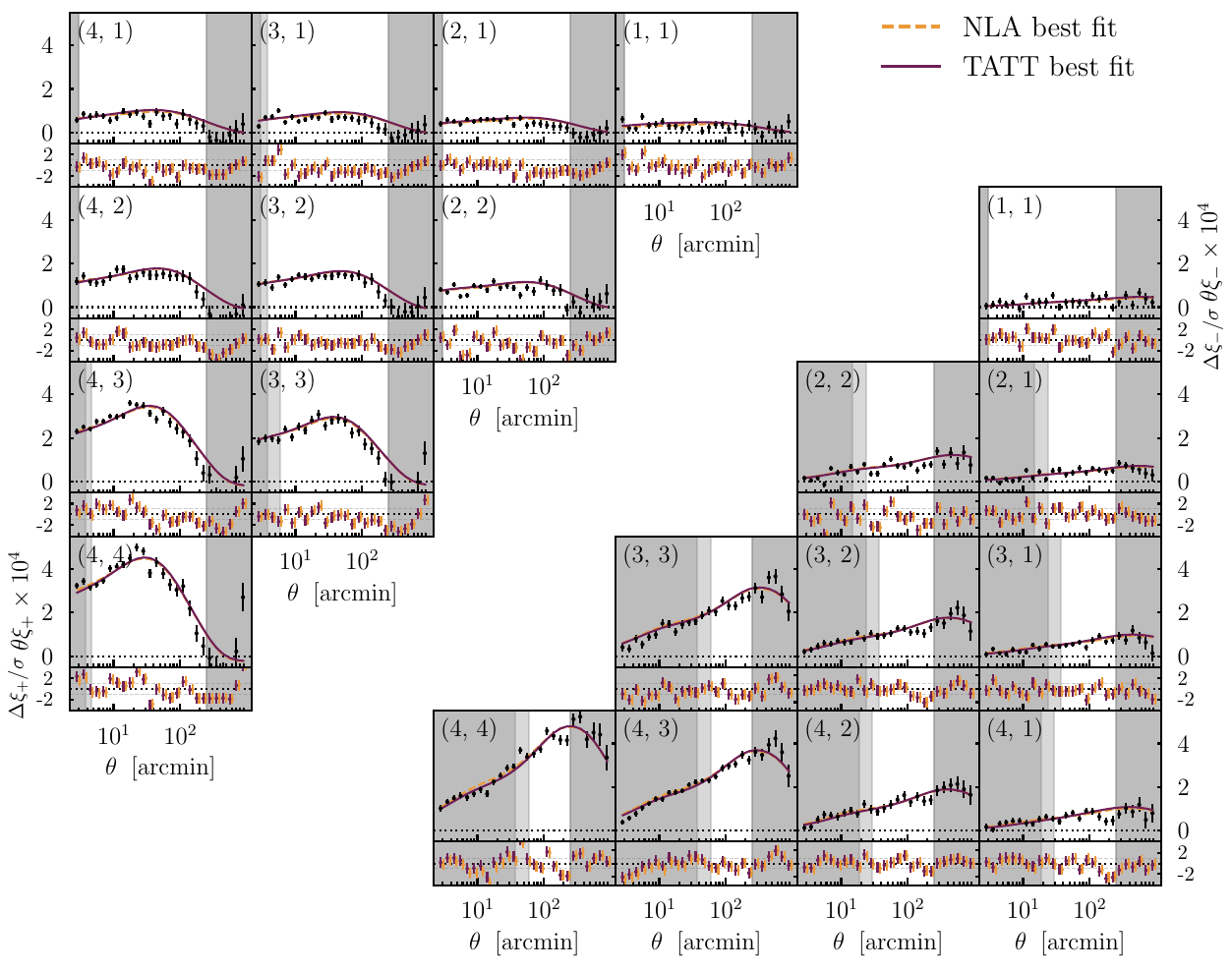}
    \end{center}
    \vspace{-0.7cm}
    \caption{Measured cosmic shear two-point correlations from DES Y6. Each panel shows a redshift bin combination (as labelled) and the upper/lower triangles represent the two types of correlation $\xi_+$ and $\xi_-$. In each case, we overlay the data with a theory prediction at the best-fit parameters from our two fiducial analyses (NLA in orange and TATT in purple). The lower panel of each correlation function shows the residual of the measured and theoretical best-fit data vector $\xi^{\rm data}_{\pm} - \xi^{\rm theory}_{\pm}$ scaled by the uncertainty in the covariance matrix. The shaded grey regions are scales excluded from the fits due to baryonic feedback and other uncertainties on very small or large scales. The goodness of fit to the $\Lambda$CDM model calculated as $p-$values are 0.06 and 0.13 for NLA and TATT, respectively.
    }
    \label{fig:2pt}
\end{figure*}

The tomographic DES Y6 cosmic shear measurement is shown in black in Figure~\ref{fig:2pt}. It is computed using the publicly available \blockfont{TreeCorr}\footnote{\url{https://rmjarvis.github.io/TreeCorr}; v5.0.1, $\mathrm{bin}\_\mathrm{slop} = 0.005$, $\mathrm{angle}\_\mathrm{slop} = 0.01$.} code \citep{jarvis04}. We opt to measure $\xi_\pm$ in 26 $\theta$ bins logarithmically spaced in the range $2.5-1000$ arcmin, yielding a total of $(2 \times N_{\rm z}(N_{\rm z}+1)/2 \times N_\theta) = 520$ data points, where $N_{\rm z}$ is the number of tomographic bins and $N_{\theta}$ is the number of angular bins. In practice, however, we conclude that there is little to no useful signal on very large scales, and so do not use scales $>250$ arcmin (see Appendix E of \citep{y6-methods} for a demonstration of this). For reasons we will come to in Section \ref{sec:method}, we also discard small scales in the cosmological inference, marked with gray shaded regions in Figure~\ref{fig:2pt}. Note that we have two sets of scale cuts for our fiducial results (refer to Section~\ref{sec:method} for more detail).

Our fiducial estimate for the data covariance is obtained from \blockfont{CosmoCov}\footnote{Note that simulated tests and initial data chains were run with a covariance matrix evaluated at the fiducial cosmology of \citep{y6-methods}. The assumed cosmology is then updated to the fiducial $3\times2$pt data MAP result \citepalias{y6-3x2pt} and the covariance matrix recomputed for all results shown in Section~\ref{sec:results}.} \citep{fang20}. Specifically, the total $\xi_\pm$ covariance is a sum of Gaussian, connected 4-point, and super-sample covariance terms. Our method here has not changed significantly since the Y3 analyses, and so we refer the reader to \cite{y3-covariances} for further details.

Our DES Y6 measurements have a signal-to-noise, S/N = 59 ($\theta \in [2.5, 250]$) and 60 ($\theta \in [2.5, 1000]$) for the best-fit fiducial models, where it is computed following
\begin{equation}
    S/N \equiv \frac{\xi^{\rm data}_{\pm}(\theta)^{T} C^{-1} \xi^{\rm model}_{\pm}(\theta)}{\sqrt{\xi^{\rm model}_{\pm}(\theta)^{T} C^{-1} \xi^{\rm model}_{\pm}(\theta)}}\, .
\end{equation}
This is a factor of $\sim$1.5 improvement compared to the DES Y3 equivalent of 40. Once the scale cuts are imposed, we find $S/N_{\rm NLA} = 40$ (267 data points) and $S/N_{\rm TATT} = 43$ (282 data points). 

Throughout our analysis, we adhere to a strict blinding protocol. There are three stages: 
catalog-level, data-vector level and parameter-level blinding. A detailed summary of the blinding and unblinding procedure is described in Section IV H of \citepalias{y6-3x2pt}. Specifically for this paper, all the pipeline-testing and null tests described in Appendix~\ref{app:data_tests} were performed using a blinded data vector via the methodology of \cite{muir_blinding} first. Only after all the null tests pass, do we update the plots with the unblinded data vector.   

\section{Modeling and Analysis Choices}\label{sec:method}

This section outlines the theoretical model for the cosmic shear correlation functions used in this analysis and discusses the evidence for its robustness.  Much of this work follows from \citep[][hereafter \citetalias{y6-methods}]{y6-methods}, where the model and its validation are described in detail. In this section, we briefly outline the model, but describe more extensively the elements that are particularly important for cosmic shear: intrinsic alignments (Section~\ref{subsec:ia}) and the non-linear power spectrum, including the impact of baryon feedback (Section~\ref{subsec:baryons}). In Sections~\ref{sec:likelihood} and ~\ref{sec:tension} we then describe the parameter inference procedure and metrics we use to assess internal and external consistency of our results.

\subsection{Cosmic shear theory}\label{subsec:cosmicshear_theory}

To theoretically model our estimator, Equation~\ref{eq:xipm_estimator}, we start with the 3D matter power spectrum $P_{\rm m}(k)$, project it into the 2D angular power spectrum $C^{ij}_{\kappa}(\ell)$, and then Fourier-transform it into its real-space counterpart, $\xi^{ij}_{\pm}(\theta)$.
 
Assuming a flat spatial geometry, we have
\begin{equation}
    C^{ij}_{\kappa}(\ell) = \int_0^{\chi_{\rm H}} {\rm d} \chi \frac{q_i(\chi) q_j (\chi)}{\chi^2} P_{\rm m}\left( k=\frac{\ell + 0.5}{\chi}, z(\chi)\right), 
\end{equation}
where $\chi_{\rm H}$ is the horizon distance. The lensing efficiency kernel $q_i(\chi)$ for a redshift bin $i$ is described as
\begin{equation}
    q_i(\chi) = \frac{3\,H_0^2\, \Omega_{\rm m}\, \chi}{2\,a(\chi)} \int^{\chi_H}_{\chi} {\rm d}\chi' \left( \frac{\chi' - \chi}{\chi'} \right) n_i(z(\chi'))\, \frac{{\rm d} z}{{\rm d} \chi'} \, ,
    \label{eq:gz}
\end{equation}
where \(n^{s}\) is the normalized redshift distribution of the source galaxies, \(H_0\) is the Hubble constant, and \(a(\chi)\) is the scale factor for a given comoving distance \(\chi\). The lensing efficiency for the source sample is shown in the lower panel of Figure~\ref{fig:nz}.

The convergence angular power spectrum $C^{ij}_{\kappa}(\ell)$ is then related to $\xi^{ij}_{\pm}(\theta)$ through Hankel transforms \citep{bartelmann01}:
\begin{equation}\label{eqn:hankel}
    \xi^{ij}_{\pm}(\theta) = \int_{0}^{\infty} \frac{\ell {\rm d}\ell}{2\pi} J_{0/4} (\ell \theta) C^{ij}_{\kappa}(\ell),
\end{equation}
where $J_{0/4}$ are Bessel functions of the first kind. Equation \ref{eqn:hankel} is valid in the limit that pre-lensing galaxy shapes are randomly oriented. The presence of intrinsic alignments (see the following section), however, breaks the equivalence of the shear and convergence spectra, and so in practice we must replace $C_{\kappa}$ with $C_{\gamma}$. 
In a full/curved sky formalism, we further decompose $C_{\gamma}$ into the E and B-mode components, and Equation~\ref{eqn:hankel} becomes
\begin{eqnarray}
    \xi^{ij}_{\pm}(\theta) = \sum_{\ell} \frac{2\ell + 1}{2\pi \ell^2 (\ell + 1)^2} \left[G^{+}_{\ell,2} (\mathrm{cos} \theta) \pm G^{-}_{\ell,2}(\mathrm{cos} \theta)\right]  \nonumber\\
    \times \left[C^{ij}_{\gamma,EE}(\ell) \pm C^{ij}_{\gamma,BB}(\ell)\right], 
\end{eqnarray}
where $G^{\pm}_{\ell}(x)$ are computed from Legendre polynomials $P_{\ell}(x)$ and averaged over angular bins \cite{y3-generalmethods}.

\subsection{Intrinsic alignments}\label{subsec:ia}

In addition to the purely cosmological lensing correlations described above, weak lensing measurements are also sensitive to more local processes. Since galaxies are astrophysical objects, situated in a particular part of the Universe, they interact with their environment. This acts to correlate intrinsic galaxy shapes, an effect known as intrinsic alignment (IA) \citep[e.g.][]{hirata04, troxel15, kiessling15}. Specifically, there are two relevant forms of correlation: first, an ``intrinsic-intrinsic (II)" contribution arises due to the correlation of intrinsic shapes in galaxies that are physically near to each other. A second, ``gravitational-intrinsic (GI)" contribution, comes from the fact that the same overdensities cause both lensing and local tidal effects. Depending on the galaxies being correlated either or both of these terms can be relevant.

The resulting two-point level contributions modify the observed angular power spectra as:

\begin{equation}
    C^{ij}_{\gamma, \mathrm{EE}} = C^{ij}_{\kappa} + C^{ij}_{\rm GI} + C^{ji}_{\rm GI} + C^{ij}_{\rm II, EE},
\end{equation}

\begin{equation}
    C^{ij}_{\gamma, \mathrm{BB}} = C^{ij}_{\rm II, BB}.
\end{equation}

\noindent
Note that there is no $BB$ signal from lensing, and we are not modeling source clustering (see \cite{linke24} and \cite{y3-generalmethods} Section B 5 for discussion) such that the only B-mode contribution in our model is that generated by IA. Assuming the Limber approximation, the IA angular power spectra can be written in the form:

\begin{equation}\label{eq:C_GI}
    C^{ij}_{\rm GI} = \int_0^{\chi_{\rm hor}} \frac{q_i(\chi) n_j(\chi)}{\chi^2} P_{\rm GI}\left (k=\frac{\ell + 0.5}{\chi}, z(\chi) \right) \mathrm{d}\chi,
\end{equation}

\begin{multline}\label{eq:C_II}
    C^{ij}_{\rm II, EE/BB} = \int_0^{\chi_{\rm hor}} \frac{n_i(\chi) n_j(\chi)}{\chi^2} \\
    \times P_{\rm II,EE/BB}\left (k=\frac{\ell + 0.5}{\chi}, z(\chi) \right) \mathrm{d}\chi.
\end{multline}

The most commonly used models for IA in cosmic shear analyses are the Non-Linear Alignment model (NLA; \citep{hirata04, bridle07}) and the Tidal Alignment and Tidal Torque model (TATT; \cite{blazek19}). 
Under the TATT model, the 3D IA power spectra can be decomposed into a sum of $k-$dependent terms modulated by amplitudes $a_1$, $a_2$ and $a_{\rm 1\delta}$ \citep{blazek19}. The amplitudes are defined:

\begin{equation}\label{eq:A1ofz}
    a_{1}(z) = - A_1 \frac{\bar{C}_1 \rho_c \omegam}{D(z)} \left ( \frac{1+z}{1+z_0} \right )^{\eta_1},
\end{equation}

\begin{equation}\label{eq:A2ofz}
    a_{2}(z) = 5 A_2 \frac{\bar{C}_1 \rho_c \omegam}{D^2(z)} \left ( \frac{1+z}{1+z_0} \right )^{\eta_2},
\end{equation}

\begin{equation}
    a_{1\delta}(z) = b_{\rm TA} a_1(z).
\end{equation}
Here $\rho_c=3H_0/8\pi G$ is the critical density of the Universe at $z=0$. The constant $\bar{C}_1$ is fixed to a value of $5\times10^{-14} M_\odot h^{-2} \mathrm{Mpc}^2$ \cite{brown02} and $z_0$ is a pivot redshift, which we chose to be $z_0=0.3$ to match the peak sensitivity of DES Y6 cosmic shear. We opt to fix $b_{\rm TA} = 1$ for TATT, guided by simulated tests presented in \citepalias{y6-methods}\footnote{The value of 1 was assumed, as an approximate galaxy bias-like value, given that this term arises from a factor of $1+\delta_g$ in the model. Pre-unblinding tests in \citepalias{y6-methods} suggested, similar to previous analyses \citep{Samuroff_Y1IA,troxel18}, that a roughly correct value was sufficient for our data without the need for marginalization. See Section \ref{subsec:robust_ia} where we return to this question \textit{aposteriori}.}. 
In this framework, NLA is a subset of TATT, with $A_2=0,\eta_2=0,b_{\rm TA} = 0$. The IA power spectra are then linear functions of the matter power spectrum, $P_{\rm GI}=a_1 P_{\rm m}$ and $P_{\rm II}=a_1^2 P_{\rm m}$. The priors for each parameter in the two cases are summarized in Table~\ref{tab:priors}.

Both NLA and TATT are consistent with direct measurements of IA on relatively large scales \citep[e.g.][]{samuroff23,Siegel_DESIIA, paus_ia}. In some of the previous work, it has been seen that the inferred cosmological parameters and uncertainties can differ when using these two IA models ($\sim1\sigma$ in $S_8$) \citep[e.g.][]{troxel18, y3-cosmicshear2, y3-cosmicshear1, des-kids}. This is, however, not always the case (see \citep{li23,dalal23}).

Pre-unblinding, there was limited \textit{a priori} evidence to justify either excluding or including the higher-order TATT contributions. In the absence of a thoroughly validated data-based selection (e.g. \cite{campos23}), we therefore opted to analyze the cosmic shear data using the NLA and TATT models in parallel, and keeping them as our fiducial results post-unblinding\footnote{Note that this approach is similar to that of HSC Y3 \cite{dalal23,li23}, who similarly follow \cite{campos23} in validating two IA model setups in parallel. Unlike those studies, however, our two fiducial models show some small differences. The IA model preferences of our data are also non-trivial (see Section \ref{subsec:insights_ia}), making it useful to consider both.}. We show the impact of these two models in Section~\ref{sec:results} and empirically assess their fits to the data.

\subsection{Matter power spectrum \& baryon feedback}\label{subsec:baryons}

Our cosmic shear analysis requires an accurate model for the matter power spectrum extending into the non-linear regime. While the evolution of matter clustering at linear scales ($k < 0.1 h \rm Mpc^{-1}$) can be accurately computed with CAMB \citep{Lewis00, Lewis02}, a model for the non-linear power spectrum is less certain. There are two effects here: (1) the non-linear evolution of dark matter and (2) the impact of energetic outflows from supernova and active galactic nuclei (AGN) feedback, known as baryon feedback. As described in \citepalias{y6-methods}, we model the non-linear power spectrum with \textsc{HMCode 2020} \citep{mead21}, an analytic halo model framework, that can include the impact of baryon feedback as it is calibrated to the \textsc{Bahamas} hydrodynamical simulation suite \citep{mccarthy17,mccarthy18}. Our fiducial model includes a baryon feedback component prescribed as a sub-grid AGN heating temperature $\Theta_{\rm AGN} = \mathrm{log}_{10} T_{\rm AGN}=7.7$. 

The amplitude and extent of the impact of baryon feedback on the matter power spectrum is not well understood \citep[e.g.][]{vanDaalen2011}; different feedback models implemented in cosmological hydrodynamical simulations produce a wide range of predictions (e.g., \cite{Dave2019, Henden2019, Hernandez-Aguayo2022, Schaye2023, Bigwood2025}). Even within a given simulation model, the `subgrid' parameters are degenerate with the impact of feedback on the matter power spectrum \citep[][]{Bigwood2025}. Furthermore, some recent direct and indirect observations have attempted to constrain the impact of feedback on the matter power spectrum, and there are hints that baryonic feedback processes could be stronger than expected from recent cosmological simulations \citep[e.g.][]{Schneider2022, amon22, preston23, Ferreira23, Bigwood2024, Dalal2025, LaPosta2025, Pandey2025, Siegel_BC25, Bigwood25_blueshear}. 

To account for this uncertainty, we adopt a conservative approach of restricting the cosmic shear measurements to larger angular scales where the impact of feedback is minimal. The scale cuts are designed to limit the bias due to feedback and non-linear growth to $<0.5\sigma$ in the marginal 1D $S_8$ constraint, compared to some choice for the strongest and weakest plausible feedback scenario. The strongest feedback scenario, or power suppression, that we consider is the prediction from the ``hi AGN" \blockfont{BAHAMAS}-8 simulation \cite{mccarthy18}; its power spectrum is more suppressed than other recent cosmological simulations (e.g. Millennium-TNG \cite{Hernandez-Aguayo2022}, FLAMINGO \citep{Schaye2023}, SIMBA \citep{Dave2019}). For the weak feedback scenario, we consider a gravity-only power spectrum. We generate 2 contaminated mock data vectors using these baryon scenarios combined with \blockfont{EuclidEmulator2} non-linear growth \citep{knabenhans21}. The procedure for determining the scale cuts from these contaminated data is detailed in \citepalias{y6-methods}. Since this procedure is sensitive to the uncertainty of our constraints on $S_8$, we derive scale cuts for both of our fiducial IA models individually. 
In the end, we use (178, 89) data points in $\xi_+$ and $\xi_-$ for NLA and (183, 99) data points for TATT, out of a total of 400 data points\footnote{When assuming the NLA IA model, since the statistical uncertainty is smaller due to the fewer free parameters, the same $\Delta \chi^2$ criteria results in removing more small-scale data points. This is the reason for the difference in the number of data points fit with the two models. See \citepalias{y6-methods} Section VI for details of the scale cut derivation process.}. Excluded angular scales for each measurement are shown as grey shaded regions in Figure~\ref{fig:2pt}.

Given the sensitivity of cosmic shear to the model for the non-linear matter power spectrum, we assess the impact on our fiducial $S_8$ constraint when an alternative model for the gravity-only non-linear matter power spectrum is used. We create simulated data using power spectra from \blockfont{EuclidEmulator2} \citep{knabenhans21} and BACCO Emulator \citep{arico21}, and analyse these using our fiducial analysis choices. We find shifts in $S_8$ relative to a baseline chain of $0.03\sigma$ and $0.08\sigma$ respectively for the two emulators. That is, given our choice of scale cuts, the uncertainty due to imperfections in the \blockfont{HMCode2020} treatment of the gravity-only power spectrum are subdominant.

In addition, the choice we make for the hydrodynamical simulation that represents the strongest feedback scenario, and therefore the extent of angular scales discarded, can impact our cosmological inference. Motivated by recent analyses that constrain the suppression of the matter power spectrum and find a strong 
suppression \citep[e.g.][]{Schneider2022, Bigwood2024, LaPosta2025, Dalal2025, Pandey2025, siegel25}, we assess the impact of underestimating the strength of baryon feedback on our cosmological constraints. We generate simulated data vectors using power spectra defined by a series of hydrodynamical simulations that exhibit a strong suppression. We then analyze them using our fiducial scale cuts. By design, if we consider the \blockfont{BAHAMAS}-8 simulation, used to determine the scale cuts, $S_8$ is biased low by 0.5$\sigma$. 
For one of the strongest feedback models, \blockfont{Illustris-1}, we find biases in $S_8$ of 0.7$\sigma$. Given that the exact amplitude of feedback is still an active research area \citep[e.g.][]{mccarthy17,Schaye2023,Dave2019,Bigwood2025}, we opt not to change scale cuts for this work. We note however that if the true feedback amplitude is indeed as extreme as these simulations, we will at most underestimate $S_8$ by $0.7\sigma$. 

\subsection{Parameter inference}
\label{sec:likelihood}

\begin{table}
    \centering
    \caption{Fiducial values and priors for the cosmological, astrophysical, and calibration parameters. Parameters are sampled using either flat priors, denoted by [min, max], or Gaussian priors, following the notation $\mathcal{N}(\mu, \, \sigma)$ for a normal distribution with mean $\mu$ and standard deviation $\sigma$. The priors on the calibration parameters of the source galaxy samples—namely the multiplicative shear bias $m$ and the source redshift distribution modes—are correlated. The width of their Gaussian priors is determined by the covariance matrix of the parameters. Note that for the NLA analysis, the IA parameters indicated with * are fixed at $A_2=\eta_2=b_{\rm TA}=0$.\\}
    \label{tab:priors}
        \begin{tabular}{ll}
            \hline \midrule
            Parameter  & Prior \\ 
            \midrule
            \multicolumn{2}{l}{\textbf{Cosmology}} \\
            $\Omega_{ \rm m}$ & [0.1, 0.6] \\ 
            $A_{\rm s} \times 10^9$  & [0.5, 5] \\ 
            $h$  & [0.58, 0.8] \\ 
            $\Omega_{\rm b}$  & [0.03, 0.07] \\ 
            $n_{\rm s}$  & [0.93, 1.00] \\ 
            $m_\nu$ [eV]  & [0.06, 0.6] \\ 
            \midrule
            \multicolumn{2}{l}{\textbf{Intrinsic alignment}} \\
            $A_1$  & [-1, 3]  \\
            $A_2$*  & [-3, 3]   \\ 
            $\eta_1$  & $\mathcal{N}(0.0,3.0)$ \\
            $\eta_2$*  &  $\mathcal{N}(0.0,3.0)$ \\
            $b_{\rm TA}$*  &  Fixed to 1 \\ 
            \midrule
            \multicolumn{2}{l}{\textbf{Source $n(z)$ modes}} \\
            $u^{j} \; (j \in [1,..., 7])$  &  $\mathcal{N}(0, 1) \in [-3,3]$ \\
            \midrule
            \multicolumn{2}{l}{\textbf{Shear calibration}} \\
                $m^1$ & $\mathcal{N}(-0.0034,0.0058)$ \\
    $m^2$ & $\mathcal{N}(0.0065,0.0066)$ \\
    $m^3$ & $\mathcal{N}(0.0159,0.0059)$ \\
    $m^4$ & $\mathcal{N}(0.0017,0.0122)$ \\
            \midrule \hline
        \end{tabular}

\end{table}

All of the modeling choices outlined above are implemented in \textsc{CosmoSIS}\footnote{\url{https://cosmosis.readthedocs.io/en/latest/}; v3.15} \citep{zuntz15}. We assume a Gaussian likelihood, such that $\mathrm{log} \mathcal{L} = - 0.5 (\mathbf{D}-\mathbf{\hat{D}}) C^{-1} (\mathbf{D}-\mathbf{\hat{D}})^T$, where $\mathbf{D}$ is our measured cosmic shear data vector, and $\mathbf{\hat{D}}$ is the theory prediction for $\mathbf{D}$, given some set of parameter values. We sample $\mathcal{L}$ within \textsc{CosmoSIS} using the \textsc{Nautilus} sampler \cite{lange23}\footnote{All runs with \textsc{Nautilus} use the following settings: $n_{\rm live} = 10,000$, $\mathrm{discard\_exploration} = \mathrm{T}$, $n_{\rm networks} = 16$}. \textsc{Nautilus} was found to be as accurate as the high-accuracy settings of the \textsc{PolyChord} nested sampling algorithm \citep{polychord} (itself previously validated in \cite*{y3-samplers}), while being considerably faster.

Following \citepalias{y6-methods} we sample 6 cosmological parameters with wide flat priors: $A_\mathrm{s}\times10^9$, $\Omega_{\rm m}$, $\Omega_{\rm b}$, $n_{\rm s}$, $h$ and $m_{\nu}$, either 2 or 4 IA parameters (for NLA and TATT), 7 mode parameters characterising the source redshift distributions $u^{1}...u^{7}$ and 4 shear calibration parameters $m^1...m^4$. For details on how these enter at the level of the predicted data vector see \citepalias{y6-methods}. We should note that our prior accounts for correlations in the $m-u$ space, and so instead of a 1D Gaussian per parameter we have a $11\times11$ covariance matrix. The shear parameters do have Gaussian priors (albeit correlated ones), with the mean/variance shown in Table \ref{tab:priors}. For the redshift modes, we sample dummy parameters from a Gaussian with unit variance, which are then transformed via the ``de-Gaussianisation" process described in \cite{y6-nzmodes} (see also \citepalias{y6-sourcepz}). In total this gives us either 19 or 21 free parameters for NLA and TATT (although considerably fewer effective parameters, as computed for best-fit $\chi^2$).

For some purposes, e.g.\ combining the likelihoods or posteriors of different probes across a common parameter space, instead of running a new chain with the joint likelihood, we use normalizing flows to combine the existing individual chains \cite{Raveri:2024dph, Raveri:2021wfz, gatti2024}. Normalizing flows learn the invertible and differentiable mapping between each individual posterior distribution and a multivariate Gaussian. Our procedure is: (i) train a normalizing flow on each posterior distribution to be combined; (ii) assuming conditionally independent data sets and a common (uniform) prior across experiments, define the joint posterior, up to normalization, as the product of the per-probe densities represented by the trained flows; (iii) sample this joint posterior within \textsc{Cobaya} \citep{Torrado2021} with \textsc{CosmoMC} or \textsc{Nautilus} \citep{lange23}. This enables combination without rerunning chains for the full joint likelihood. The normalizing flow model is implemented in the
 {\tt tensiometer}\footnote{\url{https://github.com/mraveri/tensiometer}} package~\cite{Raveri:2021wfz, RaveriHu}. 

\subsection{Goodness-of-fit and tension metrics}
\label{sec:tension}

From each Markov chain, we extract three summary statistics to express the parameter constraints and the goodness of fit to the model.
\begin{itemize}
\item 1D posterior mean and uncertainty: mean of the projected posterior with $68\%$ confidence intervals (CL).   
\item \textit{Maximum a posteriori} (MAP) and uncertainty: derived from posterior-weighted average of 20 MAP searches using the \textsc{MaxLike} algorithm\footnote{\url{https://cosmosis.readthedocs.io/en/latest/reference/samplers/maxlike.html}}. 
For an estimate of uncertainty on MAP point, we compute and show the projected joint highest posterior density (PJ-HPD) of parameters. The PJ-HPD finds the highest posterior density region that contains 68\% of a parameter's marginal posterior mass \cite{joachimi21} (see their Section 6.4). By definition, it includes the MAP and large differences between the marginal 1D posteriors and the PJ-HPD indicate where projection or volume effects may be significant.
\item Best $\chi^2$: for each of our fiducial models we select the 1 MAP search out of 20 that obtains the highest posterior value. The corresponding $\chi^2$ is referred to as the best fit. When calculating reduced $\chi^2$ and $p-$values based on this point, we evaluate an effective number of parameters as $N_{\rm p, eff} = \mathrm{Tr}\left [ C^{-1}_{\Pi} C_{\rm \mathcal{P}} \right ]$, where $C_{\Pi}$ and $C_{\mathcal{P}}$ are the covariance matrix of the posterior and prior respectively in parameter space \citep{RaveriHu}. The number of degrees of freedom is defined as $\nu = \text{(length of data vector)} - N_{\rm p, eff}$. 
\end{itemize}
As each statistic provides different information, we use their complementarity to help disentangle projection effects (see the discussion in Section \ref{sec:results}).

We take several approaches for comparing more than one posterior: 
\begin{itemize}
\item For assessing consistency between our results and uncorrelated external data, we use a full parameter space posterior-based metric. For further discussion and validation see \cite*{y3-tensions}. A key aspect of the full parameter-space-based method is the way we evaluate the probability of a shift in parameters, defined as:
\begin{equation} \label{Eq:ParamShiftProbability} \Delta = \int_{P(\Delta \mathbf{p})\geq P(0)} P(\Delta \mathbf{p}) \, d\Delta\mathbf{p}, 
\end{equation} 
where $P(\Delta \mathbf{p})$ is the parameter difference probability density as defined in ~\cite{Raveri:2021wfz, RaveriHu} 
such that
\begin{equation} \label{Eq:ParameterDifferencePDF}
P(\Delta \mathbf{p}) = \int P(\mathbf{p} | A)\, P(\mathbf{p} -\Delta \mathbf{p}|B) \,d\mathbf{p}.
\end{equation}
Here $\Delta \mathbf{p}\equiv \mathbf{p}_A - \mathbf{p}_B$ are the samples differences between two independent data sets, A and B. The quantity defined in Equation~\ref{Eq:ParamShiftProbability} corresponds to the posterior probability above the iso-density contour associated with no parameter shift ($\Delta \mathbf{p} = 0$). Because the parameter difference distribution, $P(\Delta \mathbf{p})$, is n-dimensional \textemdash with n corresponding to the total number of parameters describing the assumed theoretical model \textemdash and because in practice we only have discrete posterior samples, evaluating Equation~\ref{Eq:ParamShiftProbability} is computationally demanding. In the DES Y3 analysis, we employed Kernel Density Estimate (KDE) method to compute $\Delta$ as a Monte-Carlo integral.
The full parameter shift is then reported in terms of the effective number of standard deviations, i.e.\ the number of standard deviations that an event with the same probability would have had if it had been drawn from a Gaussian distribution \citep{RaveriHu}:
\begin{equation} 
n_\sigma \equiv \sqrt{2}\,{\rm erf}^{-1}(\Delta)\,,
\label{Eq:EffectiveSigmas}
\end{equation}
where ${\rm erf}^{-1}$ is the inverse error function. We set $n_\sigma < 3$ as the criterion for different data sets being consistent.

In this analysis, we adopt a novel approach that has already been applied to several previous analyses within the DES collaboration \cite{gatti2024, DES:2025bxy}. Specifically, we use normalizing flows to perform the tension integral, as outlined in~\cite{Raveri:2021wfz}. In summary, we first train a normalizing flow on samples from $P(\Delta \mathbf{p})$, then we evaluate the tension integral with a Monte Carlo integral estimator (see also Section IV.F in \citepalias{y6-3x2pt}). It has been shown that normalizing flows outperform KDE approaches, since they do not suffer from the curse of dimensionality and do not depend on a kernel bandwidth choice \citep{Raveri:2021wfz}.

\item In addition to the full-dimensional metric, we also consider a simple shift in marginalized mean $S_8$ in terms of the projected error bar $\sigma$, $\Delta S_8 \equiv (S_{8,1} - S_{8,2})/\sqrt{\sigma_{S_8, 1}^2 + \sigma_{S_8, 2}^2}$. As for the full parameter space metric, we set $\Delta S_8 < 3$ as the criterion for consistency.

 \item For comparing between models on the Y6 data later in the paper, we will use Bayesian evidence estimates ($\mathcal{Z}$) given by \textsc{Nautilus} (see \cite{lange23} for details and validation). 

\end{itemize}

\begin{figure*}
    \begin{center}
    \includegraphics[width=0.9\columnwidth]{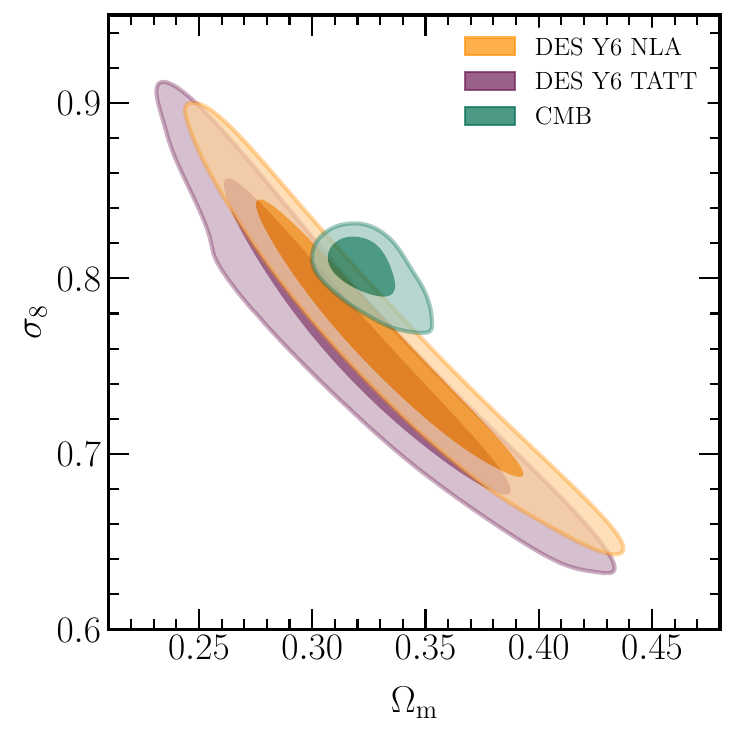}
    \includegraphics[width=0.9\columnwidth]{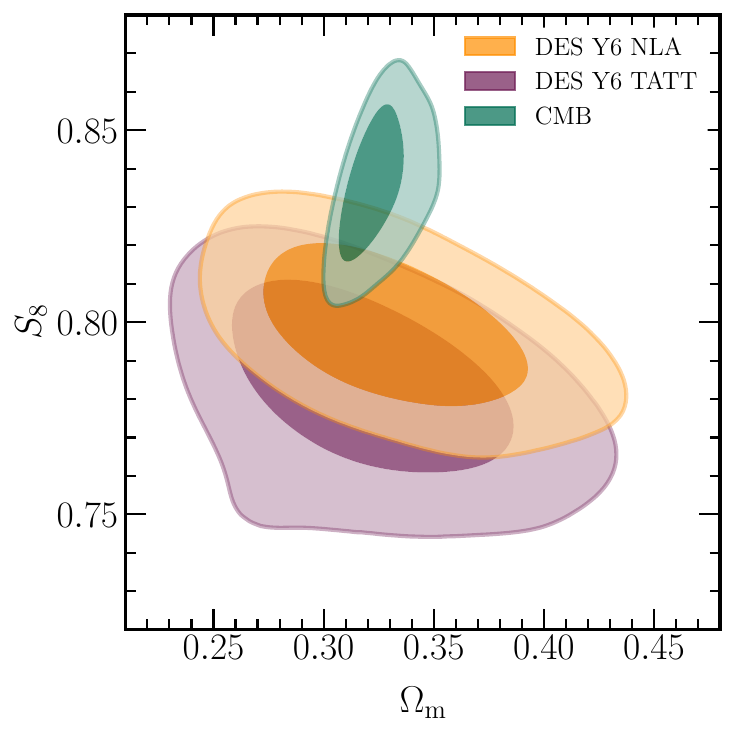}
    \end{center}
    \vspace{-0.75cm}
    \caption{
    Posteriors on cosmological parameters from DES Y6 and a combination of CMB experiments (\textit{Planck} 2018, ACT DR6 and SPT-3G, TT+TE+EE+lowE no lensing; green). The orange and purple contours represent our fiducial results with NLA and TATT analysis choices (described in Section \ref{sec:method}). In all cases, we show $68\%$ and $95\%$ confidence levels. In the full parameter space, the difference between DES Y6 cosmic shear and CMB is 1.1$\sigma$ for NLA and 1.7$\sigma$ for TATT.
    }
    \label{fig:results_fiducial}
\end{figure*}

\section{External data}\label{sec:extdata}

The following describes a selection of contemporary external data sets, with which we will compare and combine our fiducial results in later sections. Where possible we reanalyze using our modeling choices, facilitating quantitative comparison. For more details see the individual references.

\begin{itemize}
    \item \textbf{Fiducial CMB:} We use a combination of modern CMB temperature and polarisation results. Specifically our CMB combination comprises \textit{Planck} 2018 \citep{planck_data}, ACT DR6 \citep{act_data} and SPT-3G DR1 \citep{spt_data}. We use the $TT$, $EE$, and $TE$ power spectra measurements from the three surveys. Cuts are applied to the \textit{Planck} spectra to ensure no covariance between ACT and \textit{Planck}, with $\ell < 1000$ in $TT$ and $\ell < 600$ in $TE$ and $EE$. For ACT we keep $\ell>600$ in $TT$, assuming no correlation between \textit{Planck} and ACT in the range $600 < \ell < 1000$ \citep{act_data}. In addition, we use low$-\ell$  $EE$ \textit{Planck} measurements to constrain $\tau$. Note that we opt to use the ``lite'' versions of the ACT and \textit{Planck} likelihoods, which avoids the need to marginalize over all the nuisance parameters. Finally, for SPT we use the cuts outlined in \citep{spt_data}: $400 < \ell < 3000$ for $TT$ and $400 < \ell < 4000$ for $TE$ and $EE$. 
    We analyse this joint CMB data set with our parameter space and priors (which includes a free neutrino mass).
    \item \textbf{CMB Lensing:} We include the combination of CMB lensing measurements presented in \citep{qu25}, which includes \textit{Planck} PR4 \citep{carron22}, ACT DR6 \citep{madhavacheril25} and SPT-3G ``M2PM" \citep{ge25}. Following \citep{qu25} we use the \blockfont{actplanckspt3g\_extended} version of the likelihood, over 40 $\leq \ell \leq$ 1300 for ACT. 
    \item \textbf{BAO:} We use measurements from the DESI Data Release 2 (DR2; \cite{desidr2}). Specifically, the data includes isotropic BAO for the Bright Galaxy Sample (BGS), and both isotropic and Alcock-Paczynski scaling parameters for the Luminous Red Galaxy (LRG), Emission Line Galaxy (ELG) and Quasar (QSO) sample -- LRG1, LRG2, LRG3+ELG1, ELG2, QSO and Lyman-$\alpha$ samples. The effective redshift range is $0.295 < z < 2.330$.
    \item \textbf{KiDS-Legacy cosmic shear:} The final Kilo-Degree Survey data (KiDS-Legacy or KiDS-DR5) contains $\sim$43M galaxies over 967 deg$^2$, post-masking, with an effective density $n_{\rm eff}=8.79$ galaxies/arcmin$^2$ \citep{wright24}. Galaxies are imaged in four optical bands ($ugri$) and five near-infrared bands ($ZYJHK_s$). Their fiducial shear analysis \citep{kids-legacy} uses Complete Orthogonal Sets of E/B-Integrals (COSEBIs, \cite{schneider_cosebi}) estimators in six redshift bins. 
    
    \item \textbf{DECADE cosmic shear:} The Dark Energy Camera All Data Everywhere (DECADE) data contains $\sim$170 M galaxies across $\sim$9,000 deg$^2$ independent of DES across $riz$ bands, split into four tomographic bins. When combined with DES Y3, in a joint dataset referred to as ``DECam 13k'', it spans $\sim$270 M galaxies across $\sim$13,000 deg$^2$ at an effective number density of 5.19 galaxies/arcmin$^2$ \citep*{decade, anbajagane25}.
    
    \item \textbf{HSC Y3 cosmic shear:} The Year 3 data release from the Hyper Suprime Camera Strategic Survey Program comprises $\sim$25 M galaxies over a patch of 416 deg$^2$, with $n_{\rm eff} = 15$ galaxies/arcmin$^{2}$. The survey takes images in five optical filters, $grizY$, at a greater depth than KiDS and DES. We compare to their $\xi_\pm$ measurements across 4 redshift bins \cite{li23}.
\end{itemize}

\section{Cosmological Constraints}\label{sec:results}

In this section, we present cosmological parameter constraints from Y6 cosmic shear. We present the results in terms of the commonly used $S_8$ parameter, defined as $S_8 \equiv \sigma_8 (\Omega_{\rm m}/0.3)^{\alpha}$, the combination of the matter density and the amplitude of matter fluctuations with $\alpha=0.5$. 
We report the mean in each parameter, along with the 68\% confidence limit (CL) of posterior volume around the mean, and \textit{maximum a posteriori} (MAP) point and PJ-HPD shown in parentheses (Section~\ref{sec:tension} for more detail). We also compute the figure-of-merit (FoM) in ($S_8, \Omega_{\rm m}$) and ($\sigma_8, \Omega_{\rm m}$) spaces, which can be described as:
\begin{eqnarray}
\text{FoM}_{\Omega_{\rm m}, \,S_8} \equiv \left( \det{\text{Cov}(\Omega_{\rm m}, S_8)} \right)^{-1/2} \nonumber \\
\text{FoM}_{\Omega_{\rm m}, \,\sigma_8} \equiv \left( \det{\text{Cov}(\Omega_{\rm m}, \sigma_8)} \right)^{-1/2} .
\end{eqnarray}

\subsection{Fiducial results}\label{subsec:fiducial}
Our fiducial analyses are shown in Figure~\ref{fig:results_fiducial} and Table~\ref{tab:results}. For these two sets of analysis choices, we find the marginalized mean (and MAP + PJ-HPD) values of $S_8, \Omega_{\rm m}, \sigma_8$ to be 

\begin{align*}
S_{8} = 0.798^{+0.014}_{-0.015}\,\quad (0.803^{+0.017}_{-0.014}) \notag \\
\textrm{NLA}: \Omega_{\rm m} = 0.332^{+0.035}_{-0.044}\,\quad (0.328^{+0.037}_{-0.042}) \notag \\
\sigma_{8} = 0.763^{+0.050}_{-0.057}\,\quad (0.768^{+0.064}_{-0.048}) \\ \notag \\  
S_{8} = 0.783^{+0.019}_{-0.015}\,\quad (0.763^{+0.032}_{-0.007}) \notag \\
\textrm{TATT}: \Omega_{\rm m} = 0.321^{+0.036}_{-0.047}\,\quad (0.377^{+0.026}_{-0.083})  \notag \\
\sigma_{8} = 0.763^{+0.053}_{-0.062}\,\quad (0.681^{+0.116}_{-0.018}) . 
\end{align*}
We find 1.8\% and 2.5\% fractional uncertainty on $S_8$ with $\rm FoM_{\Omega_{\rm m}, \,S_8} = $ 2080 (NLA) and 1532 (TATT), and $\rm FoM_{\Omega_{\rm m}, \,\sigma_8} = $ 1903 (NLA) and 1450 (TATT). The TATT model has additional parameters and flexibility, which weakens the cosmological constraints. Comparing the two results, the TATT model prefers a slightly lower but consistent $S_8$ value, with $\Delta S_8 = 0.6\sigma$. Note this small downward shift is similar to that seen in some previous comparisons \citep[e.g.][]{des-kids, y3-cosmicshear1, y3-cosmicshear2}. 

\begin{table*}
\centering
\begin{tabular}{lcccccccc}

\hline \hline 
Data & $S_8$ & $\hat{S_8}$ & $\Omega_{\rm m}$ & $\sigma_8$ & FoM$_{S_8,\Omega_{\rm m}}$ & FoM$_{\sigma_8,\Omega_{\rm m}}$ & $\chi^2 / \rm dof$ \\ [0.2cm]
\hline \\

DES Y6 fiducial (NLA) & $ 0.798^{+0.014}_{-0.015} $ & $ 0.803^{+0.017}_{-0.014} $ &  $ 0.332^{+0.035}_{-0.044} $  & $ 0.763^{+0.050}_{-0.057} $  & 2080 & 1903 & 299.43 / 263.49 = 1.14  \\ [0.2cm] 
DES Y6 fiducial (TATT) & $ 0.783^{+0.019}_{-0.015} $ & $ 0.763^{+0.032}_{-0.007} $ &  $ 0.321^{+0.036}_{-0.047} $  & $ 0.763^{+0.053}_{-0.062} $  & 1532 & 1450 & 304.95 / 278.72 = 1.09  \\ [0.2cm] 
\hline \\
CMB & $ 0.836^{+0.012}_{-0.013} $ & $0.834^{+0.015}_{-0.009}$ &  $ 0.324^{+0.007}_{-0.012} $  & $ 0.805^{+0.014}_{-0.007} $  & 8527  & 8994 & --- \\ [0.2cm]
CMB Lensing & $ 0.865^{+0.048}_{-0.044} $ & $ 0.886^{+0.058}_{-0.047} $  &  $ 0.378^{+0.064}_{-0.096} $  & $ 0.780^{+0.044}_{-0.048} $ & 983 & 1069 & ---  \\ [0.2cm] 
\hline \\
DECADE + DES Y3* & $ 0.805^{+0.019}_{-0.019} $ & --- &  $ 0.262^{+0.023}_{-0.036} $  & $ 0.867^{+0.063}_{-0.063} $  & 1872 & --- & 774.3/681 = 1.14  \\ [0.2cm] 
KiDS-Legacy* & $ 0.815^{+0.016}_{-0.021} $ & $ 0.811^{+0.022}_{-0.015} $ & ---  & --- & ---  & --- & 127.8/120.5 = 1.06  \\ [0.2cm]
HSC Y3* & $ 0.769^{+0.031}_{-0.034}  $ & ---  & $ 0.256^{+0.056}_{-0.044} $  & $  0.818^{+0.089}_{-0.091} $  & --- & --- & 150/134 = 1.12  \\ [0.2cm] 
DES Y3* & $ 0.759^{+0.025}_{-0.023} $ & --- &  $  0.290^{+0.039}_{-0.063} $  & $ 0.783^{+0.073}_{-0.092} $  & 927 & --- & 237.7/222 = 1.07  \\ [0.2cm] 
\hline\hline
\end{tabular}
\caption{Summary of marginalized parameter constraints in \lcdm\ derived from DES Y6 cosmic shear measurements in comparison with other data. The mean and 68\% CL are provided for each cosmological parameter. $\hat{S_8}$ represents \textit{maximum a posteriori} point and the projected joint highest posterior density interval (PJ-HPD).  Note that the constraints from previous cosmic shear analyses are as published, rather than reanalyzed, indicated with $*$, and the analysis choice for each survey can be found in the corresponding text.}
\label{tab:results}
\end{table*}

\subsection{Comparison with external data \& probes}\label{subsec:ext}

In the following, we assess the consistency of our results with a range of other cosmological probes. Where relevant, we compute tension metrics and adopt the same $p-$value threshold of 0.01 as in DES Y3 \citep*{y3-cosmicshear2} to define consistency.

\subsubsection{CMB}\label{subsec:CMB}

We begin by quantifying the difference between DES Y6 cosmic shear and the combination of CMB measurements described in Section \ref{sec:extdata}\footnote{Note that this CMB constraint is reanalyzed with our parameter space and priors, including a free neutrino mass. Our DES Y6 results where we assume a normal hierarchy with the heaviest neutrino mass fixed to 0.06 eV are $S_8=0.803^{+0.014}_{-0.014}$ and $S_8=0.789^{+0.020}_{-0.014}$, for NLA and TATT, respectively (shown in Row 21 of Figure~\ref{fig:results_lensing}).}. As is apparent from Figure \ref{fig:results_fiducial}, there is some offset in certain parameters. For NLA vs CMB $\Lambda$CDM, in the full parameter space and 1D $S_8$ we find:

\begin{eqnarray*}
\textrm{Full space shift:} & \qquad\quad\qquad\quad\quad 1.1\sigma \,\, (p=0.28)\\
\textrm{1D $\Delta S_8$ shift:} & \qquad\quad 2.0\sigma \,.
\end{eqnarray*} 
\noindent
And for TATT vs CMB $\Lambda$CDM we obtain:
\begin{eqnarray*}
\textrm{Full space shift:} & \qquad\quad\qquad\quad\quad 1.7\sigma \,\, (p=0.09)\\
\textrm{1D $\Delta S_8$ shift:} & \qquad\quad 2.3\sigma \, .
\end{eqnarray*}
\noindent

Overall, DES prefers a lower $S_8$ than the CMB primary constraints, but the differences are only at the level of about $\lesssim 2\sigma$. For this comparison with the CMB, we also compute an alternative evidence-based tension metric known as `Suspiciousness' \citep{y3-tensions,handley19}. This gives very similar results to the parameter shift statistic, with $0.9\sigma$ ($p=0.35$) and $1.4\sigma$ ($p=0.17$) respectively for NLA and TATT. Both of our fiducial analyses comfortably meet the $p-$value threshold to be considered consistent with the external CMB data, and therefore we can combine our two fiducial shear results with the CMB. The combination gives us a final joint constraint of: 
\begin{eqnarray*}
\textrm{NLA}: S_{8} & = & 0.817^{+0.009}_{-0.008},\,\\
\textrm{TATT}: S_{8} & = & 0.812^{+0.009}_{-0.009}.
\end{eqnarray*}

As one might expect from Figure \ref{fig:results_fiducial}, the joint results lie between those obtained from analyzing the CMB and cosmic shear data independently. We obtain a $1\%$ constraint on $S_8$, with an error bar that is roughly independent of the choice of IA model. The two fiducial models are offset by $\Delta S_8 = 0.4\sigma$, which is similar to the $0.6\sigma$ difference discussed in Section \ref{subsec:fiducial}.

\subsubsection{Low-$z$ measurements: CMB lensing, BAO and RSD}\label{cmb_lens}

We compare our cosmic shear constraints with those from other low-$z$ observations, such as CMB lensing, baryon acoustic oscillations (BAO) and redshift space distortion (RSD). 

\begin{figure}
    \centering
    \includegraphics[width=0.9\columnwidth]{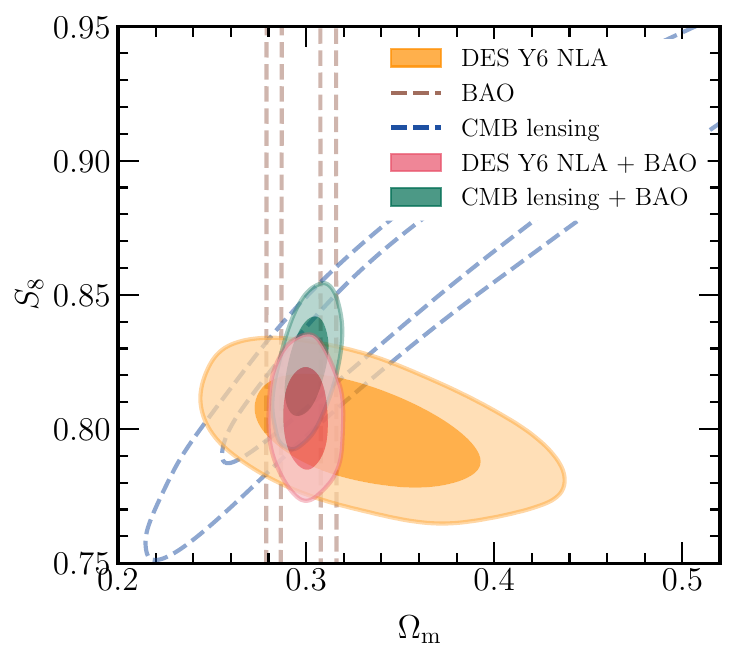}
  %  \vspace{-0.75cm}
    \caption{DES Y6 cosmic shear (orange) compared with DESI DR2 BAO result (brown, dashed), CMB lensing (blue, dashed), the combination of our fiducial Y6 cosmic shear with NLA analysis with DESI BAO (pink), and the combination of CMB lensing (\textit{Planck} + ACT + SPT) with DESI BAO (green). This tests the consistency of cosmological constraints probed with galaxy lensing and CMB lensing, when anchored to the same $\Omega_{\rm m}$, as constrained by the BAO. }
    \label{fig:ext_lowz}
\end{figure}

First, we compare our constraints to those from CMB lensing as a probe of the same physical process at different redshifts. As these measurements have different parameter degeneracies (blue and orange contours in Figure \ref{fig:ext_lowz}), we combine both DES cosmic shear and CMB lensing (see Section \ref{sec:extdata} for details) with an external BAO measurement from Dark Energy Spectroscopic Instrument (DESI; \citep{desi}) DR2 \citep{desi_dr2_cosmo}. The BAO provides an anchor on $\Omega_{\rm m}\sim0.3$ (brown dashed lines in Figure~\ref{fig:ext_lowz}), allowing us to compare the $S_8$ constraints. The pink and green contours in Figure~\ref{fig:ext_lowz} show the $S_8-\Omega_{\rm m}$ constraints from the combined probes. We avoid computing the full tension metrics for these combined probes since our tension metrics assume the datasets are independent, and we expect that cosmic shear and CMB lensing are covariant. Hence, we only compute the difference in marginalized mean $S_8$. Our fiducial shear analyses compared with CMB lensing are consistent:
\begin{eqnarray*}
\textrm{1D $\Delta S_8$ shift:} & \qquad\quad 1.5\sigma \qquad[\textrm{NLA}], \\
\textrm{1D $\Delta S_8$ shift:} & \qquad\quad 1.7\sigma \qquad[\textrm{TATT}].
\end{eqnarray*} 

When combined with DESI DR2 BAO, the difference becomes
\begin{eqnarray*}
\textrm{1D $\Delta S_8$ shift:} & \qquad\quad 1.1\sigma \qquad[\textrm{NLA}], \\
\textrm{1D $\Delta S_8$ shift:} & \qquad\quad 1.7\sigma \qquad[\textrm{TATT}].
\end{eqnarray*} 

Overall, considering low-$z$ measures of the structure growth using the BAO and CMB lensing, we find consistency. Comparing with the recent full shape analysis of the clustering of galaxy, quasar and Lyman-$\alpha$ forest tracers in the DESI Data Release 1 (DR1) combined with its BAO measurement \citep[$S_8=0.836 \pm 0.035$;][]{desi_fs, desi_bao_dr1}, our DES Y6 results are consistent with $\Delta S_8 = 1.0\sigma$ and $1.3\sigma$, for our NLA and TATT analysis, respectively. Comparing our cosmic shear results with the published constraints from the cross-correlation of CMB lensing maps with galaxy positions of unWISE \citep[$S_8=0.810 \pm 0.015$;][]{farren24} and DESI \citep[$S_8=0.775^{+0.019}_{-0.022}$;][]{sailer25}, we also find consistency at $0.6\sigma$ and $1.0\sigma$ for our NLA analysis and $1.1\sigma$ and $0.3\sigma$ for our TATT analysis.

\subsubsection{Weak lensing surveys}\label{subsec:lensing}

\begin{figure}
    \centering
    \includegraphics[width=\columnwidth]{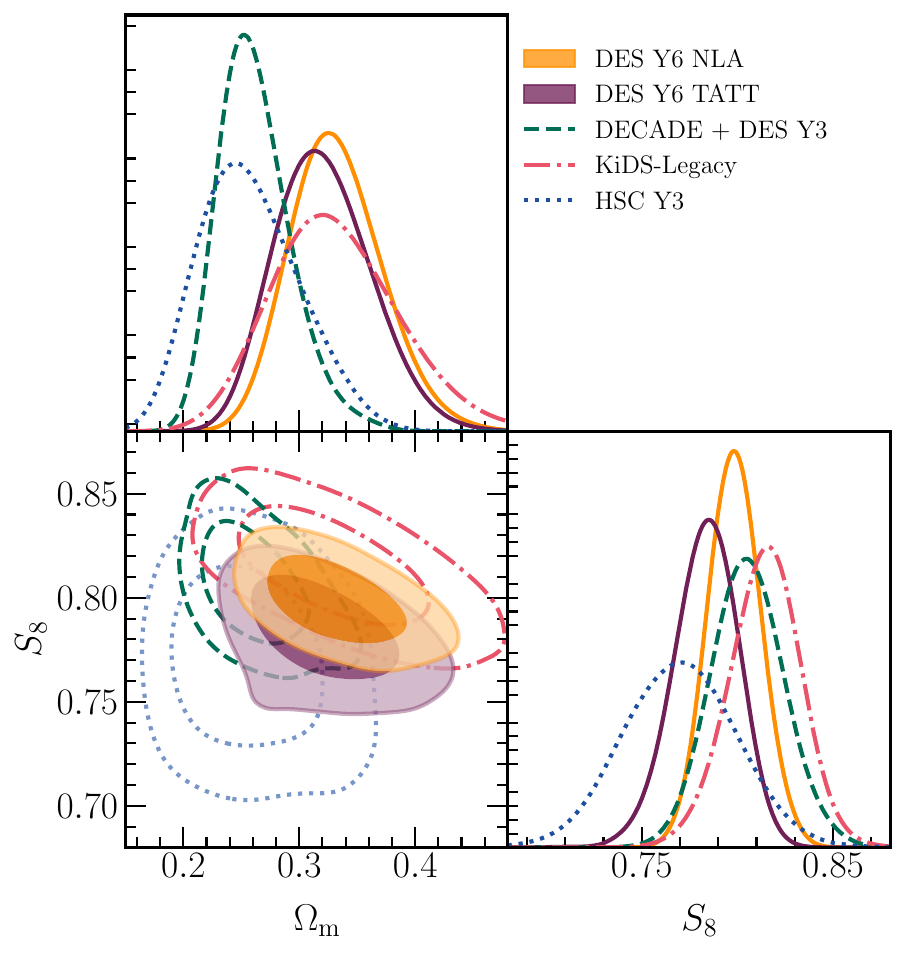}
    \caption{ DES Y6 results (filled orange; NLA and purple; TATT) compared with other lensing survey results - DECADE+DES Y3 (dashed green, \cite{decade}), KiDS-Legacy (dash-dot pink, \cite{kids-legacy}), and HSC Y3 (dotted blue, \cite{li23}). }
    \label{fig:results_lensing}
\end{figure}

\begin{figure}
    \centering
    \includegraphics[width=0.9\columnwidth]{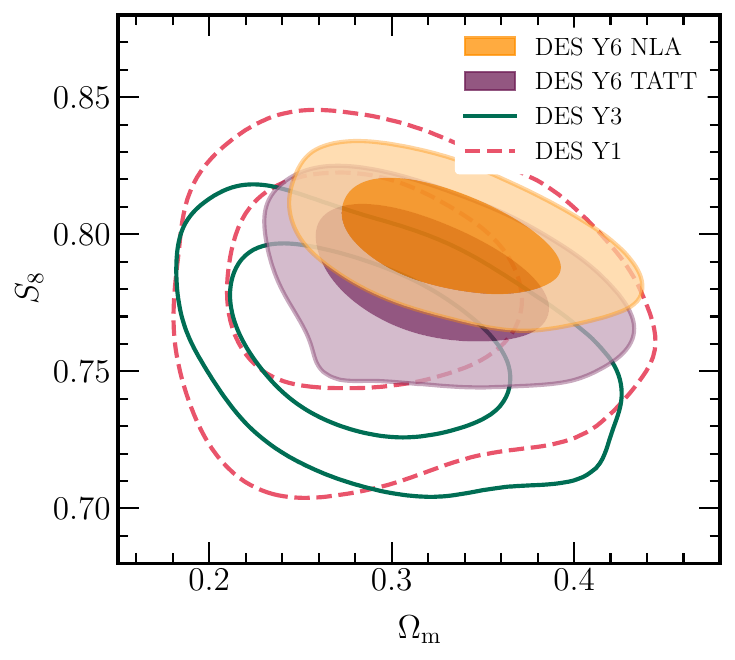}
    \caption{DES Y6 fiducial results (orange and purple) compared with previous DES results from Y1 (dashed pink, \cite{troxel18}) and Y3 (solid green, \cite{y3-cosmicshear1, y3-cosmicshear2}) cosmological constraints. }
    \label{fig:des_results}
\end{figure}

We now compare our Y6 results with other weak lensing surveys that are described in Section~\ref{sec:extdata}. 
Figure~\ref{fig:results_lensing} shows the Y6 results compared with the published DECADE+DES Y3 \citep{decade}, KiDS-Legacy \citep{kids-legacy}, and HSC Y3 real-space cosmic shear analyses \citep{li23}. Since different analyses differ in their priors and other analysis choices (e.g. measurement statistics, scale cuts), direct comparison is complicated, and we caution readers against drawing conclusions on consistency from visual examination of contours. Specifically, the fiducial DECADE+DES Y3 analysis uses TATT IA model and non-linear matter power spectrum computed with \textsc{HMCode2020}. The KiDS-Legacy utilizes the COSEBIs estimator with a NLA variant model that depends on galaxy sample's physical property along with \textsc{HMCode2020} non-linear matter power spectrum with free AGN feedback strength parameter, as their fiducial analysis choices. Finally, the fiducial HSC Y3 analysis uses \textsc{HMCode2016} non-linear matter power spectrum with free baryonic feedback strength parameter, and TATT IA model. All the analyses derive their scale cuts according to their fiducial analysis model choices. Despite these differences, there is still a general tendency amongst lensing surveys towards a lower $S_8$ than the CMB result. Although the individual differences have narrowed to $\lesssim 2\sigma$ in recent years, assigning a significance to this collective tendency is beyond our current scope for the reasons listed above.

\subsubsection{Previous DES weak lensing results}\label{subsec:des_lensing}

Figure~\ref{fig:des_results} shows our Y6 NLA and TATT Y6 results compared with the published DES Y1 and Y3 cosmic shear constraints \cite{y3-cosmicshear1, y3-cosmicshear2, troxel18}. Compared to DES Y3 ($S_8=0.759^{+0.025}_{-0.023}$)\footnote{Note that we are comparing with the DES Y3 ``$\Lambda$CDM fiducial" shear analysis from \cite{y3-cosmicshear1,y3-cosmicshear2}.}, which also used the TATT IA model, our DES Y6 TATT analysis finds a higher value of $S_8$ and $\Omega_{\rm m}$. There are a number of changes in our analysis choices that contribute to this shift. The most important of these is the model for the non-linear matter power spectrum, and the associated scale cuts. For the gravity-only power spectrum, the change from \textsc{Halofit} \citep{takahashi12} to the more accurate \textsc{HMCode2020} has been demonstrated to shift the Y3 constraint by $\sim$1-2$\sigma$ \citep{des-kids}. In addition, the DES Y3 analysis used the OWLS-AGN hydrodynamical simulation to define the angular scale cuts, while DES Y6 chooses the BAHAMAS-8 simulation. The latter has a more strongly suppressed matter power spectrum, resulting in more conservative scale cuts. We conclude that the combination of a more accurate non-linear matter power spectrum estimator, which also includes the impact of baryonic feedback, and scale cuts that account for a wider range of feedback scenarios push our $S_8$ constraints higher than our previous analyses. 

Despite the shift to higher values of $S_8$ from the DES Y6 data, the TATT analysis gives the same $\Delta S_8=2.3 \sigma$ parameter shift compared to the CMB value as found in DES Y3. We attribute this to two factors. First,  we observe $\sim \times 2$ improvement in the constraining power in Y6 in terms of the figure-of-merit in ($S_8, \Omega_{\rm m}$) resulting from the increased depth, effective number density and redshift.
Second, the DES Y3 result was compared to the \textit{Planck} 2018 constraint, while here we use the combined CMB result from \textit{Planck}+ACT+SPT (see Section \ref{sec:extdata}), which has a smaller uncertainty and also shifts to higher values of $S_8$.

\subsection{Contribution to the DES Y6 uncertainty}\label{subsec:limiting_sys}

\begin{figure}
    \begin{center}
    \includegraphics[width=\columnwidth]{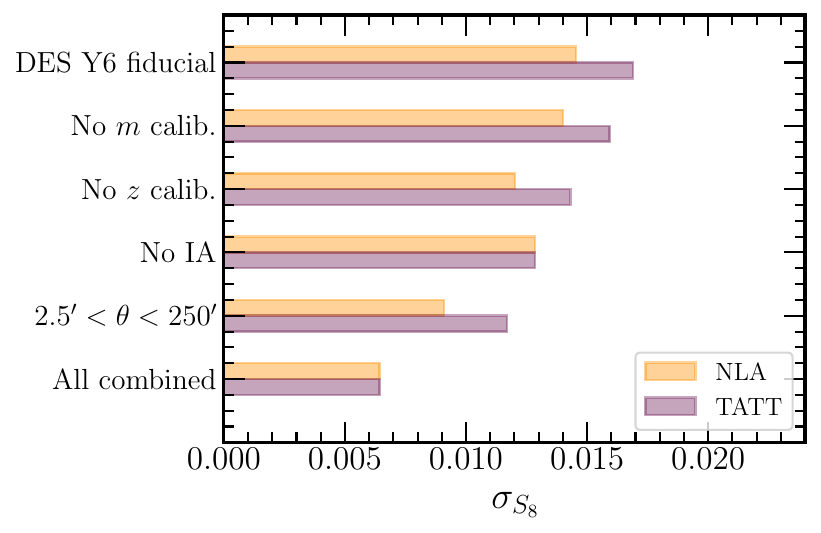}~
    \end{center}
    \vspace{-0.75cm}
    \caption{\label{fig:limiting_2d_1d} Estimating the cosmological constraining power lost in the DES Y6 cosmic shear analyses due to systematic uncertainties. The fiducial Y6 $\Lambda$CDM uncertainty (symmetric) in $S_8$, $\sigma_{S_8}$ for NLA (orange) and TATT (purple) are compared to constraints obtained when the nuisance parameters for each systematic are fixed (i.e., IA modeling, and calibrations of shear measurement and redshift distributions), and/or when the scale cuts, designed to mitigate the impact of baryon feedback, are removed. The uncertainty from the DES Y3 analysis, $\sigma_{S_8}=0.024$ \citep{y3-cosmicshear1}, is shown as the right edge of the plot.
    }
\end{figure}

Cosmic shear cosmological analyses demand robust mitigation of systematic effects to avoid biases in constraints. In this section, we assess the contribution to the uncertainty in our fiducial results. For the future surveys to take advantage of their low measurement noise, it is helpful to understand what systematic biases and mitigation strategies prevent us from achieving the maximum precision on cosmological parameters. To explore this, we repeat the cosmological inference while fixing the nuisance parameters one at a time or changing the scale cuts (assuming all systematic effects on the additional scales can be modelled perfectly and without additional free parameters). We then compute the changes in the symmetric uncertainty in the 1D $S_8$ marginalized posterior shown in Figure~\ref{fig:limiting_2d_1d}, compared to the fiducial NLA (orange, $\sigma=0.015$) and TATT (purple, $\sigma=0.017$) constraints.

First, we assess the impact of fixing several nuisance parameters in our analysis. If we assume perfect knowledge of our shear measurements, by fixing the shear calibration parameters (`No $m$-calib.'), we find a minimal change in the constraining power, at $\sigma=0.014$ and $\sigma=0.016$ for NLA and TATT, respectively. If we assume to have perfect knowledge of our redshift distribution by fixing the redshift uncertainty parameters (`No $z$-calib.'), we find a $\times 1.2 $ gain in constraining power for NLA ($\sigma=0.012$) and TATT ($\sigma=0.014$).

If we assume a perfect knowledge of modeling IA by fixing the IA parameters (`no IA'), we find $\sigma=0.013$, a factor of $\sigma_{\rm fid, NLA}/\sigma=1.1$. The relative gain in fixing IA is larger for the TATT case, as the more flexible IA model, improving by $\sigma_{\rm fid, TATT}/\sigma=1.3$.

Next, we consider the case where we remove the scale cuts to include measured points down to 2.5 arcmin (`$2.5'<\theta<250'$'). As discussed in Section~\ref{subsec:baryons}, the primary motivation for the scale cuts is driven by the uncertainty in the impact of baryonic effects on the small-scale matter power spectrum. Therefore, this simulates the scenario where we can perfectly model the small scales. This means perfect understanding of the baryonic impact on the matter power spectrum, as well as small-scale IA and other observational and astrophysical effects which become more significant on small scales and we do not consider here. We find that including these scales results in a large gain (a factor of $\sigma_{\rm fid, NLA}/\sigma=1.6$ and $\sigma_{\rm fid, TATT}/\sigma=1.4$) in constraining power, with $\sigma=0.009$ and $\sigma=0.012$, respectively. We also note that extending the maximum scale cut to 1000 arcmin results in a negligible change in constraining power for both the NLA and TATT cases.

Finally, we consider the case where all three classes of nuisance parameters are fixed, such that only cosmological parameters are varied, and the small-scale cuts are removed (`All combined'). We find $\sigma=0.006$, a gain of a factor of $\sigma_{\rm fid, NLA}/\sigma=2.3, \sigma_{\rm fid, TATT}/\sigma=2.6$. 
 
To summarize, if we use angular scales between 2.5 and 250 arcmin without incurring systematic uncertainty, there is the potential to gain a factor of $2.3-2.6$ in the constraint on $S_8$ from our cosmic shear measurements. Among the systematics considered in this section, the imposed small-scale causes the most significant loss of constraining power. Though our ability to model the impact of baryonic feedback on the non-linear matter power spectrum is widely recognized as a major model uncertainty, on small physical scales, uncertainties on the dark-matter-only non-linear matter power spectrum, IA, and higher-order lensing distortions will also require further study. We note that for our Y6 analysis, which includes substantially deeper data than in Y3, the uncertainties in modeling of IA and the redshift distributions are also important in setting the overall constraining power. 

\begin{figure*}
    \centering
    \includegraphics[width=\textwidth]{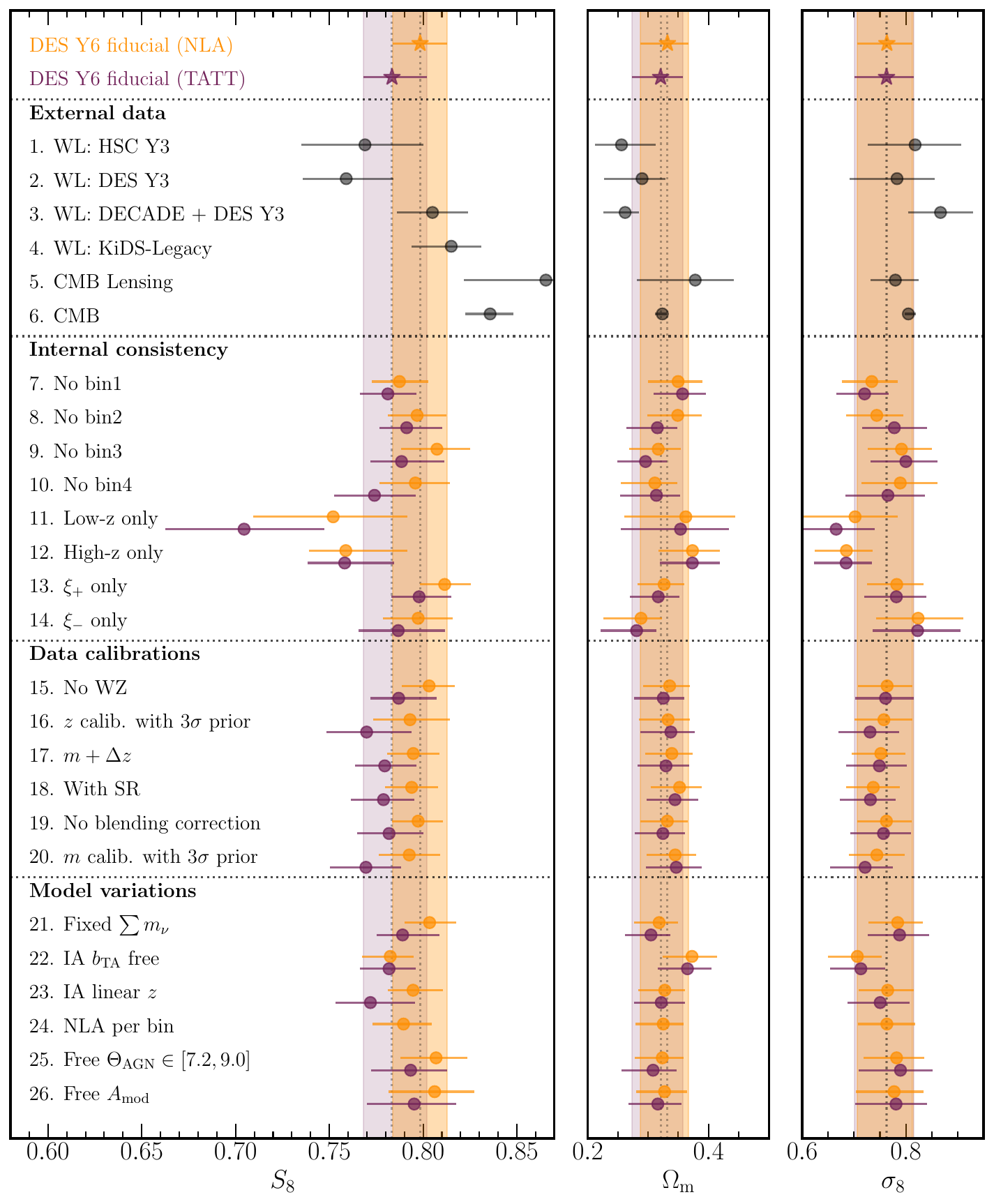}
    \caption{Constraints on $S_8$, $\Omega_{\rm m}$ and $\sigma_8$ from the fiducial cosmic shear results (NLA; orange, TATT; purple), compared to external data described in Section~\ref{sec:extdata} and a number of robustness tests performed in Section~\ref{sec:robustness} for both NLA and TATT analysis choices. All constraints are shown as marginal $1\sigma$ constraints. From top to bottom, we show the comparisons with external data, and we show our tests for including/excluding part of the data vector (removing one bin at a time, splitting the data by redshift, and splitting the data by $\xi_\pm$), different approaches to calibrate the shear measurement and redshift distributions, and astrophysical effects (IA and baryonic feedback).
    }
    \label{fig:results_lensing}
\end{figure*}

\section{Robustness tests}\label{sec:robustness}

In this section, we test the robustness of the DES Y6 cosmological constraints. A summary of the validation tests is shown in Figure~\ref{fig:results_lensing} for both the fiducial analyses that use the NLA IA model (orange) and the TATT model (purple).   

\subsection{Internal consistency}\label{sec:internal_consistency}
We first assess the overall consistency between different parts of our data. We consider 1) removing the two-point correlation function measurements from each tomographic bin, 2) using only $\xi_+$ or $\xi_-$, and 3) using $\xi_{\pm}$ from only low-$z$ (bins 1 and 2) or high-$z$ (bins 3 and 4). The resulting marginalized $S_8, \Omega_{\rm m}, \sigma_8$ constraints are shown in Rows 7-14 of Figure~\ref{fig:results_lensing}. In these 1D projected constraints, we find that for our NLA analysis, all three of these tests give results that are consistent within 1$\sigma$, while for TATT, the low-$z$ analysis compared with the full analysis appears to be more discrepant at around $2\sigma$. We ascribe the slightly larger shift in this case to a combination of factors. First, the redshift split significantly reduces the constraining power of the data, and since IA is more constraining at low-$z$, the split can potentially create or worsen projection effects due to IA parameters that correlate with $S_8$. Second, it is feasible a shift of this kind can result from a random noise fluctuation: given that we perform $\sim 20$ tests, having one result in a $2\sigma$ shift is expected, statistically speaking. 

\subsection{Observational systematics}\label{subsec:obssys_variants}

\subsubsection{Redshift calibration}\label{subsec:obssys_nz}

Our redshift distributions are calibrated using two methods based on photometry (SOMPZ) and clustering cross-correlations (WZ), and we marginalize over their associated calibration uncertainty using mode projection amplitudes, preserving full shape information and tomographic bin correlations (Section~\ref{subsec:redshifts}). Rows 15-18 of Figure~\ref{fig:results_lensing} show how the posteriors for $S_8, \Omega_{\rm m}$ and $\sigma_8$ are impacted by choices in the redshift calibration. 

First, we test the impact of the two calibration strategies. We compare our fiducial result to that obtained when the redshift distributions are calibrated with SOMPZ only, neglecting WZ information (``No WZ'', Row 15). The posteriors are consistent, and we see that the addition of the WZ information slightly reduces the uncertainty, highlighting the value of constraining $n(z)$ with complementary information.

We also explore the case where we have substantially under-estimated the uncertainty in our redshift distribution, characterized as having the mode amplitude $u_i$ to have a uniform prior with 3$\sigma$ width (``$z$ calib. with 3$\sigma$ prior''; Row 16). Here TATT shows a larger shift than NLA. We can understand this shift as a projection in 1D $S_8$; in 2D ($S_8, \Omega_{\rm m}$) plane, the shift is $\sim 0.4\sigma$. Although this significantly reduces our constraining power, the constraints are consistent with our fiducial analysis.

Next, we study the impact of the way we account for uncertainties in the redshift distributions. 
We compare our fiducial result using the new mode sampling methodology to the result when sampling the shift in the mean redshift, $\Delta z_i$, for each bin (``$m+\Delta z$'', Row 17). To use this method, we recompute $\Delta z_i$ from image simulations, and its covariance with the shear multiplicative bias $m$. We see no noticeable impact. For DES Y6, using higher-order redshift information with mode sampling does not significantly affect our cosmological constraints. Nonetheless, we expect it to be a useful tool for the Stage-IV surveys, where redshift uncertainties are more important.

Finally, we further validate our redshift and shear calibration with the shear ratio (SR) technique \citep{y3-shearratio}. The implementation of SR used here is described and validated in more detail in \citep{y6-gglens} (see their Section 6). In brief, we take the ratio of galaxy-galaxy lensing measurements (i.e. the cross correlation of lens galaxy positions with source galaxy shapes) using two different source bins and the same lens bin. To avoid double counting information, we only consider SR on small scales ($<6 \mathrm{Mpc}/h$), which are not used in the $2 \times 2$pt and $3 \times 2$pt cosmology analyses. In the limit where the lens redshift distribution is infinitely narrow and there is no overlap between the lens and source bins, sensitivity to the distribution of matter around lens galaxies cancels away, providing a powerful independent cross-check on our understanding of the source sample. In this analysis, because of the uncertainty on modeling IA and matter power spectrum on those small scales, we use the ratio of two highest source bins and one lowest lens bin\footnote{No lens bin 2 was used in $2 \times 2$pt and $3 \times 2$pt analyses. Readers may refer to \citepalias{y6-2x2pt, y6-3x2pt} for more detail.} to test our redshift calibration instead of using SR in the fiducial analyses. As in DES Y3, SR is incorporated as an additional likelihood, which is added to the one described in Section \ref{sec:likelihood} at each step in parameter space. Our SR result (Row 18) is consistent with our fiducial result within $\sim 0.3\sigma$ for both the NLA and TATT. The inclusion of this information constrains the calibration parameters, resulting in a 1\% (10\%) improvement in $S_8$ constraining power, for the NLA (TATT) case. We note that the addition of the extra redshift information via the SR likelihood does not appreciably change the posteriors on IA parameters in either fiducial setup.

\subsubsection{Shear calibration}\label{subsec:obssys_shear_psf}

We calibrate the multiplicative bias, $m$, in our galaxy shape measurements, in particular that due to redshift-dependent blending effects, using a multi-band image simulation suite (Section~\ref{subsec:shapes}). Rows 19-20 of Figure~\ref{fig:results_lensing} shows how our fiducial results are impacted by choices in the shear calibration.

We test the case where we ignore the contribution of redshift-dependent blending to the multiplicative bias $m$ (``No blending correction''; Row 19). To do this, we use image simulations where the known constant shear is applied across all redshifts, and we turn off the correlation between $m$ and $u$. Importantly, this analysis tests the effects of neglecting the 0.034 mean shift in redshift of the highest tomographic bin (see Section~\ref{subsec:imagesims}). Our result indicates that our $m$ calibration from image simulations, where we model the coupling between shear and redshifts, has little impact on overall cosmology for the statistical power of our data. Nonetheless, we expect this correction to be more critical for future surveys as the blending fraction increases with deeper surveys.

We also test the impact of inflating the width of the $m$ distribution prior from 1 to 3$\sigma$, as if the uncertainty was substantially underestimated (``$m$ calib. with 3$\sigma$ prior''; Row 20). As is the case with redshift uncertainty, though TATT shows a larger shift than NLA, we also check that the shift in 2D ($S_8, \Omega_{\rm m}$) plane is less than $0.5\sigma$ and conclude this shift is due to projecting the posterior onto $S_8$. Inflating the uncertainty degrades the cosmological precision, but the posterior is consistent with the fiducial result, demonstrating that the analysis is robust to this additional freedom. 

\subsection{Astrophysical systematics}\label{subsec:astro}
\subsubsection{Intrinsic alignments}\label{subsec:robust_ia}

In Section~\ref{sec:results}, we presented results using two fiducial IA models: NLA and TATT. These posteriors differ by 0.6$\sigma$, with TATT preferring lower values of $S_8$. In Table~\ref{tab:model_comparison}, we report $\Delta \chi^2/\Delta {\rm dof}$\footnote{For practical reasons, the model comparison in this section uses the \blockfont{Nautilus} best $\chi^2$ for each model. Note that these $\chi^2$ estimates have some noise, though this is smaller than that on the associated parameter values.} 
(difference in the best-fit per unit difference in degrees of freedom) and the Bayes ratio $R = \mathcal{Z_{\rm model\ 1}}/\mathcal{Z_{\rm model\ 2}}$ ($\mathcal{Z}$: Bayesian evidence) to quantify the preference and fit to the data between model choices\footnote{To facilitate comparison in Table~\ref{tab:model_comparison} and the discussion in Section \ref{subsec:robust_ia} we unify the scale cuts to those validated with NLA. These remove (slightly) more angular scales than the TATT cuts and should be robust for all more complex models than NLA. We confirm that this change does not significantly change either the mean or size of the TATT contours shown in Figure \ref{fig:results_fiducial}.}. When comparing the fiducial variants of NLA and TATT, we find a Bayes ratio of $R=\mathcal{Z}_{\rm NLA}/\mathcal{Z}_{\rm TATT} = 16.9\pm0.2$; according to the Jeffrey's scale \citep{jeffreys39} this constitutes a ``strong" preference for the NLA model. The difference in the best-fit is $\Delta \chi^2/\Delta\mathrm{dof}=2.1$. We note, however, that this preference for NLA is a product of our choice to fix $b_{\rm TA}=1$ in our fiducial TATT model, and not an indication that complexity beyond 2-parameter NLA model is unwarranted. As we will discuss below, the overall data-preferred model is a 3-parameter parameterization ($A_1,\eta_1,b_{\rm TA}$), which is somewhere between the two fiducial models in flexibility. 
One caveat to the evidence ratio is that it depends on the width of the priors assigned to the parameters introduced in more complex models. The prior bounds placed on $A_2$ and $\eta_2$, for example, are not strongly specified by any theory, and if they were each to be tightened or loosened by a factor $\sqrt{2}$, then the evidence ratio for the fiducial TATT vs NLA comparison would change by a factor of 2. This caveat also applies to the evidence for other variant IA models discussed below.

We also consider a number of alternative IA models to test the robustness of our result. We note that our fiducial variants of the NLA and TATT models make particular assumptions which can impact the resulting IA and cosmological constraints. Table~\ref{tab:model_comparison} lists these model variations, as well as the preference of each model relative to the `no IA' case (in terms of $\Delta \chi^2/\Delta {\rm dof}$, $p$-value and Bayesian evidence ratio $R$), and the resulting $S_8$ constraint. We have two forms of model variants here. In the upper half of Table~\ref{tab:model_comparison}, we alter the flexibility of the TATT model by increasing and decreasing the number of free parameters assumed, including the $b_{\rm TA}$ parameter, which we have fixed in our fiducial analyses. In the lower half, we consider different forms of redshift evolution in the IA parameters beyond the power law form in Equations \ref{eq:A1ofz} - \ref{eq:A2ofz}. `NLA per bin' allows for one free IA amplitude per source tomographic bin; `NLA/TATT linear $z$' assumes a linear redshift evolution of the NLA/TATT parameters in the form $z$, $a_{1,2}(z) \propto A_{1,2} + \beta_{1,2} (z - z_0)$ (where $\beta_{1,2}$ are free parameters that replace $\eta_{1,2}$ in Equations \ref{eq:A1ofz} and \ref{eq:A2ofz}). The $S_8-\Omega_{\rm m}$ posteriors for a subset of these variants that add model complexity are shown in the left panel of Figure~\ref{fig:variants_ia_baryons}, and they are also summarized in Rows 22-24 of Figure~\ref{fig:results_lensing}. Overall, we find that the DES Y6 constraint can vary at the $\sim$1$\sigma$ level in $S_8-\Omega_{\rm m}$ plane, depending on the IA model choice, consistent with previous findings \citep[e.g][]{troxel18, y3-cosmicshear1, y3-cosmicshear2, des-kids, li23, dalal23, Samuroff_Y1IA}. 

{When considering the preferences of the data with respect to these different IA models, 
there is no simple trend with model complexity, but we have a number of interesting observations.

\begin{table}
	\centering	 
	\vspace{-0.2cm}
	\begin{tabular}{l|ccccc}
        \hline
        IA Model & $b_{\rm TA}$ & $\Delta \chi^2/\Delta \mathrm{dof}$ & $p-$val. & $R$ & $S_8$ \\
        \hline
        \hline
        None & 0 & 0/0 & 0.06 & 1 & $0.792^{+0.013}_{-0.013}$  \\
        NLA-1 ($A_1$) & 0 &$3.8/0.3$ & 0.07 & $71$ & $0.799^{+0.014}_{-0.014}$ \\
        $\rm NLA^*$ ($A_1$, $\eta_1$) & 0 & $3.5/0.6$ & 0.07 & $41 $ & $0.798^{+0.014}_{-0.015}$ \\
        TA ($A_1$, $\eta_1$) & 1 & $0.24/0.7$ & 0.05 & $1$ & $0.802^{+0.014}_{-0.014}$ \\
        TA no $z$ ($A_1$, $b_{\rm TA}$) & [-2,2]  & $11.9/0.7$ & 0.13 & $143$ & $0.788^{+0.014}_{-0.015}$ \\
        TA-3 ($A_1$, $\eta_1$, $b_{\rm TA}$) & [-2,2] &  $13.9/1.9$ & 0.15 & $232$ & $0.782^{+0.014}_{-0.015}$ \\
        TATT-3 ($A_1$, $A_2$, $\eta_1$) & 1 & $6.6/1.2$ & 0.08 & $3$ & $0.789^{+0.017}_{-0.016}$ \\
        TATT no $z$ ($A_1$, $A_2$) & 1 &   $5.0/1.1$ & 0.08 & $4$ & $0.792^{+0.015}_{-0.015}$ \\
        TATT no $z$ ($A_1$, $A_2$) & 0 &   $11.9/0.9$ & 0.13 & $95$ & $0.787^{+0.015}_{-0.015}$ \\
        $\rm TATT^*$  & 1 &  $5.3/1.5$ & 0.07 & $2$ & $0.786^{+0.018}_{-0.018}$ \\
        TATT  & 0  & $14.0/1.5$ & 0.14 & $114$ & $0.779^{+0.016}_{-0.016}$ \\
        TATT + $b_{\rm TA}$ & [-2,2] &  $15.4/2.7$ & 0.14 & $104$ & $0.782^{+0.016}_{-0.016}$ \\
        \hline
        NLA per bin  & 0  & $5.0/2.1$ & 0.07 & $5$ & $0.790^{+0.015}_{-0.017}$ \\
        NLA linear $z$  & 0  & $4.0/0.9$ & 0.07 & $31$ & $0.795^{+0.016}_{-0.014}$ \\
        TATT linear $z$  & 1  & $6.0/1.6$ & $0.08$ & $2$ & $0.773^{+0.023}_{-0.018}$ \\
        \hline
	\end{tabular}
        \caption{A summary of IA complexity tests presented Section \ref{subsec:insights_ia}. A * next to the IA model name represents our fiducial model choices. All analyses were run using the NLA scale cuts, which use fewer scales than the TATT ones. In each case we show (left to right) the value of $b_{\rm TA}$ assumed in the model; the reductions in the best $\chi^2$ and effective degrees of freedom (both relative to the no IA case $\chi^2/$dof=300.6/263.2); the associated $p-$value; the Bayesian evidence ratio with respect to the no IA case $R$; the marginalized mean value $S_8$ $\pm 1 \sigma$ confidence intervals. Note that the models in the upper section are nested sub-spaces of TATT, whereas the lower three are reparametrisations. 
        }\label{tab:model_comparison}
\end{table}

\begin{figure*}
    \begin{center}
    \includegraphics[width=0.4\textwidth]{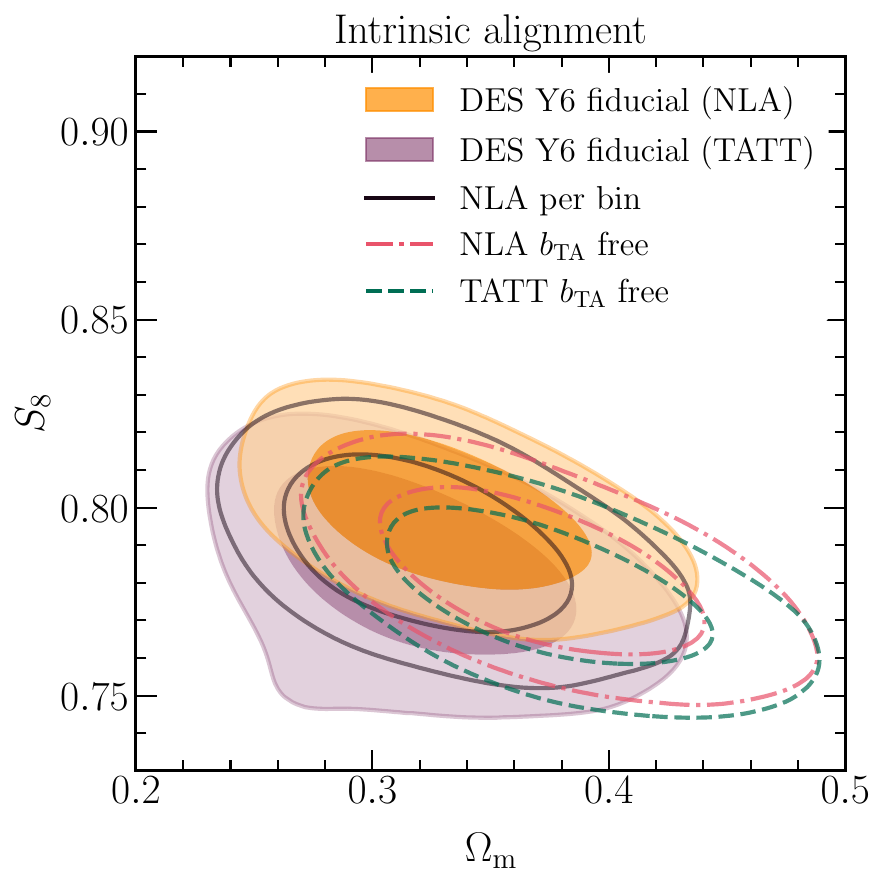}
    \includegraphics[width=0.4\textwidth]{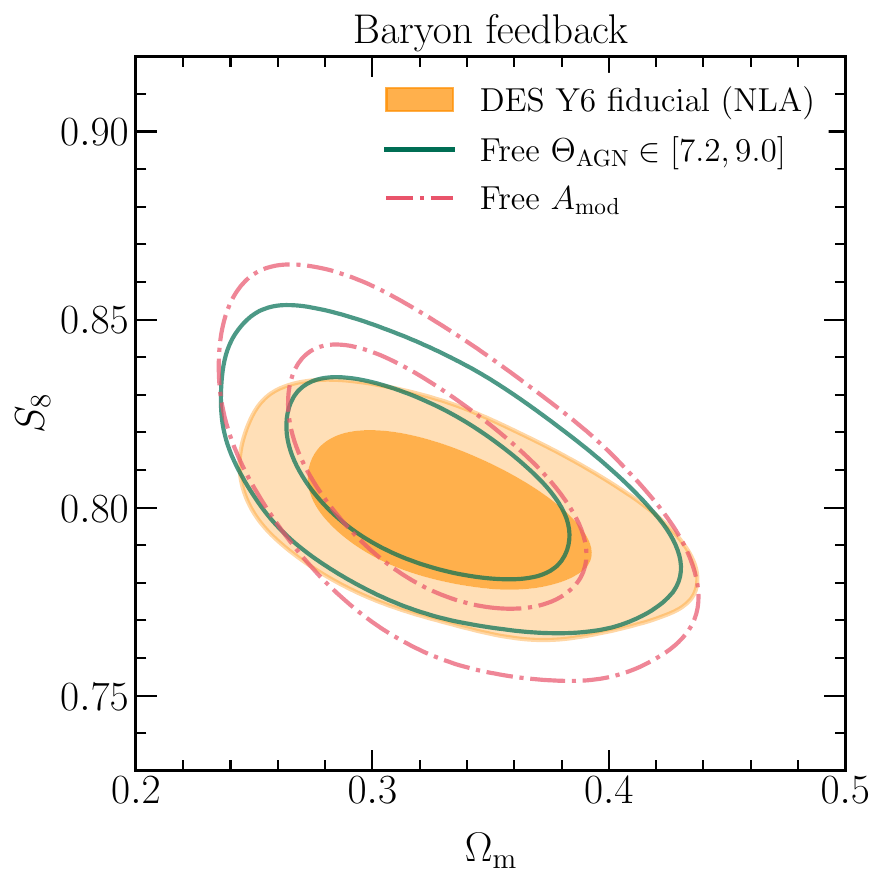}
    \end{center}
    \vspace{-0.75cm}
    \caption{\textit{\textbf{Left}}: Fiducial NLA and TATT analyses compared with other IA variants (solid black: NLA per bin, dashed-dot pink: NLA with $b_{\rm TA} \in [-2,2]$ and dashed green: TATT with $b_{\rm TA} \in [-2,2]$). \textit{\textbf{Right}}: Our fiducial NLA analysis with a fixed baryon component $\Theta_{\rm AGN}=7.7$ (orange) compared with analyses with $\Theta_{\rm{AGN}}$ varied (solid green), and the $A_{\rm{mod}}$ model, a parametric model for capturing power suppression (dashed-dotted pink). 
    }
    \label{fig:variants_ia_baryons}
\end{figure*}

\begin{itemize}

    \item The data have a clear preference for a non-zero IA model, with the IA parameters that are small, but significantly non-zero\footnote{Note that this is different from Y3, where zero IA was weakly favored by the data \citep{y3-cosmicshear2}. }.

    \item Under the assumption of NLA, allowing for a redshift dependence is disfavored. The evidence decreases, and there is no appreciable improvement in $\chi^2$; $\eta_{1}$ is consistent with zero, and the implied $A_1(z)$ does not show a significant trend with redshift, which will be discussed more later in Section~\ref{subsec:insights_ia}. Allowing additional flexibility either by reparametrizing the assumed evolution to a linear function of $z$ or by varying an independent amplitude per redshift bin, does not significantly change this picture, although in the latter case we see some hints of non-zero evolution; see Section~\ref{subsec:insights_ia}. The same holds in our fiducial TATT variant: fixing $\eta_1=\eta_2=0$ improves the evidence by a factor of $\sim 2$. This is not universally true, however. We consider the  TA model, which is an extension to NLA where the $b_{\rm TA}$ parameter is allowed to deviate from 0. For both the TA model and TATT with $b_{\rm TA}=0$, we see significant increases in evidence when adding redshift power-law parameters.

    \item A fixed density weighting contribution with $b_{\rm TA}=1$ is substantially disfavored compared to either assuming $b_{\rm TA}=0$ or varying the parameter. The choice to fix $b_{\rm TA}=1$ is the primary cause of the above noted preference for our fiducial NLA variant over fiducial TATT, with the data strongly preferring some variants of TATT over NLA once deviation from $b_{\rm TA}=1$ is allowed (see Table \ref{tab:model_comparison}). Consequently, the best-fit $b_{\rm TA}$ is relatively well constrained at $b_{\rm TA} = -0.73 ^{+0.21}_{-0.25}$ for the TA-3 model.

    \item Varying $b_{\rm TA}$ brings NLA and TATT into better agreement: in both cases, this results in a lower value of $S_8=0.782$ and higher values of $\Omega_{\rm m}$. We show the $S_8$-$\Omega_{\rm m}$ constraints in the left panel of Figure~\ref{fig:variants_ia_baryons}, compared to our fiducial constraints. The data prefer these models with free $b_{\rm TA}$ (the TA model).

    \item We check the degeneracy of the IA model parameters with redshift mode parameters $u_j$ within the two fiducial and two data-preferred $b_{\rm TA}$ free models and find that they do not show evidence for absorbing the redshift calibration errors.
\end{itemize}

In summary, as described in Section~\ref{subsec:ia}, before unblinding we defined two fiducial IA models, variants of NLA and TATT. Between these models, the fiducial NLA was preferred by the Y6 data. However, the model favored overall by the Y6 data is a 3-parameter version of the model with an alignment amplitude with redshift dependence and varying $b_{\rm TA}$: $A_1\sim[-1,3]$, $\eta_1\sim\mathcal{N}(0.0,3.0)$, $b_{\rm TA}\sim[-2,2]$. Our IA model constraints bear further investigation. The IA parameters may reflect a combination of a real underlying signal, absorption of residual systematics (such as photo-$z$ errors), and compensation for model misspecification (e.g., imposing a power-law redshift evolution). In Section~\ref{sec:insights} we investigate the constraints on the IA model parameters and discuss strategies for model selection in future weak lensing analyses. 

\subsubsection{Baryon feedback}\label{subsubsec:robust_baryons}

To mitigate baryonic feedback effects, our fiducial model implements both scale cuts and a fixed feedback prescription using \textsc{HMCode2020}, at a moderate feedback strength of $\Theta_{\mathrm{AGN}}=7.7$. 
In the right panel of Figure~\ref{fig:variants_ia_baryons}, and Rows 25 and 26 of Figure~\ref{fig:results_lensing}, we demonstrate the impact of marginalizing over this baryon feedback model parameter, while retaining our scale cuts, on our cosmological constraints. We vary this parameter in the range ($\Theta_{\mathrm{AGN}} \in [7.2, 9.0]$), where the prior choice on $\Theta_{\mathrm{AGN}}$ is largely motivated by \cite{Bigwood2024} to allow for strong feedback scenarios. For the NLA case, the $S_8$ posterior, shown in green, is consistent within $0.4\sigma$ with the fiducial result ($S_8=0.807^{+0.017}_{-0.019}$), with the posterior allowing for higher values of $S_8$ with degraded precision. A similar behavior is seen for TATT (see Table~\ref{tab:giant_results_table} in Appendix~\ref{app:res}).

In addition, while other ways to model small scales exist (e.g. \cite{Huang21, Arico2023, Chen23, Bigwood2024, garcia24,decade, siegel25, Bigwood25_blueshear}), we constrain the non-linear power suppression through the phenomenological $A_{\mathrm{mod}}$ model \citep{amon22, preston23}. We choose this approach as the model is not calibrated to any specific simulation, and so is agnostic to the source of the suppression, shown in pink in Figure~\ref{fig:variants_ia_baryons}. The power spectrum is modulated according to the formula
\begin{equation}
P_{\rm m}(k, z) =  P^{\rm L}_{\rm m}(k, z) + A_{\rm mod}[P^{\rm NL}_{\rm m} (k, z) - P^{\rm L}_{\rm m}(k, z)] \,. \label{equ:amod}
\end{equation} 
We vary this parameter in the range ($A_{\rm mod} \in [0.5, 1.5]$), where the prior choice is motivated by \cite{preston23}.
This result is also consistent within $0.3\sigma$ ($S_8=0.806^{+0.021}_{-0.025}$), again with the posterior allowing for higher values of $S_8$ and with further degraded constraints. 

Overall, we find that our scale cuts are effectively removing the parts of the measurement that are sensitive to the non-linear effects. In Section~\ref{subsec:insights_feedback} we analyze the DES Y6 measurements without scale cuts to constrain the suppression of the matter power spectrum.

\section{Intrinsic Alignment \& Matter Power Spectrum Constraints}\label{sec:insights}

In Section~\ref{sec:robustness}, we demonstrated that our fiducial constraints are robust to astrophysical and observational systematic effects to within 1$\sigma$. In this section we discuss the constraints on the intrinsic alignment parameters (Section~\ref{subsec:insights_ia}), and the matter power spectrum, including baryonic feedback (Section~\ref{subsec:insights_feedback}) from our cosmic shear data. 

\subsection{Intrinsic alignment model parameters}\label{subsec:insights_ia}

\begin{figure}
    \centering
    \includegraphics[width=0.9\columnwidth]{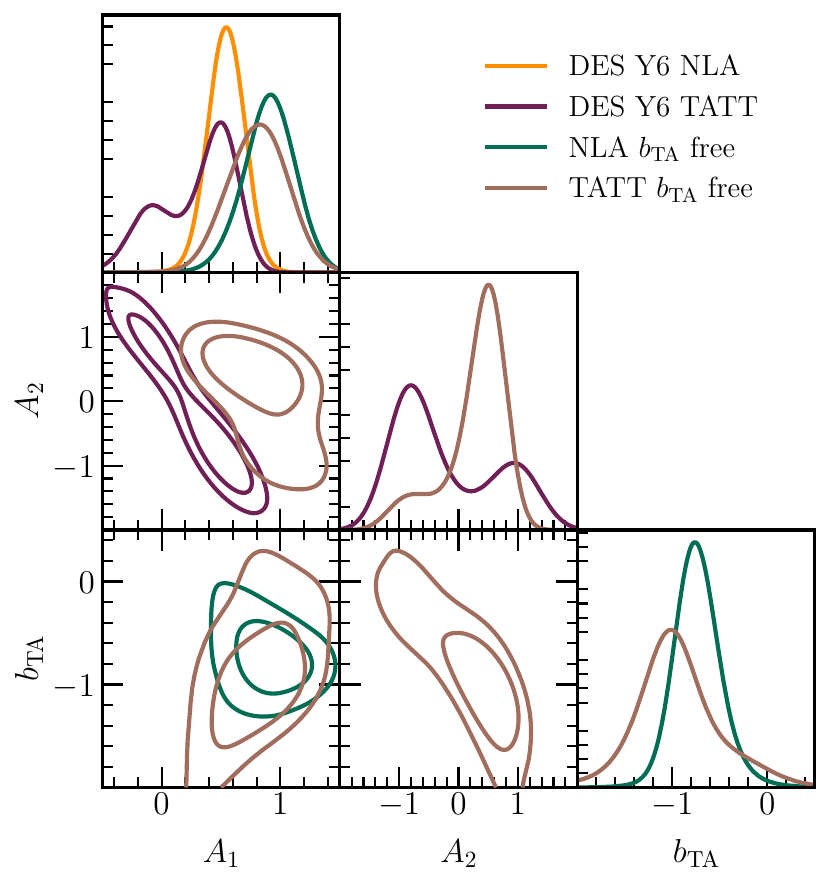}
    \includegraphics[width=\columnwidth]{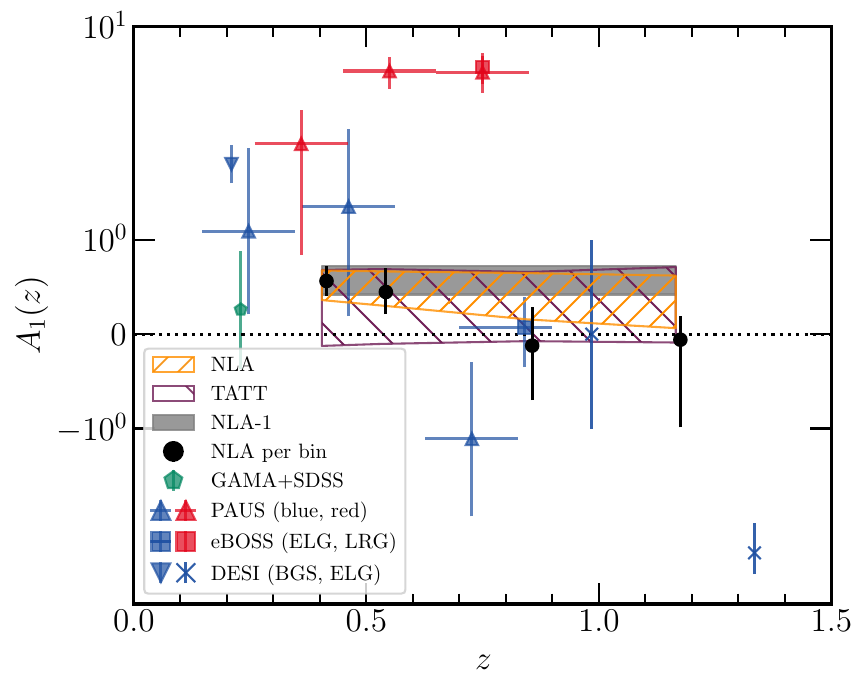}
    \caption{\textit{\textbf{Top:}} Posteriors of intrinsic alignment parameters obtained from the two fiducial analyses (NLA and TATT, orange and purple) and those varying $b_{\rm TA}$ (green and brown). The two sets of contours are $68\%$ and $95\%$ confidence levels. 
    \textit{\textbf{Bottom:}} Effective intrinsic alignment amplitude $A_1$ as a function of redshift. The four redshifts shown for DES Y6 NLA per-bin constraints represent the means of our four redshift bins (shown in Figure \ref{fig:nz}). For the NLA and TATT cases, which assume a power law, we use a pivot redshift at $z_0=0.3$. We additionally show a selection of published direct IA measurements using spectroscopic data from: GAMA+SDSS \citep{Johnston19}, PAUS \citep{paus_ia}, eBOSS \citep{samuroff23}, and DESI \citep{Siegel_DESIIA}. Note that the samples for the direct measurements are qualitatively different from those used for lensing in terms of galaxy populations.
    \label{fig:ia_z}
    }
\end{figure}
Figure \ref{fig:ia_z} shows the constrained IA amplitudes, $A_1$, $A_2$ and $b_{\rm TA}$ at our chosen pivot redshift $z=0.3$ (top panel) and as a function of redshift  (bottom panel). We show the results using both NLA (orange) and TATT (purple), both of which assume a power-law redshift dependence. We also show analyses using NLA with a non-parametric free amplitude per redshift bin (NLA per bin, black) and NLA without redshift dependence (NLA-1, shaded grey). The picture here is largely consistent across models -- at low redshift, where IA is well constrained, the data prefer a positive linear amplitude at $A_1\sim0.5$ for our fiducial models. Due to the degeneracy between $A_1$ and $A_2$ in the TATT model, the marginalized constraint on $A_1$ is weaker here but consistent with small positive values. Freeing $b_{\rm TA}$ raises the best-fit $A_1$ to around $\sim 1$ and reduces the bimodality in $A_2$, as we can see in the top panel of Figure~\ref{fig:ia_z}.  
Beyond $A_1$, $b_{\rm{TA}}$ is constrained at mildly negative values for both NLA$+b_{\rm{TA}}$ and TATT$+b_{\rm{TA}}$ models, which interestingly deviates from our fiducial choice of fixing $b_{\rm{TA}}=0$ for NLA and $b_{\rm{TA}}=1$ for TATT (see Section \ref{sec:method}). This $b_{\rm{TA}}$ constraint could be the result of approximations or incompleteness in our models, of residual systematics, or of a physical IA signal. The physical implications of this is an open question for the field (see e.g. \cite{ichikawa26,paviot26}, which show hints of negative $b_{\rm TA}$ in different contexts).
At higher redshift, the IA parameters are not well constrained and are consistent with no linear alignment, $A_1=0$. Under the assumption that any evolution follows a power law, the fiducial models are consistent with a redshift-independent $A_1$ (although this may be in part due to the size of the error bars at high $z$). In the more flexible case (NLA per bin) the data point towards a gradual reduction in amplitude at high-$z$, which would be consistent with the expected qualitative trend from changes in population (i.e. the increasing prevalence of fainter bluer galaxies; see, for example, \cite{krause16}). We note, however, the trend is weak, and the flexibility in redshift is not needed to describe the data in a statistical sense; if we compare NLA-1 and NLA per bin we find a Bayes ratio of $\mathcal{Z_{\rm NLA-1}}/\mathcal{Z_{\rm NLA per\ bin}} = 14$, preferring a flat, non-evolving $A_1$. We note that under all of these different models, the inferred $S_8$ is stable to $\sim 0.5 \sigma$, again suggesting that our power law approximation is sufficient for current datasets. 

We also include a selection of direct IA measurements in Figure \ref{fig:ia_z} (colored points; GAMA+SDSS \citep{Johnston19}, PAUS \citep{Navarro25}, eBOSS \citep{samuroff23}, and DESI \citep{Siegel_DESIIA}). While the comparison of uncertainties is useful, we should be careful not to over-interpret here: the direct IA measurements use specific galaxy populations (e.g., LRG, BGS, ELG), which do not resemble our Y6 \mdet\ sample and typically have larger IA amplitude. Note also that the composition of the DES sample will change with $z$, and we do not explicitly model IA luminosity dependence. Thus, any evolution detected will naturally combine population effects with any inherent change in IA strength, and we should not expect agreement either in the amplitude $A_1$ or in its trend with redshift. 
The measurements to which we compare are derived from both blue and red galaxies, with red galaxies consistently demonstrating stronger alignments. Our IA amplitude is comparatively small, consistent with the fact that our DES Y6 sample has a high blue fraction and is fainter on average then direct IA samples.

Understanding the IA of galaxies as a function of their galaxy properties is paramount for progress ahead of next-generation lensing surveys. Specifically, this task includes making direct IA measurements for galaxy populations in which they are currently lacking, but which contribute to lensing samples (magnitude limited, blue, and high redshift) \citep[e.g.][]{Johnston19, samuroff23, Siegel_DESIIA, paus_ia}, as well as testing the need for additional model complexity to fit these direct measurements \citep[e.g.][]{fortuna21, vlah20, Pandya2025}. In addition, given the constraining power of current and future cosmic shear data, it is interesting to investigate and exploit the IA parameter constraints when splitting the sample by galaxy properties \citep{Samuroff_Y1IA, mccullough2024, KiDSLegacy_consistency, Bigwood25_blueshear}. 

\begin{figure}
    \begin{center}
    \includegraphics[width=\columnwidth]{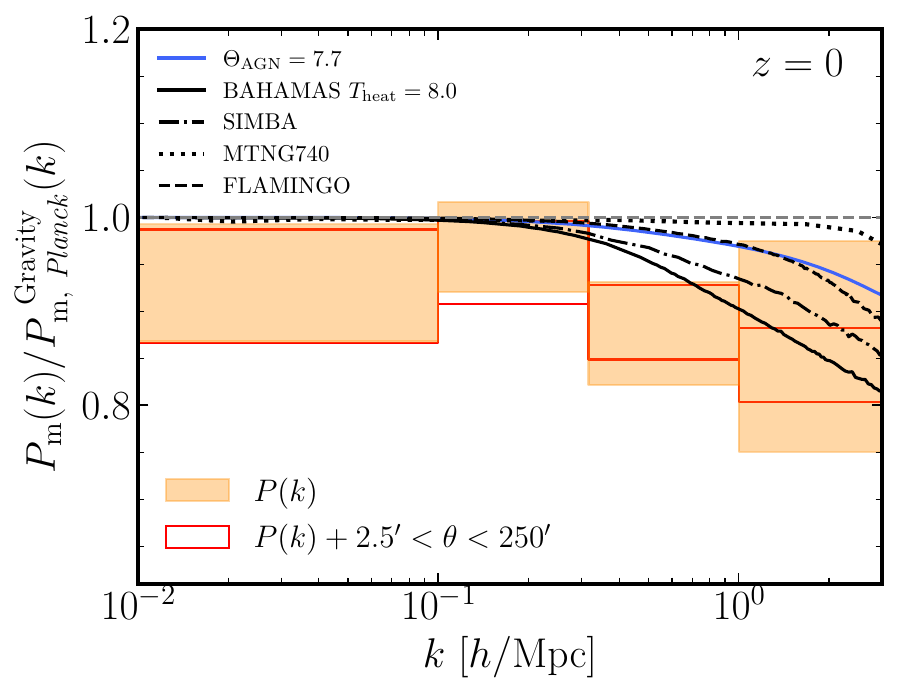}
    \end{center}
    \caption{Marginalized constraints in the suppression of the matter power spectrum (reconstructed with and without small-scale data shown as filled orange and unfilled red boxes, respectively), relative to a gravity-only power spectrum. They are compared to four different hydrodynamical simulations (\textsc{BAHAMAS} $\Delta T_{\rm heat}=8.0$, \textsc{SIMBA}, \textsc{MTNG740}, and \textsc{FLAMINGO}. We also show our fiducial \textsc{HMCode2020} $\Theta_{\rm{AGN}}=7.7$ model (see Section~\ref{subsec:baryons}), plotted in solid blue.}
    \label{fig:variants_ia_baryons_2}
\end{figure}

\begin{figure*}
\centering 
     \includegraphics[width=1.5\columnwidth]{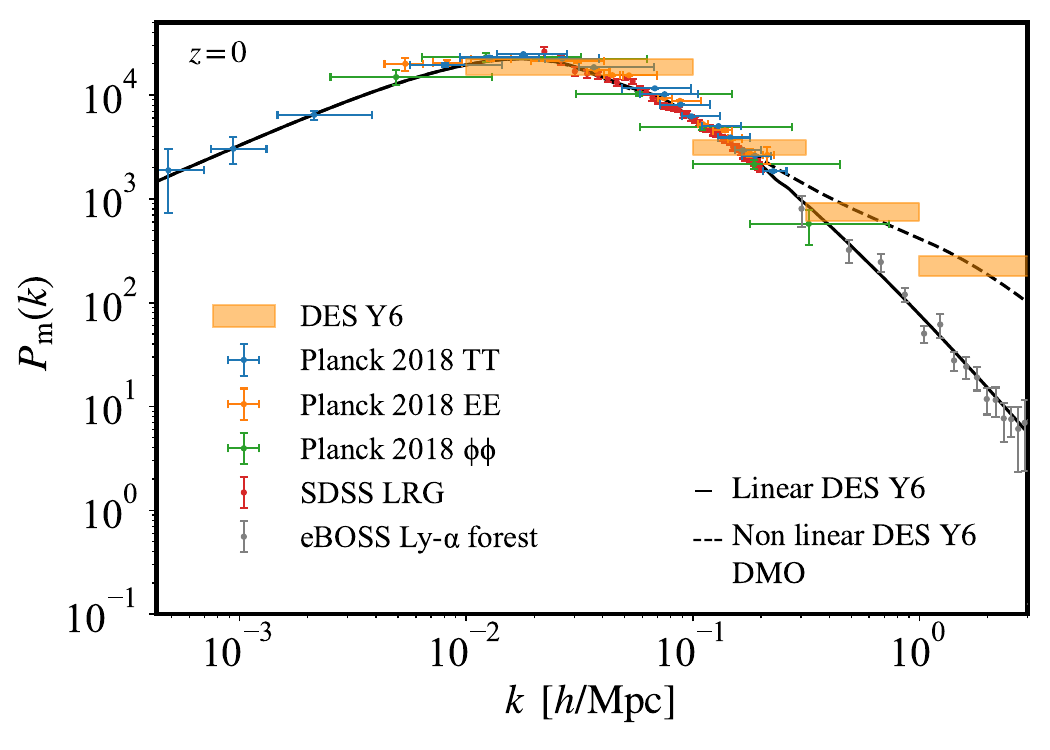} 
     \vspace{-0.75cm}
\caption{Full matter power spectrum constraints from DES Y6 power spectrum reconstruction technique using fiducial analysis choices. We compare to results from the literature: \textit{Planck} 2018 TT, EE and lensing spectra $\phi \phi $ \citep{Planck_legacy}, constraints from the BAO analysis from SDSS in red \citep{Reid_SDSS_LRG} and the Lyman-$\alpha$ forest constraints from the eBOSS analysis in grey \citep{eBOSS_DR14, Chabanier_lya}. The constraints contain linear modes consistent with those from CMB and BAO measurements of the matter power spectrum, and extend into the non-linear regime. The solid and dashed lines show the linear and dark-matter only (DMO) non-linear power spectrum predictions from \textsc{HMCode2020} assuming the DES Y6 cosmic shear best-fit cosmology (NLA). Note that the differences at high $k$ arise from the fact that the Lyman-$\alpha$ forest probes small scales but at high redshift ($z \gtrsim 2$), where the Universe is still linear to good approximation. Thus, unlike cosmic shear, the grey points are not sensitive to non-linearities that emerge at low redshift.}
\label{fig:full_spectru_with_scalecuts}
\end{figure*}

\subsection{The shape of the matter power spectrum}
\label{subsec:insights_feedback}

In this section, we constrain the suppression of the matter power spectrum and reconstruct the total matter power spectrum from our DES Y6 cosmic shear data, which is sensitive to scales extending into the non-linear regime, where galaxy formation processes and non-standard physics may imprint signatures \cite{bechtol_DM}. This allows us to assess deviations of the DES Y6 data relative to a fixed dark-matter only \LCDM cosmology. We also compare our constraints to those from other cosmological probes as a function of scale, following \citep{Planck_legacy}.

We follow the methodology of \cite{preston24}, built upon \cite{Tegmark_2002} and similar to \cite{Doux2022}, to reconstruct the non-linear matter power spectrum. To do this, we first vary the shape of the non-linear matter power spectrum assuming a fixed \textit{Planck} 2018 cosmology (and therefore a fixed expansion history). We bin the spectrum in $N$ wavenumber bins of $\log(k)$, extending over scales $k=10^{-2}-10^1$, such that
\begin{equation}
    \frac{P_{\rm{m}}(k,z)}{P^{{\rm{~Gravity}}}_{\rm{m},~\textit{Planck} }(k,z)}  = C(k_{i}) \hspace{1mm} , \hspace{5mm} i=[1,N] 
\label{equ:BinningPkz}
\end{equation}
where $P^{{\rm{~Gravity}}}_{\rm{m},~\textit{Planck} }(k,z)$ is the power spectrum from HMCode2020 at a fixed Planck cosmology \citep{planck_data} with no feedback prescription, $P_{\rm{m}}(k,z)$ is the true matter power spectrum and $C(k_{i})$ are the ratios of observed to gravity-only matter power spectra in each of the $N$ bins, centred on $k_{i}$. These scaling coefficients for each bin are introduced as free parameters in the analysis, rather than the standard cosmological parameters. The result is a constraint on the total amplitude of the fractional suppression of the non-linear matter power spectrum in each wavenumber bin. Our treatment assumes this quantity is constant with redshift. 
By fixing the cosmology, the redshift evolution of the suppression is assumed to be described by a \textit{Planck} \LCDM model. For future surveys, more complex models could be introduced to capture the specific redshift dependence of the suppression at the same time as assessing the scale dependence of the suppression.

We first study the suppression of the matter power spectrum relative to the dark matter-only power spectrum (Equation~\ref{equ:BinningPkz}) computed from \textsc{HMCode2020} with \textit{Planck} 2018 cosmology. This is shown in Figure~\ref{fig:variants_ia_baryons_2}. A suppression could arise due to baryon feedback \citep{vanDaalen2011}, new physics, such as massive neutrinos, warm dark matter or axions \citep[e.g.][]{Viel, Liu:2018, Rogers:2023}, or unknown systematics. The reconstruction approach that we employ using weak lensing data is agnostic to the source of the suppression. Reconstructing the shape of the power spectrum in a model-independent way may provide clues to the physics responsible \citep{disentanglingDM}. 

The orange constraints in Figure~\ref{fig:variants_ia_baryons_2} are obtained using the fiducial analysis choices with the NLA IA model. Also shown in red are constraints obtained from analyzing the cosmic shear measurements without the scale cuts, including small angular scales ($2.5' < \theta < 250'$). We overlay a selection of suppressed power spectra from cosmological hydrodynamical simulations (\textsc{BAHAMAS-8}: \cite{mccarthy17}, \textsc{SIMBA}: \cite{Dave2019}, \textsc{MTNG740}: \cite{Hernandez-Aguayo2022} and \textsc{FLAMINGO}: \cite{Schaye2023}). We note that although we compare with predictions from the hydrodynamic simulations at $z=0$ where results are readily available, our results are not constrained by data at $z = 0$, but rather by the lensing field at higher redshift. The power suppression obtained with DES Y6 data is consistent with the \textsc{BAHAMAS-8} simulation, which was used as the upper limit feedback strength to define the scale cuts used in this analysis (see Section~\ref{subsec:baryons}).

Although the bulk of this suppression is likely due to baryonic feedback, which motivates this comparison, we note that this reconstruction is sensitive to any physics that alters the underlying power spectrum as well as other effects that impact the lensing measurement, including mis-modelled astrophysical effects, higher-order lensing, or systematics not removed from the data. In Appendix~\ref{app:data_tests}, we demonstrate that the small-scale data vector is robust to a mis-modelled PSF and does not contain significant B-mode contamination, providing confidence that the measurement is reliable on these scales. Simulated tests in \citep{preston24} demonstrated that, at the current level of constraining power, it is unlikely that mis-modelled IA would account for the bulk of the 10-20\% suppression amplitude, although we note that a fully non-linear IA model on these small scales could contribute more than NLA or TATT.  

Next, in order to achieve a reconstructed matter power spectrum that does not assume a set of cosmological parameters, we perform the same analysis that implements equation~\ref{equ:BinningPkz}, but varying cosmological parameters in the gravity-only spectrum, rather than remaining fixed to a \textit{Planck} cosmology.  We then multiply these constraints by the best-fit DES Y6 non-linear dark-matter only matter power spectrum predicted by \textsc{HMCode2020} to report a constraint on the full matter power spectrum from DES Y6 data. We maintain the fiducial analysis choices and scale cuts, opting for the NLA IA model.

This result is shown in Figure~\ref{fig:full_spectru_with_scalecuts}, with constraints on the full matter power spectrum reported at $z=0$. The solid and dashed lines show the dark-matter only predictions from \textsc{HMCode2020} assuming the DES Y6 cosmic shear best-fit cosmology for the linear and non-linear power spectrum, respectively. The orange constraints represent the reconstruction where deviations from a dark-matter only scenario are allowed. We compare to linear constraints from the primary CMB from the \textit{Planck} 2018 TT and EE spectra \citep{Planck_legacy}, as well as lensing ($\phi\phi$) from \textit{Planck} 2018 \citep{planck_data}. We also show constraints on the linear scales from the BAO analysis from SDSS \citep{Reid_SDSS_LRG} in red.  Note that the Lyman$-\alpha$ forest constraints from the eBOSS analysis \citep{eBOSS_DR14, Chabanier_lya} extend to small scales at high redshift, therefore constraining the matter power spectrum before any significant non-linearities have begun to form. Our DES constraints extend to the scales $k>1~[h/\rm{Mpc}]$ at $z<1$, where non-linear physics has an important impact, but also dovetail linear modes and constraints from CMB experiments on larger scales. On linear scales, we find good agreement with external data. This highlights the complementarity of weak lensing as a unique window to the low redshift ($z<2$), non-linear scales.

\section{Conclusions}\label{sec:con}
This paper presents the cosmic shear cosmological analysis from the Dark Energy Survey using all six years of data (DES Y6). This analysis follows the decade-long DES weak-lensing program, which started with the Science Verification stage \citep{sv-cosmicshear}. Over this period, DES has pioneered various data processing and analysis methods across the pixels-to-cosmology pipeline to support robust constraints. This analysis is supported by a number of papers \citep{y6-gold, y6-piff, y6-balrog, y6-metadetect, y6-imagesims, y6-sourcepz, y6-wz, y6-nzmodes, y6-methods, y6-gglens}, ranging from extracting galaxy information from images to modeling the cosmic shear. We present cosmological constraints assuming a flat $\Lambda$CDM model. Given that previous literature does not provide decisive evidence for a single intrinsic alignment (IA) model, we assume two IA models and analyze the data in parallel: the non-linear alignment (NLA) model and the tidal alignment and tidal torquing (TATT) model. 

We use the weak lensing galaxy shape catalog presented in \citep{y6-metadetect}, containing 139,662,173 galaxies over 4031 deg$^2$ in the Southern sky (8.29 galaxies/arcmin$^2$), binned in four redshift bins with the mean of the highest bin reaching $z = 1.2$. We measure the real-space cosmic shear signal in the angular range of ($\theta \in [2.5', 250']$). The signal-to-noise ($S/N$) of the data vector is $S/N=60$ using all the data points, and 40 / 43 with derived scale cuts using the NLA / TATT intrinsic alignment model analyses, respectively. 

For the flat $\Lambda$CDM model, we find marginalized cosmological constraints on $S_8\equiv \sigma_8 (\Omega_{\rm m}/0.3)^{0.5}$ from our two fiducial analyses to be $S_8 = 0.798^{+0.014}_{-0.015}$ (NLA) and $S_8 = 0.783^{+0.019}_{-0.015}$ (TATT), achieving 1.8\% and 2.5\% uncertainty on $S_8$, respectively. These two fiducial analysis choices differ by $0.6\sigma$ in the marginalized mean $S_8$. The two fiducial analyses are consistent with the CMB primary anisotropy constraint from \textit{Planck}+ACT+SPT at the $ 1.1\sigma$ (NLA) and $1.7\sigma$ (TATT) level (in full parameter space). Using a simple $\Delta S_8$ metric, the differences are $2.0\sigma$ (NLA) and $2.3\sigma$ (TATT). The results are also consistent with latest constraints from CMB lensing \citep{qu25} and redshift space distortions \citep{desi_fs}.

We study the main contributors to the uncertainties in our fiducial results. We find that the uncertainty on shear calibration, redshift calibration and the IA amplitude all degrade the precision of $S_8$ inference by, at most, 30\%. Conversely, we also look at the potential gain in statistical power if we are able to use and model all the data points measured (removing the small-scale scale cut). We find that these additional data points translate to an improvement in uncertainties by up to a factor of 2.6. 

We perform extensive tests with our data and model to ensure the DES Y6 cosmic shear results are robust. First, we check that our data is internally consistent when inferring cosmology with different subsets of the data vector. Second, we check that the cosmological parameters are stable to alternative calibration strategies for the redshift distributions and our shear measurement. Finally, we test the impact of using alternative models for either IA or the non-linear power spectrum. We find that with a large range of IA models, our results on $S_8$ are stable within 1$\sigma$. Interestingly while between the two fiducial models, our data prefers NLA with an evidence ratio of $Z_{\rm NLA}/Z_{\rm TATT} = 16.9$, our data \textit{most} prefers an IA model that is in-between TATT and NLA (a model with three parameters, $A_1$, $\eta_1$ and $b_{\rm TA}$). For the non-linear power spectrum, we find that $S_8$ does not shift by more than $0.5\sigma$ when we marginalize over a wide range of baryonic feedback strength. 

With the high signal-to-noise cosmic shear data, we also perform two extended analyses beyond our fiducial cosmological inference. First, we examine the IA model parameter constraints and compare with those derived from a wide range of external direct IA measurements. We find that with the DES Y6 data, IA is well constrained at low redshift and the data prefer a positive but small linear amplitude. Next, we perform an approximate reconstruction of the 3D matter power spectrum directly from our cosmic shear data vector. We find that, on large scales, our data agree with other reconstructions based on the CMB, the Ly-$\alpha$ forest, and galaxy clustering. On small scales, the weak lensing data clearly contain information on the non-linear power spectrum, probing scales that are not accessible to the other datasets. We also look at the inferred suppression of the non-linear power spectrum mostly due to baryonic feedback and compare it to hydrodynamical simulations.  

Looking ahead, we are entering the era of Stage-IV imaging surveys, which includes \textit{Euclid}, \textit{Rubin}'s LSST, and \textit{Roman}'s HLIS. Cosmic shear will continue to be one of the main cosmology pillars for all these surveys, where the cosmic structure is measured directly via purely gravitational effects. With the expected increase in statistical power, we imagine meeting the requirements on various data and modeling accuracies will be the focus of the community in the coming years. 
Our work highlights that in addition to the continued effort needed to calibrate shear measurements and redshift distributions for deeper data, understanding astrophysical systematic effects such as IA and baryonic feedback is also critical. Taken together, our Y6 analysis marks a milestone in the DES cosmic shear program. The data products that we produced, the methodological advancements in data calibration that we developed, the pixels-to-cosmology pipeline we built and the results we presented are a step towards next-generation cosmic shear surveys.

\section{Data Availability}

The Y6KP data products in this analysis, including catalogues, data-vector measurements, masks and chains, are publicly available at \url{https://des.ncsa.illinois.edu/releases/y6a2}, see \cite{y6-das}. The DES Y6 Gold catalogue derived from the DES Data Release 2 (DR2) \cite{DESDR2} is publicly available at \url{https://des.ncsa.illinois.edu/releases} and described in \cite{y6-gold}.

\section*{Acknowledgments}

{We would like to thank Angus H. Wright for providing KiDS-Legacy cosmic shear posteriors used in Section~\ref{subsec:lensing}. }
\newline

{\bf Author contributions: } All authors contributed to this paper and/or carried out infrastructure work that made this analysis possible. Some highlighted contributions (in alphabetical order) include:

\begin{itemize}
    \itemsep0em
    
    \item Source sample selection and calibrations: A.~Amon, K.~Bechtol, M.~R.~Becker, M.~Gatti, R.~A.~Gruendl, M.~Jarvis, F.~Menanteau, A.~Roodman, E.~S.~Rykoff, T.~Schutt, E.~Sheldon, M.~Yamamoto
    
    \item Photometric redshifts: A.~Amon, A.~Alarcon, G.~M.~Bernstein, G.~Camacho-Ciurana, A.~Campos, W.~d'Assignies, J.~Myles, C.~Sanchez, M.~A.~Troxel, B.~Yin

    \item Simulations and derived calibrations: D.~Anbajagane, J.~Beas-Gonzalez, M.~R.~Becker, S.~Everett, S.~Mau, T.~Schutt, M.~Tabbutt, M.~Yamamoto, B.~Yanny
    
    \item Data-vector measurements, validations and Cosmological inference: A.~Amon, M.~R.~Becker, J.~Blazek, G.~Camacho-Ciurana, G.~Campailla, J.~M.~Coloma-Nadal, A.~Fert{\'e}, M.~Jarvis, E.~Krause, J.~Muir, A.~Porredon, J.~Prat, C.~Preston, M.~Raveri, S.~Samuroff, D.~Sanchez-Cid, T.~Schutt, M.~Yamamoto, B.~Yin

    \item Paper writing and figures: S.~Samuroff, M.~Yamamoto (Paper Coordination), A.~Amon, G.~Camacho-Ciurana, G.~Campailla, A.~Fert{\'e}, S.~Mau, J.~Muir, C.~Preston, S.~Samuroff, D.~Sanchez-Cid, M.~Yamamoto, B.~Yin (Text and Figures)

    \item Comments to manuscript: G.~M.~Bernstein, J.~Blazek, C.~Chang, M.~Crocce, W.~d'Assignies, T.~Diehl, E.~Krause, M.~Raveri, M.~A.~Troxel 

    \item Coordination and scientific management: M.~R.~Becker and M.~Crocce (Year Six Key Project Coordinators), C.~Chang and M.~A.~Troxel (Science Committee Chairs), A.~Alarc{\'o}n, A.~Amon, S.~Avila, J.~Blazek, A.~Fert{\'e}, G.~Giannini, A.~Porredon, J.~Prat, M.~Rodr{\'i}guez-Monroy, C.~Sanchez, D.~Sanchez-Cid, T.~Schutt, N.~Weaverdyck, M.~Yamamoto, B.~Yin (Working Group and Analysis Team leads)

\end{itemize}

The remaining authors have made contributions to this paper that include, but are not limited to, the construction of DECam and other aspects of collecting the data; data processing and calibration; developing broadly used methods, codes, and simulations; running the pipelines and validation tests; and promoting the science analysis. 
\newline

{\bf Funding and Institutional Support:} 

Funding for the DES Projects has been provided by the U.S. Department of Energy, the U.S. National Science Foundation, the Ministry of Science and Education of Spain, the Science and Technology Facilities Council of the United Kingdom, the Higher Education Funding Council for England, the National Center for Supercomputing Applications at the University of Illinois at Urbana-Champaign, the Kavli Institute of Cosmological Physics at the University of Chicago, the Center for Cosmology and Astro-Particle Physics at the Ohio State University, the Mitchell Institute for Fundamental Physics and Astronomy at Texas A\&M University, Financiadora de Estudos e Projetos, Funda{\c c}{\~a}o Carlos Chagas Filho de Amparo {\`a} Pesquisa do Estado do Rio de Janeiro, Conselho Nacional de Desenvolvimento Cient{\'i}fico e Tecnol{\'o}gico and the Minist{\'e}rio da Ci{\^e}ncia, Tecnologia e Inova{\c c}{\~a}o, the Deutsche Forschungsgemeinschaft and the Collaborating Institutions in the Dark Energy Survey. 

The Collaborating Institutions are Argonne National Laboratory, the University of California at Santa Cruz, the University of Cambridge, Centro de Investigaciones Energ{\'e}ticas, Medioambientales y Tecnol{\'o}gicas-Madrid, the University of Chicago, University College London, the DES-Brazil Consortium, the University of Edinburgh, the Eidgen{\"o}ssische Technische Hochschule (ETH) Z{\"u}rich, Fermi National Accelerator Laboratory, the University of Illinois at Urbana-Champaign, the Institut de Ci{\`e}ncies de l'Espai (IEEC/CSIC), the Institut de F{\'i}sica d'Altes Energies, Lawrence Berkeley National Laboratory, the Ludwig-Maximilians Universit{\"a}t M{\"u}nchen and the associated Excellence Cluster Universe, the University of Michigan, NSF NOIRLab, the University of Nottingham, The Ohio State University, the University of Pennsylvania, the University of Portsmouth, SLAC National Accelerator Laboratory, Stanford University, the University of Sussex, Texas A\&M University, and the OzDES Membership Consortium.

Based in part on observations at NSF Cerro Tololo Inter-American Observatory at NSF NOIRLab (NOIRLab Prop. ID 2012B-0001; PI: J. Frieman), which is managed by the Association of Universities for Research in Astronomy (AURA) under a cooperative agreement with the National Science Foundation.

The DES data management system is supported by the National Science Foundation under Grant Numbers AST-1138766 and AST-1536171. This work used Jetstream2 and OSN at Indiana University through allocation PHY240006: Dark Energy Survey from the Advanced
Cyberinfrastructure Coordination Ecosystem: Services $\&$ Support (ACCESS) program, which is supported by U.S. National Science Foundation grants
2138259, 2138286, 2138307, 2137603, and 2138296.

The DES Spanish institutions are partially supported by MICINN/MICIU/AEI (/10.13039/501100011033) under grants PID2021-123012NB, PID2021-128989NB, PID2022-141079NB, PID2023-153229NA, PID2024-159420NB, PID2024-156844NB, 
CEX2020-001058-M, CEX2020-001007-S and CEX2024-001441-S, some of which include ERDF/FEDER funds from the European Union, and a grant by LaCaixa Foundation (ID 100010434) code LCF/BQ/PI23/11970028. IFAE is partially funded by the CERCA program of the Generalitat de Catalunya. We acknowledge the use of Spanish Supercomputing Network (RES) resources provided by the Barcelona Supercomputing Center (BSC) in MareNostrum 5 under allocations AECT-2024-3-0020, 2025-1-0045 and 2025-2-0046. 

We acknowledge support from the Brazilian Instituto Nacional de Ci\^encia e Tecnologia (INCT) do e-Universo (CNPq grant 465376/2014-2).

This document was prepared by the DES Collaboration using the resources of the Fermi National Accelerator Laboratory (Fermilab), a U.S. Department of Energy, Office of Science, Office of High Energy Physics HEP User Facility. Fermilab is managed by Fermi Forward Discovery Group, LLC, acting under Contract No. 89243024CSC000002.

This research used resources of the National Energy Research Scientific Computing Center (NERSC), a Department of Energy User Facility using NERSC award HEP-ERCAP0035553.

\bibliography{refs_short,y1kp, y3kp, y6kp}

@PREAMBLE{
 "\providecommand{\noopsort}[1]{}" 
 # "\providecommand{\singleletter}[1]{#1}%" 
}

@ARTICLE{paviot26,
       author = {{Paviot}, R. and {Joachimi}, B. and others},
        title = "{Euclid preparation. Testing analytic models of galaxy intrinsic alignments in the Euclid Flagship simulation}",
      journal = {arXiv e-prints},
         year = 2026,
        month = jan,
          eid = {arXiv:2601.07784},
        pages = {arXiv:2601.07784},
          doi = {10.48550/arXiv.2601.07784},
archivePrefix = {arXiv},
       eprint = {2601.07784},
 primaryClass = {astro-ph.CO},
       adsurl = {https://ui.adsabs.harvard.edu/abs/2026arXiv260107784E}
}

@ARTICLE{ichikawa26,
       author = {{Ichikawa}, Masaya and {Shi}, Jingjing and {Kurita}, Toshiki and {Takada}, Masahiro and {Nishimichi}, Takahiro and {Blot}, Linda},
        title = "{The Environmental Dependence of Halo Intrinsic Alignments: Stronger Signals in Underdense Regions}",
      journal = {arXiv e-prints},
         year = 2026,
        month = jul,
          eid = {arXiv:2607.11114},
        pages = {arXiv:2607.11114},
          doi = {10.48550/arXiv.2607.11114},
archivePrefix = {arXiv},
       eprint = {2607.11114},
 primaryClass = {astro-ph.CO},
       adsurl = {https://ui.adsabs.harvard.edu/abs/2026arXiv260711114I}
}

@ARTICLE{handley19,
       author = {{Handley}, Will and {Lemos}, Pablo},
        title = "{Quantifying dimensionality: Bayesian cosmological model complexities}",
      journal = {\prd},
         year = 2019,
        month = jul,
       volume = {100},
       number = {2},
          eid = {023512},
        pages = {023512},
          doi = {10.1103/PhysRevD.100.023512},
archivePrefix = {arXiv},
       eprint = {1903.06682},
 primaryClass = {astro-ph.CO},
       adsurl = {https://ui.adsabs.harvard.edu/abs/2019PhRvD.100b3512H}
}

@ARTICLE{planck2018,
       author = {{Planck Collaboration} and others},
        title = "{Planck 2018 results. VI. Cosmological parameters}",
      journal = {\aap},
         year = 2020,
        month = sep,
       volume = {641},
          eid = {A6},
        pages = {A6},
          doi = {10.1051/0004-6361/201833910},
archivePrefix = {arXiv},
       eprint = {1807.06209},
 primaryClass = {astro-ph.CO},
       adsurl = {https://ui.adsabs.harvard.edu/abs/2020A&A...641A...6P}
}

@article{schneider_cosebi,
  title = {{{cosebis}}: {{Extracting}} the Full {{E-}}/{{B-mode}} Information from Cosmic Shear Correlation Functions},
  shorttitle = {{{cosebis}}},
  author = {Schneider, P. and Eifler, T. and Krause, E.},
  year = {2010},
  month = sep,
  journal = {Astronomy and Astrophysics},
  volume = {520},
  pages = {A116},
  issn = {0004-6361},
  doi = {10.1051/0004-6361/201014235},
  urldate = {2024-05-23}
}

@ARTICLE{mead21,
       author = {{Mead}, A.~J. and {Brieden}, S. and {Tr{\"o}ster}, T. and {Heymans}, C.},
        title = "{HMCODE-2020: improved modelling of non-linear cosmological power spectra with baryonic feedback}",
      journal = {\mnras},
         year = 2021,
        month = mar,
       volume = {502},
       number = {1},
        pages = {1401-1422},
          doi = {10.1093/mnras/stab082},
archivePrefix = {arXiv},
       eprint = {2009.01858},
 primaryClass = {astro-ph.CO},
       adsurl = {https://ui.adsabs.harvard.edu/abs/2021MNRAS.502.1401M}
}

@ARTICLE{spt_data,
       author = {{Camphuis}, E. and {Quan}, W. and {Balkenhol}, L. and {Khalife}, A.~R. and {Ge}, F. and others},
        title = "{SPT-3G D1: CMB temperature and polarization power spectra and cosmology from 2019 and 2020 observations of the SPT-3G Main field}",
      journal = {arXiv e-prints},
         year = 2025,
        month = jun,
          eid = {arXiv:2506.20707},
        pages = {arXiv:2506.20707},
          doi = {10.48550/arXiv.2506.20707},
archivePrefix = {arXiv},
       eprint = {2506.20707},
 primaryClass = {astro-ph.CO},
       adsurl = {https://ui.adsabs.harvard.edu/abs/2025arXiv250620707C}
}

@article{Rogers:2023,
       author = {{Rogers}, Keir K. and {Hlo{\v{z}}ek}, Ren{\'e}e and {Lagu{\"e}}, Alex and {Ivanov}, Mikhail M. and {Philcox}, Oliver H.~E. and {Cabass}, Giovanni and {Akitsu}, Kazuyuki and {Marsh}, David J.~E.},
        title = "{Ultra-light axions and the S $_{8}$ tension: joint constraints from the cosmic microwave background and galaxy clustering}",
      journal = {\jcap},
         year = 2023,
        month = jun,
       volume = {2023},
       number = {6},
          eid = {023},
        pages = {023},
          doi = {10.1088/1475-7516/2023/06/023},
archivePrefix = {arXiv},
       eprint = {2301.08361},
 primaryClass = {astro-ph.CO},
       adsurl = {https://ui.adsabs.harvard.edu/abs/2023JCAP...06..023R}
}

@ARTICLE{Viel,
       author = {{Viel}, M. and {Markovi{\v{c}}}, K. and {Baldi}, M. and {Weller}, J.},
        title = "{The non-linear matter power spectrum in warm dark matter cosmologies}",
      journal = {\mnras},
         year = 2012,
        month = mar,
       volume = {421},
       number = {1},
        pages = {50-62},
          doi = {10.1111/j.1365-2966.2011.19910.x},
archivePrefix = {arXiv},
       eprint = {1107.4094},
 primaryClass = {astro-ph.CO},
       adsurl = {https://ui.adsabs.harvard.edu/abs/2012MNRAS.421...50V}
}

@ARTICLE{Liu:2018,
       author = {{Liu}, Jia and {Bird}, Simeon and {Zorrilla Matilla}, Jos{\'e} Manuel and {Hill}, J. Colin and {Haiman}, Zolt{\'a}n and {Madhavacheril}, Mathew S. and {Petri}, Andrea and {Spergel}, David N.},
        title = "{MassiveNuS: cosmological massive neutrino simulations}",
      journal = {\jcap},
         year = 2018,
        month = mar,
       volume = {2018},
       number = {3},
          eid = {049},
        pages = {049},
          doi = {10.1088/1475-7516/2018/03/049},
archivePrefix = {arXiv},
       eprint = {1711.10524},
 primaryClass = {astro-ph.CO},
       adsurl = {https://ui.adsabs.harvard.edu/abs/2018JCAP...03..049L}
}

@ARTICLE{disentanglingDM,
       author = {{Preston}, Calvin and {Rogers}, Keir K. and {Amon}, Alexandra and {Efstathiou}, George},
        title = "{Prospects for disentangling dark matter with weak lensing}",
      journal = {\mnras},
         year = 2025,
        month = oct,
       volume = {542},
       number = {4},
        pages = {2698-2713},
          doi = {10.1093/mnras/staf1321},
archivePrefix = {arXiv},
       eprint = {2505.02233},
 primaryClass = {astro-ph.CO},
       adsurl = {https://ui.adsabs.harvard.edu/abs/2025MNRAS.542.2698P}
}

@ARTICLE{act_data,
       author = {{Louis}, Thibaut and {La Posta}, Adrien and {Atkins}, Zachary and {Jense}, Hidde T. and {Abril-Cabezas}, Irene and others},
        title = "{The Atacama Cosmology Telescope: DR6 power spectra, likelihoods and {\ensuremath{\Lambda}}CDM parameters}",
      journal = {\jcap},
         year = 2025,
        month = nov,
       volume = {2025},
       number = {11},
          eid = {062},
        pages = {062},
          doi = {10.1088/1475-7516/2025/11/062},
archivePrefix = {arXiv},
       eprint = {2503.14452},
 primaryClass = {astro-ph.CO},
       adsurl = {https://ui.adsabs.harvard.edu/abs/2025JCAP...11..062L}
}

@ARTICLE{planck_data,
       author = {{Planck Collaboration}},
        title = "{Planck 2018 results. V. CMB power spectra and likelihoods}",
      journal = {\aap},
         year = 2020,
        month = sep,
       volume = {641},
          eid = {A5},
        pages = {A5},
          doi = {10.1051/0004-6361/201936386},
archivePrefix = {arXiv},
       eprint = {1907.12875},
 primaryClass = {astro-ph.CO},
       adsurl = {https://ui.adsabs.harvard.edu/abs/2020A&A...641A...5P}
}

@ARTICLE{buchs19,
       author = {{Buchs}, R. and {Davis}, C. and {Gruen}, D. and {DeRose}, J. and {Alarcon}, A. and others},
        title = "{Phenotypic redshifts with self-organizing maps: A novel method to characterize redshift distributions of source galaxies for weak lensing}",
      journal = {\mnras},
         year = 2019,
        month = oct,
       volume = {489},
       number = {1},
        pages = {820-841},
          doi = {10.1093/mnras/stz2162},
archivePrefix = {arXiv},
       eprint = {1901.05005},
 primaryClass = {astro-ph.CO},
       adsurl = {https://ui.adsabs.harvard.edu/abs/2019MNRAS.489..820B}
}

@ARTICLE{giri21,
       author = {{Giri}, Sambit K. and {Schneider}, Aurel},
        title = "{Emulation of baryonic effects on the matter power spectrum and constraints from galaxy cluster data}",
      journal = {\jcap},
         year = 2021,
        month = dec,
       volume = {2021},
       number = {12},
          eid = {046},
        pages = {046},
          doi = {10.1088/1475-7516/2021/12/046},
archivePrefix = {arXiv},
       eprint = {2108.08863},
 primaryClass = {astro-ph.CO},
       adsurl = {https://ui.adsabs.harvard.edu/abs/2021JCAP...12..046G}
}

@ARTICLE{arico21,
       author = {{Aric{\`o}}, Giovanni and {Angulo}, Raul E. and {Contreras}, Sergio and {Ondaro-Mallea}, Lurdes and {Pellejero-Iba{\~n}ez}, Marcos and {Zennaro}, Matteo},
        title = "{The BACCO simulation project: a baryonification emulator with neural networks}",
      journal = {\mnras},
         year = 2021,
        month = sep,
       volume = {506},
       number = {3},
        pages = {4070-4082},
          doi = {10.1093/mnras/stab1911},
archivePrefix = {arXiv},
       eprint = {2011.15018},
 primaryClass = {astro-ph.CO},
       adsurl = {https://ui.adsabs.harvard.edu/abs/2021MNRAS.506.4070A}
}

@ARTICLE{Tegmark_2002,
       author = {{Tegmark}, Max and {Zaldarriaga}, Matias},
        title = "{Separating the early universe from the late universe: Cosmological parameter estimation beyond the black box}",
      journal = {\prd},
         year = 2002,
        month = nov,
       volume = {66},
       number = {10},
          eid = {103508},
        pages = {103508},
          doi = {10.1103/PhysRevD.66.103508},
archivePrefix = {arXiv},
       eprint = {astro-ph/0207047},
 primaryClass = {astro-ph},
       adsurl = {https://ui.adsabs.harvard.edu/abs/2002PhRvD..66j3508T}
}

@ARTICLE{vlah20,
       author = {{Vlah}, Zvonimir and {Chisari}, Nora Elisa and {Schmidt}, Fabian},
        title = "{An EFT description of galaxy intrinsic alignments}",
      journal = {\jcap},
         year = 2020,
        month = jan,
       volume = {2020},
       number = {1},
          eid = {025},
        pages = {025},
          doi = {10.1088/1475-7516/2020/01/025},
archivePrefix = {arXiv},
       eprint = {1910.08085},
 primaryClass = {astro-ph.CO},
       adsurl = {https://ui.adsabs.harvard.edu/abs/2020JCAP...01..025V}
}

@ARTICLE{fortuna21,
       author = {{Fortuna}, Maria Cristina and {Hoekstra}, Henk and {Joachimi}, Benjamin and {Johnston}, Harry and {Chisari}, Nora Elisa and {Georgiou}, Christos and {Mahony}, Constance},
        title = "{The halo model as a versatile tool to predict intrinsic alignments}",
      journal = {\mnras},
         year = 2021,
        month = feb,
       volume = {501},
       number = {2},
        pages = {2983-3002},
          doi = {10.1093/mnras/staa3802},
archivePrefix = {arXiv},
       eprint = {2003.02700},
 primaryClass = {astro-ph.CO},
       adsurl = {https://ui.adsabs.harvard.edu/abs/2021MNRAS.501.2983F}
}

@ARTICLE{maion24,
       author = {{Maion}, Francisco and {Angulo}, Raul E. and {Bakx}, Thomas and {Chisari}, Nora Elisa and {Kurita}, Toshiki and {Pellejero-Ib{\'a}{\~n}ez}, Marcos},
        title = "{HYMALAIA: a hybrid lagrangian model for intrinsic alignments}",
      journal = {\mnras},
         year = 2024,
        month = jun,
       volume = {531},
       number = {2},
        pages = {2684-2700},
          doi = {10.1093/mnras/stae1331},
archivePrefix = {arXiv},
       eprint = {2307.13754},
 primaryClass = {astro-ph.CO},
       adsurl = {https://ui.adsabs.harvard.edu/abs/2024MNRAS.531.2684M}
}

@ARTICLE{takahashi12,
       author = {{Takahashi}, Ryuichi and {Sato}, Masanori and {Nishimichi}, Takahiro and {Taruya}, Atsushi and {Oguri}, Masamune},
        title = "{Revising the Halofit Model for the Nonlinear Matter Power Spectrum}",
      journal = {\apj},
         year = 2012,
        month = dec,
       volume = {761},
       number = {2},
          eid = {152},
        pages = {152},
          doi = {10.1088/0004-637X/761/2/152},
archivePrefix = {arXiv},
       eprint = {1208.2701},
 primaryClass = {astro-ph.CO},
       adsurl = {https://ui.adsabs.harvard.edu/abs/2012ApJ...761..152T}
}

@ARTICLE{knabenhans21,
       author = {{Knabenhans}, M. and {Stadel}, J. and {Potter}, D. and {Dakin}, J. and {Hannestad}, S. and others},
        title = "{Euclid preparation: IX. EuclidEmulator2 - power spectrum emulation with massive neutrinos and self-consistent dark energy perturbations}",
      journal = {\mnras},
         year = 2021,
        month = aug,
       volume = {505},
       number = {2},
        pages = {2840-2869},
          doi = {10.1093/mnras/stab1366},
archivePrefix = {arXiv},
       eprint = {2010.11288},
 primaryClass = {astro-ph.CO},
       adsurl = {https://ui.adsabs.harvard.edu/abs/2021MNRAS.505.2840E}
}

@ARTICLE{bartelmann01,
       author = {{Bartelmann}, M. and {Schneider}, P.},
        title = "{Weak gravitational lensing}",
      journal = {PhysRep},
         year = 2001,
        month = jan,
       volume = {340},
       number = {4-5},
        pages = {291-472},
          doi = {10.1016/S0370-1573(00)00082-X},
archivePrefix = {arXiv},
       eprint = {astro-ph/9912508},
 primaryClass = {astro-ph},
       adsurl = {https://ui.adsabs.harvard.edu/abs/2001PhR...340..291B}
}

@ARTICLE{linke24,
       author = {{Linke}, L. and {Unruh}, S. and {Wittje}, A. and {Schrabback}, T. and {Grandis}, S. and others},
        title = "{Euclid and KiDS-1000: Quantifying the impact of source-lens clustering on cosmic shear analyses}",
      journal = {\aap},
         year = 2025,
        month = jan,
       volume = {693},
          eid = {A210},
        pages = {A210},
          doi = {10.1051/0004-6361/202451494},
archivePrefix = {arXiv},
       eprint = {2407.09810},
 primaryClass = {astro-ph.CO},
       adsurl = {https://ui.adsabs.harvard.edu/abs/2025A&A...693A.210L}
}

@ARTICLE{blazek19,
       author = {{Blazek}, Jonathan A. and {MacCrann}, Niall and {Troxel}, M.~A. and {Fang}, Xiao},
        title = "{Beyond linear galaxy alignments}",
      journal = {\prd},
         year = 2019,
        month = nov,
       volume = {100},
       number = {10},
          eid = {103506},
        pages = {103506},
          doi = {10.1103/PhysRevD.100.103506},
archivePrefix = {arXiv},
       eprint = {1708.09247},
 primaryClass = {astro-ph.CO},
       adsurl = {https://ui.adsabs.harvard.edu/abs/2019PhRvD.100j3506B}
}

@ARTICLE{troxel15,
       author = {{Troxel}, M.~A. and {Ishak}, Mustapha},
        title = "{The intrinsic alignment of galaxies and its impact on weak gravitational lensing in an era of precision cosmology}",
      journal = {physrep},
         year = 2015,
        month = feb,
       volume = {558},
        pages = {1-59},
          doi = {10.1016/j.physrep.2014.11.001},
archivePrefix = {arXiv},
       eprint = {1407.6990},
 primaryClass = {astro-ph.CO},
       adsurl = {https://ui.adsabs.harvard.edu/abs/2015PhR...558....1T}
}

@article{alonso_namaster,
  title = {A Unified Pseudo-{{C$\ell$}} Framework},
  author = {Alonso, David and Sanchez, Javier and Slosar, An{\v z}e and {LSST Dark Energy Science Collaboration}},
  year = {2019},
  month = apr,
  journal = {Monthly Notices of the Royal Astronomical Society},
  volume = {484},
  pages = {4127--4151},
  publisher = {OUP},
  issn = {0035-8711},
  doi = {10.1093/mnras/stz093},
  urldate = {2024-05-23}
}

@article{hikage_pseudocell,
  title = {Shear Power Spectrum Reconstruction Using the Pseudo-Spectrum Method},
  author = {Hikage, Chiaki and Takada, Masahiro and Hamana, Takashi and Spergel, David},
  year = {2011},
  month = mar,
  journal = {Monthly Notices of the Royal Astronomical Society},
  volume = {412},
  pages = {65--74},
  publisher = {OUP},
  issn = {0035-8711},
  doi = {10.1111/j.1365-2966.2010.17886.x},
  urldate = {2024-05-23}
}

@article{becker_bmode2013,
  title = {Cosmic Shear {{E}}/{{B-mode}} Estimation with Binned Correlation Function Data},
  author = {Becker, Matthew R.},
  year = {2013},
  month = oct,
  journal = {Monthly Notices of the Royal Astronomical Society},
  volume = {435},
  pages = {1547--1562},
  publisher = {OUP},
  issn = {0035-8711},
  doi = {10.1093/mnras/stt1396},
  urldate = {2024-05-23}
}

@article{becker_bandpower2016,
  title = {Fourier Band-Power {{E}}/{{B-mode}} Estimators for Cosmic Shear},
  author = {Becker, Matthew R. and Rozo, Eduardo},
  year = {2016},
  month = mar,
  journal = {\mnras},
  volume = {457},
  pages = {304--312},
  publisher = {OUP},
  issn = {0035-8711},
  doi = {10.1093/mnras/stv3018},
  urldate = {2024-05-23}
}

@ARTICLE{jarvis04,
       author = {{Jarvis}, M. and {Bernstein}, G. and {Jain}, B.},
        title = "{The skewness of the aperture mass statistic}",
      journal = {\mnras},
         year = 2004,
        month = jul,
       volume = {352},
       number = {1},
        pages = {338-352},
          doi = {10.1111/j.1365-2966.2004.07926.x},
archivePrefix = {arXiv},
       eprint = {astro-ph/0307393},
 primaryClass = {astro-ph},
       adsurl = {https://ui.adsabs.harvard.edu/abs/2004MNRAS.352..338J}
}

@ARTICLE{campos23,
       author = {{Campos}, A. and {Samuroff}, S. and {Mandelbaum}, R.},
        title = "{An empirical approach to model selection: weak lensing and intrinsic alignments}",
      journal = {\mnras},
         year = 2023,
        month = oct,
       volume = {525},
       number = {2},
        pages = {1885-1901},
          doi = {10.1093/mnras/stad2213},
archivePrefix = {arXiv},
       eprint = {2211.02800},
 primaryClass = {astro-ph.CO},
       adsurl = {https://ui.adsabs.harvard.edu/abs/2023MNRAS.525.1885C}
}

@ARTICLE{des-kids,
       author = {{DES and KiDS Collaboration}},
        title = "{DES Y3 + KiDS-1000: Consistent cosmology combining cosmic shear surveys}",
      journal = {The Open Journal of Astrophysics},
         year = 2023,
        month = oct,
       volume = {6},
          eid = {36},
        pages = {36},
          doi = {10.21105/astro.2305.17173},
archivePrefix = {arXiv},
       eprint = {2305.17173},
 primaryClass = {astro-ph.CO},
       adsurl = {https://ui.adsabs.harvard.edu/abs/2023OJAp....6E..36D}
}

@ARTICLE{brown02,
       author = {{Brown}, M.~L. and {Taylor}, A.~N. and {Hambly}, N.~C. and {Dye}, S.},
        title = "{Measurement of intrinsic alignments in galaxy ellipticities}",
      journal = {\mnras},
         year = 2002,
        month = jul,
       volume = {333},
       number = {3},
        pages = {501-509},
          doi = {10.1046/j.1365-8711.2002.05354.x},
archivePrefix = {arXiv},
       eprint = {astro-ph/0009499},
 primaryClass = {astro-ph},
       adsurl = {https://ui.adsabs.harvard.edu/abs/2002MNRAS.333..501B}
}

@ARTICLE{zuntz15,
       author = {{Zuntz}, J. and {Paterno}, M. and {Jennings}, E. and {Rudd}, D. and {Manzotti}, A. and {Dodelson}, S. and {Bridle}, S. and {Sehrish}, S. and {Kowalkowski}, J.},
        title = "{CosmoSIS: Modular cosmological parameter estimation}",
      journal = {Astronomy and Computing},
         year = 2015,
        month = sep,
       volume = {12},
        pages = {45-59},
          doi = {10.1016/j.ascom.2015.05.005},
archivePrefix = {arXiv},
       eprint = {1409.3409},
 primaryClass = {astro-ph.CO},
       adsurl = {https://ui.adsabs.harvard.edu/abs/2015A&C....12...45Z}
}

@ARTICLE{lange23,
       author = {{Lange}, Johannes U.},
        title = "{NAUTILUS: boosting Bayesian importance nested sampling with deep learning}",
      journal = {\mnras},
         year = 2023,
        month = oct,
       volume = {525},
       number = {2},
        pages = {3181-3194},
          doi = {10.1093/mnras/stad2441},
archivePrefix = {arXiv},
       eprint = {2306.16923},
 primaryClass = {astro-ph.IM},
       adsurl = {https://ui.adsabs.harvard.edu/abs/2023MNRAS.525.3181L}
}

@ARTICLE{muir_blinding,
       author = {{Muir}, J. and {Bernstein}, G.~M. and {Huterer}, D. and {Elsner}, F. and {Krause}, E. and others},
        title = "{Blinding multiprobe cosmological experiments}",
      journal = {\mnras},
         year = 2020,
        month = may,
       volume = {494},
       number = {3},
        pages = {4454-4470},
          doi = {10.1093/mnras/staa965},
archivePrefix = {arXiv},
       eprint = {1911.05929},
 primaryClass = {astro-ph.CO},
       adsurl = {https://ui.adsabs.harvard.edu/abs/2020MNRAS.494.4454M}
}

@article{sheldon20,
	adsurl = {https://ui.adsabs.harvard.edu/abs/2020ApJ...902..138S},
	archiveprefix = {arXiv},
	author = {{Sheldon}, Erin S. and {Becker}, Matthew R. and {MacCrann}, Niall and {Jarvis}, Michael},
	eid = {138},
	eprint = {1911.02505},
	journal = {\apj},
	month = {October},
	number = {2},
	pages = {138},
	primaryclass = {astro-ph.CO},
	title = {{Mitigating Shear-dependent Object Detection Biases with Metacalibration}},
	volume = {902},
	year = {2020}}

@ARTICLE{schneider02b,
       author = {{Schneider}, P. and {van Waerbeke}, L. and {Kilbinger}, M. and {Mellier}, Y.},
        title = "{Analysis of two-point statistics of cosmic shear. I. Estimators and covariances}",
      journal = {\aap},
         year = 2002,
        month = dec,
       volume = {396},
        pages = {1-19},
          doi = {10.1051/0004-6361:20021341},
archivePrefix = {arXiv},
       eprint = {astro-ph/0206182},
 primaryClass = {astro-ph},
       adsurl = {https://ui.adsabs.harvard.edu/abs/2002A&A...396....1S}
}

@ARTICLE{paus_ia,
       author = {{Navarro-Giron{\'e}s}, D. and {Crocce}, M. and {Gazta{\~n}aga}, E. and {Wittje}, A. and {Siudek}, M. and others},
        title = "{The PAU Survey: measuring intrinsic galaxy alignments in deep wide fields as a function of colour, luminosity, stellar mass, and redshift}",
      journal = {\mnras},
         year = 2026,
        month = jan,
       volume = {545},
       number = {2},
          eid = {staf1630},
        pages = {staf1630},
          doi = {10.1093/mnras/staf1630},
archivePrefix = {arXiv},
       eprint = {2505.15470},
 primaryClass = {astro-ph.CO},
       adsurl = {https://ui.adsabs.harvard.edu/abs/2026MNRAS.545f1630N}
}

@article{Samuroff_Y1IA,
       author = {{Samuroff}, S. and {Blazek}, J. and {Troxel}, M.~A. and others},
        title = "{Dark Energy Survey Year 1 results: constraints on intrinsic alignments and their colour dependence from galaxy clustering and weak lensing}",
      journal = {\mnras},
         year = 2019,
        month = nov,
       volume = {489},
       number = {4},
        pages = {5453-5482},
          doi = {10.1093/mnras/stz2197},
archivePrefix = {arXiv},
       eprint = {1811.06989},
 primaryClass = {astro-ph.CO},
       adsurl = {https://ui.adsabs.harvard.edu/abs/2019MNRAS.489.5453S}
}

@article{KiDSLegacy_consistency,
   title={KiDS-Legacy: Consistency of cosmic shear measurements and joint cosmological constraints with external probes},
   volume={702},
   ISSN={1432-0746},
   url={http://dx.doi.org/10.1051/0004-6361/202554893},
   DOI={10.1051/0004-6361/202554893},
   journal={Astronomy and Astrophysics},
   publisher={EDP Sciences},
   author={Stölzner, Benjamin and Wright, Angus H. and Asgari, Marika and others},
   year={2025},
   month=oct, pages={A169} }

@ARTICLE{vanDaalen2011,
       author = {{van Daalen}, Marcel P. and {Schaye}, Joop and {Booth}, C.~M. and {Dalla Vecchia}, Claudio},
        title = "{The effects of galaxy formation on the matter power spectrum: a challenge for precision cosmology}",
      journal = {MNRAS},
         year = 2011,
        month = aug,
       volume = {415},
       number = {4},
        pages = {3649-3665},
          doi = {10.1111/j.1365-2966.2011.18981.x},
archivePrefix = {arXiv},
       eprint = {1104.1174},
 primaryClass = {astro-ph.CO},
       adsurl = {https://ui.adsabs.harvard.edu/abs/2011MNRAS.415.3649V}
}

@ARTICLE{Siegel_DESIIA,
       author = {{Siegel}, J. and {McCullough}, J. and {Amon}, A. and {Lamman}, C. and {Jeffrey}, N. and others},
        title = "{Intrinsic alignment demographics for next-generation lensing: Revealing galaxy property trends with DESI Y1 direct measurements}",
      journal = {\mnras},
         year = 2026,
        month = sep,
          doi = {10.1093/mnras/stag1691},
archivePrefix = {arXiv},
       eprint = {2507.11530},
 primaryClass = {astro-ph.CO},
       adsurl = {https://ui.adsabs.harvard.edu/abs/2026MNRAS.tmp.1558S}
}

@misc{mccullough2024,
      author={J. McCullough and A. Amon and E. Legnani and D. Gruen and A. Roodman and others},
        title = "{Dark Energy Survey Year 3: Blue shear}",
      journal = {\prd},
         year = 2026,
        month = may,
       volume = {113},
       number = {10},
          eid = {103509},
        pages = {103509},
          doi = {10.1103/wmp3-qc3p},
archivePrefix = {arXiv},
       eprint = {2410.22272},
 primaryClass = {astro-ph.CO},
       adsurl = {https://ui.adsabs.harvard.edu/abs/2026PhRvD.113j3509M}
}

@ARTICLE{schaller25,
       author = {{Schaller}, Matthieu and {Schaye}, Joop},
        title = "{An analytic redshift-independent formulation of baryonic effects on the matter power spectrum}",
      journal = {\mnras},
         year = 2025,
        month = jul,
       volume = {540},
       number = {3},
        pages = {2322-2330},
          doi = {10.1093/mnras/staf871},
archivePrefix = {arXiv},
       eprint = {2504.15633},
 primaryClass = {astro-ph.CO},
       adsurl = {https://ui.adsabs.harvard.edu/abs/2025MNRAS.540.2322S}
}

@ARTICLE{samuroff23,
       author = {{Samuroff}, S. and {Mandelbaum}, R. and {Blazek}, J. and {Campos}, A. and {MacCrann}, N. and others},
        title = "{The Dark Energy Survey Year 3 and eBOSS: constraining galaxy intrinsic alignments across luminosity and colour space}",
      journal = {\mnras},
         year = 2023,
        month = sep,
       volume = {524},
       number = {2},
        pages = {2195-2223},
          doi = {10.1093/mnras/stad2013},
archivePrefix = {arXiv},
       eprint = {2212.11319},
 primaryClass = {astro-ph.CO},
       adsurl = {https://ui.adsabs.harvard.edu/abs/2023MNRAS.524.2195S}
}

@ARTICLE{blazek15,
       author = {{Blazek}, Jonathan and {Vlah}, Zvonimir and {Seljak}, Uro{\v{s}}},
        title = "{Tidal alignment of galaxies}",
      journal = {\jcap},
         year = 2015,
        month = aug,
       volume = {2015},
       number = {8},
        pages = {015-015},
          doi = {10.1088/1475-7516/2015/08/015},
archivePrefix = {arXiv},
       eprint = {1504.02510},
 primaryClass = {astro-ph.CO},
       adsurl = {https://ui.adsabs.harvard.edu/abs/2015JCAP...08..015B}
}

@ARTICLE{bakx23,
       author = {{Bakx}, Thomas and {Kurita}, Toshiki and {Chisari}, Nora Elisa and {Vlah}, Zvonimir and {Schmidt}, Fabian},
        title = "{Effective field theory of intrinsic alignments at one loop order: a comparison to dark matter simulations}",
      journal = {\jcap},
         year = 2023,
        month = oct,
       volume = {2023},
       number = {10},
          eid = {005},
        pages = {005},
          doi = {10.1088/1475-7516/2023/10/005},
archivePrefix = {arXiv},
       eprint = {2303.15565},
 primaryClass = {astro-ph.CO},
       adsurl = {https://ui.adsabs.harvard.edu/abs/2023JCAP...10..005B}
}

@ARTICLE{kiessling15,
       author = {{Kiessling}, Alina and {Cacciato}, Marcello and {Joachimi}, Benjamin and {Kirk}, Donnacha and {Kitching}, Thomas D. and others},
        title = "{Galaxy Alignments: Theory, Modelling \& Simulations}",
      journal = {SSR},
         year = 2015,
        month = nov,
       volume = {193},
       number = {1-4},
        pages = {67-136},
          doi = {10.1007/s11214-015-0203-6},
archivePrefix = {arXiv},
       eprint = {1504.05546},
 primaryClass = {astro-ph.GA},
       adsurl = {https://ui.adsabs.harvard.edu/abs/2015SSRv..193...67K}
}

@ARTICLE{mccarthy18,
       author = {{McCarthy}, Ian G. and {Bird}, Simeon and {Schaye}, Joop and {Harnois-Deraps}, Joachim and {Font}, Andreea S. and {van Waerbeke}, Ludovic},
        title = "{The BAHAMAS project: the CMB-large-scale structure tension and the roles of massive neutrinos and galaxy formation}",
      journal = {\mnras},
         year = 2018,
        month = may,
       volume = {476},
       number = {3},
        pages = {2999-3030},
          doi = {10.1093/mnras/sty377},
archivePrefix = {arXiv},
       eprint = {1712.02411},
 primaryClass = {astro-ph.CO},
       adsurl = {https://ui.adsabs.harvard.edu/abs/2018MNRAS.476.2999M}
}

@ARTICLE{hirata04,
       author = {{Hirata}, Christopher M. and {Seljak}, Uro{\v{s}}},
        title = "{Intrinsic alignment-lensing interference as a contaminant of cosmic shear}",
      journal = {\prd},
         year = 2004,
        month = sep,
       volume = {70},
       number = {6},
          eid = {063526},
        pages = {063526},
          doi = {10.1103/PhysRevD.70.063526},
archivePrefix = {arXiv},
       eprint = {astro-ph/0406275},
 primaryClass = {astro-ph},
       adsurl = {https://ui.adsabs.harvard.edu/abs/2004PhRvD..70f3526H}
}

@ARTICLE{bridle07,
       author = {{Bridle}, Sarah and {King}, Lindsay},
        title = "{Dark energy constraints from cosmic shear power spectra: impact of intrinsic alignments on photometric redshift requirements}",
      journal = {New Journal of Physics},
         year = 2007,
        month = dec,
       volume = {9},
       number = {12},
        pages = {444},
          doi = {10.1088/1367-2630/9/12/444},
archivePrefix = {arXiv},
       eprint = {0705.0166},
 primaryClass = {astro-ph},
       adsurl = {https://ui.adsabs.harvard.edu/abs/2007NJPh....9..444B}
}

@ARTICLE{joachimi21,
       author = {{Joachimi}, B. and {Lin}, C.-A. and {Asgari}, M. and {Tr{\"o}ster}, T. and {Heymans}, C. and others},
        title = "{KiDS-1000 methodology: Modelling and inference for joint weak gravitational lensing and spectroscopic galaxy clustering analysis}",
      journal = {\aap},
         year = 2021,
        month = feb,
       volume = {646},
          eid = {A129},
        pages = {A129},
          doi = {10.1051/0004-6361/202038831},
archivePrefix = {arXiv},
       eprint = {2007.01844},
 primaryClass = {astro-ph.CO},
       adsurl = {https://ui.adsabs.harvard.edu/abs/2021A&A...646A.129J}
}

@ARTICLE{ge25,
       author = {{Ge}, F. and {Millea}, M. and {Camphuis}, E. and {Daley}, C. and {Huang}, N. and others},
        title = "{Cosmology from CMB lensing and delensed EE power spectra using 2019-2020 SPT-3G polarization data}",
      journal = {\prd},
         year = 2025,
        month = apr,
       volume = {111},
       number = {8},
          eid = {083534},
        pages = {083534},
          doi = {10.1103/PhysRevD.111.083534},
archivePrefix = {arXiv},
       eprint = {2411.06000},
 primaryClass = {astro-ph.CO},
       adsurl = {https://ui.adsabs.harvard.edu/abs/2025PhRvD.111h3534G}
}

@ARTICLE{madhavacheril25,
       author = {{Madhavacheril}, Mathew S. and {Qu}, Frank J. and {Sherwin}, Blake D. and {MacCrann}, Niall and {Li}, Yaqiong and others},
        title = "{The Atacama Cosmology Telescope: DR6 Gravitational Lensing Map and Cosmological Parameters}",
      journal = {\apj},
         year = 2024,
        month = feb,
       volume = {962},
       number = {2},
          eid = {113},
        pages = {113},
          doi = {10.3847/1538-4357/acff5f},
archivePrefix = {arXiv},
       eprint = {2304.05203},
 primaryClass = {astro-ph.CO},
       adsurl = {https://ui.adsabs.harvard.edu/abs/2024ApJ...962..113M}
}

@ARTICLE{qu25,
       author = {{Qu}, Frank J. and {Ge}, Fei and {Kimmy Wu}, W.~L. and {Abril-Cabezas}, Irene and {Madhavacheril}, Mathew S. and others},
        title = "{Unified and Consistent Structure Growth Measurements from Joint ACT, SPT, and Planck CMB Lensing}",
      journal = {\prl},
         year = 2026,
        month = jan,
       volume = {136},
       number = {2},
          eid = {021001},
        pages = {021001},
          doi = {10.1103/k5yr-3h6d},
archivePrefix = {arXiv},
       eprint = {2504.20038},
 primaryClass = {astro-ph.CO},
       adsurl = {https://ui.adsabs.harvard.edu/abs/2026PhRvL.136b1001Q}
}

@ARTICLE{carron22,
       author = {{Carron}, Julien and {Mirmelstein}, Mark and {Lewis}, Antony},
        title = "{CMB lensing from Planck PR4 maps}",
      journal = {\jcap},
         year = 2022,
        month = sep,
       volume = {2022},
       number = {9},
          eid = {039},
        pages = {039},
          doi = {10.1088/1475-7516/2022/09/039},
archivePrefix = {arXiv},
       eprint = {2206.07773},
 primaryClass = {astro-ph.CO},
       adsurl = {https://ui.adsabs.harvard.edu/abs/2022JCAP...09..039C}
}

@ARTICLE{desidr2,
       author = {{DESI Collaboration}},
        title = "{DESI DR2 results. I. Baryon acoustic oscillations from the Lyman alpha forest}",
      journal = {\prd},
         year = 2025,
        month = oct,
       volume = {112},
       number = {8},
          eid = {083514},
        pages = {083514},
          doi = {10.1103/2wwn-xjm5},
archivePrefix = {arXiv},
       eprint = {2503.14739},
 primaryClass = {astro-ph.CO},
       adsurl = {https://ui.adsabs.harvard.edu/abs/2025PhRvD.112h3514A}
}

@ARTICLE{desi_dr2_cosmo,
        title = "{DESI DR2 results. II. Measurements of baryon acoustic oscillations and cosmological constraints}",
      journal = {\prd},
         year = 2025,
        month = oct,
       volume = {112},
       number = {8},
          eid = {083515},
        pages = {083515},
          doi = {10.1103/tr6y-kpc6},
archivePrefix = {arXiv},
       eprint = {2503.14738},
 primaryClass = {astro-ph.CO},
       adsurl = {https://ui.adsabs.harvard.edu/abs/2025PhRvD.112h3515A}
}

@ARTICLE{dalal23,
       author = {{Dalal}, Roohi and {Li}, Xiangchong and {Nicola}, Andrina and {Zuntz}, Joe and {Strauss}, Michael A. and others},
        title = "{Hyper Suprime-Cam Year 3 results: Cosmology from cosmic shear power spectra}",
      journal = {\prd},
         year = 2023,
        month = dec,
       volume = {108},
       number = {12},
          eid = {123519},
        pages = {123519},
          doi = {10.1103/PhysRevD.108.123519},
archivePrefix = {arXiv},
       eprint = {2304.00701},
 primaryClass = {astro-ph.CO},
       adsurl = {https://ui.adsabs.harvard.edu/abs/2023PhRvD.108l3519D}
}

@ARTICLE{fang20,
       author = {{Fang}, Xiao and {Eifler}, Tim and {Krause}, Elisabeth},
        title = "{2D-FFTLog: efficient computation of real-space covariance matrices for galaxy clustering and weak lensing}",
      journal = {\mnras},
         year = 2020,
        month = sep,
       volume = {497},
       number = {3},
        pages = {2699-2714},
          doi = {10.1093/mnras/staa1726},
archivePrefix = {arXiv},
       eprint = {2004.04833},
 primaryClass = {astro-ph.CO},
       adsurl = {https://ui.adsabs.harvard.edu/abs/2020MNRAS.497.2699F}
}

@ARTICLE{li23,
       author = {{Li}, Xiangchong and {Zhang}, Tianqing and {Sugiyama}, Sunao and {Dalal}, Roohi and {Terasawa}, Ryo and others},
        title = "{Hyper Suprime-Cam Year 3 results: Cosmology from cosmic shear two-point correlation functions}",
      journal = {\prd},
         year = 2023,
        month = dec,
       volume = {108},
       number = {12},
          eid = {123518},
        pages = {123518},
          doi = {10.1103/PhysRevD.108.123518},
archivePrefix = {arXiv},
       eprint = {2304.00702},
 primaryClass = {astro-ph.CO},
       adsurl = {https://ui.adsabs.harvard.edu/abs/2023PhRvD.108l3518L}
}

@ARTICLE{Planck_legacy,
       author = {{Planck Collaboration} and others},
        title = "{Planck 2018 results. I. Overview and the cosmological legacy of Planck}",
      journal = {\aap},
         year = 2020,
        month = sep,
       volume = {641},
          eid = {A1},
        pages = {A1},
          doi = {10.1051/0004-6361/201833880},
archivePrefix = {arXiv},
       eprint = {1807.06205},
 primaryClass = {astro-ph.CO},
       adsurl = {https://ui.adsabs.harvard.edu/abs/2020A&A...641A...1P}
}

@ARTICLE{wittman00,
       author = {{Wittman}, David M. and {Tyson}, J. Anthony and {Kirkman}, David and {Dell'Antonio}, Ian and {Bernstein}, Gary},
        title = "{Detection of weak gravitational lensing distortions of distant galaxies by cosmic dark matter at large scales}",
      journal = {\nat},
         year = 2000,
        month = may,
       volume = {405},
       number = {6783},
        pages = {143-148},
          doi = {10.1038/35012001},
archivePrefix = {arXiv},
       eprint = {astro-ph/0003014},
 primaryClass = {astro-ph},
       adsurl = {https://ui.adsabs.harvard.edu/abs/2000Natur.405..143W}
}

@ARTICLE{schneider25,
       author = {{Schneider}, Aurel and {Kova{\v{c}}}, Michael and {Bucko}, Jozef and {Nicola}, Andrina and {Reischke}, Robert and {Giri}, Sambit K. and {Teyssier}, Romain and {Tr{\"o}ster}, Tilman and {Refregier}, Alexandre and {Schaller}, Matthieu and {Schaye}, Joop},
        title = "{Baryonification: an alternative to hydrodynamical simulations for cosmological studies}",
      journal = {\jcap},
         year = 2025,
        month = dec,
       volume = {2025},
       number = {12},
          eid = {043},
        pages = {043},
          doi = {10.1088/1475-7516/2025/12/043},
archivePrefix = {arXiv},
       eprint = {2507.07892},
 primaryClass = {astro-ph.CO},
       adsurl = {https://ui.adsabs.harvard.edu/abs/2025JCAP...12..043S}
}

@ARTICLE{vanWaerbeke00,
       author = {{\PRF{Waerbeke}{Van}{Van} Waerbeke}, L. and {Mellier}, Y. and {Erben}, T. and {Cuillandre}, J.~C. and others},
        title = "{Detection of correlated galaxy ellipticities from CFHT data: first evidence for gravitational lensing by large-scale structures}",
      journal = {\aap},
         year = 2000,
        month = jun,
       volume = {358},
        pages = {30-44},
          doi = {10.48550/arXiv.astro-ph/0002500},
archivePrefix = {arXiv},
       eprint = {astro-ph/0002500},
 primaryClass = {astro-ph},
       adsurl = {https://ui.adsabs.harvard.edu/abs/2000A&A...358...30V}
}

@ARTICLE{kaiser00,
       author = {{Kaiser}, Nick and {Wilson}, Gillian and {Luppino}, Gerard A.},
        title = "{Large-Scale Cosmic Shear Measurements}",
      journal = {arXiv e-prints},
         year = 2000,
        month = mar,
          eid = {astro-ph/0003338},
        pages = {astro-ph/0003338},
          doi = {10.48550/arXiv.astro-ph/0003338},
archivePrefix = {arXiv},
       eprint = {astro-ph/0003338},
 primaryClass = {astro-ph},
       adsurl = {https://ui.adsabs.harvard.edu/abs/2000astro.ph..3338K}
}

@ARTICLE{bacon00,
       author = {{Bacon}, David J. and {Refregier}, Alexandre R. and {Ellis}, Richard S.},
        title = "{Detection of weak gravitational lensing by large-scale structure}",
      journal = {\mnras},
         year = 2000,
        month = oct,
       volume = {318},
       number = {2},
        pages = {625-640},
          doi = {10.1046/j.1365-8711.2000.03851.x},
archivePrefix = {arXiv},
       eprint = {astro-ph/0003008},
 primaryClass = {astro-ph},
       adsurl = {https://ui.adsabs.harvard.edu/abs/2000MNRAS.318..625B}
}

@book{jeffreys39,
  title={The theory of probability},
  author={Jeffreys, Harold},
  year={1939},
  publisher={OUP Oxford}
}

@article{ACT_DR6_extended_models,
       author = {{Calabrese}, Erminia and {Hill}, J. Colin and {Jense}, Hidde T. and {La Posta}, Adrien and {Abril-Cabezas}, Irene and others},
      journal = {\jcap},
         year = 2025,
        month = nov,
       volume = {2025},
       number = {11},
          eid = {063},
        pages = {063},
          doi = {10.1088/1475-7516/2025/11/063},
archivePrefix = {arXiv},
       eprint = {2503.14454},
 primaryClass = {astro-ph.CO},
       adsurl = {https://ui.adsabs.harvard.edu/abs/2025JCAP...11..063C}
}

@ARTICLE{eBOSS_DR14,
       author = {{Abolfathi}, Bela and {Aguado}, D.~S. and {Aguilar}, Gabriela and {Allende Prieto}, Carlos and {Almeida}, Andres and others},
        title = "{The Fourteenth Data Release of the Sloan Digital Sky Survey: First Spectroscopic Data from the Extended Baryon Oscillation Spectroscopic Survey and from the Second Phase of the Apache Point Observatory Galactic Evolution Experiment}",
      journal = {\apjs},
         year = 2018,
        month = apr,
       volume = {235},
       number = {2},
          eid = {42},
        pages = {42},
          doi = {10.3847/1538-4365/aa9e8a},
archivePrefix = {arXiv},
       eprint = {1707.09322},
 primaryClass = {astro-ph.GA},
       adsurl = {https://ui.adsabs.harvard.edu/abs/2018ApJS..235...42A}
}

@ARTICLE{Chabanier_lya,
       author = {{Chabanier}, Sol{\`e}ne and {Millea}, Marius and {Palanque-Delabrouille}, Nathalie},
        title = "{Matter power spectrum: from Ly {\ensuremath{\alpha}} forest to CMB scales}",
      journal = {\mnras},
         year = 2019,
        month = oct,
       volume = {489},
       number = {2},
        pages = {2247-2253},
          doi = {10.1093/mnras/stz2310},
archivePrefix = {arXiv},
       eprint = {1905.08103},
 primaryClass = {astro-ph.CO},
       adsurl = {https://ui.adsabs.harvard.edu/abs/2019MNRAS.489.2247C}
}

@ARTICLE{deJong13,
       author = {{\PRF{Jong}{de}{de} Jong}, J.~T.~A. and {Kuijken}, K. and {Applegate}, D. and {Begeman}, K. and {Belikov}, A. and others},
        title = "{The Kilo-Degree Survey}",
      journal = {The Messenger},
         year = 2013,
        month = dec,
       volume = {154},
        pages = {44-46},
       adsurl = {https://ui.adsabs.harvard.edu/abs/2013Msngr.154...44D}
}

@ARTICLE{aihara22,
       author = {{Aihara}, Hiroaki and {AlSayyad}, Yusra and {Ando}, Makoto and {Armstrong}, Robert and {Bosch}, James and others},
        title = "{Third data release of the Hyper Suprime-Cam Subaru Strategic Program}",
      journal = {PASJ},
         year = 2022,
        month = apr,
       volume = {74},
       number = {2},
        pages = {247-272},
          doi = {10.1093/pasj/psab122},
archivePrefix = {arXiv},
       eprint = {2108.13045},
 primaryClass = {astro-ph.IM},
       adsurl = {https://ui.adsabs.harvard.edu/abs/2022PASJ...74..247A}
}

@ARTICLE{des_proposal,
       author = {{Dark Energy Survey Collaboration}},
        title = "{The Dark Energy Survey}",
      journal = {arXiv e-prints},
         year = 2005,
        month = oct,
          eid = {astro-ph/0510346},
        pages = {astro-ph/0510346},
          doi = {10.48550/arXiv.astro-ph/0510346},
archivePrefix = {arXiv},
       eprint = {astro-ph/0510346},
 primaryClass = {astro-ph},
       adsurl = {https://ui.adsabs.harvard.edu/abs/2005astro.ph.10346T}
}

@ARTICLE{MiraldaEscude91,
       author = {{Miralda-Escude}, Jordi},
        title = "{The Correlation Function of Galaxy Ellipticities Produced by Gravitational Lensing}",
      journal = {\apj},
         year = 1991,
        month = oct,
       volume = {380},
        pages = {1},
          doi = {10.1086/170555},
       adsurl = {https://ui.adsabs.harvard.edu/abs/1991ApJ...380....1M}
}

@ARTICLE{sv-cosmicshear,
       author = {{Dark Energy Survey Collaboration}},
        title = "{Cosmology from cosmic shear with Dark Energy Survey Science Verification data}",
      journal = {\prd},
         year = 2016,
        month = jul,
       volume = {94},
       number = {2},
          eid = {022001},
        pages = {022001},
          doi = {10.1103/PhysRevD.94.022001},
archivePrefix = {arXiv},
       eprint = {1507.05552},
 primaryClass = {astro-ph.CO},
       adsurl = {https://ui.adsabs.harvard.edu/abs/2016PhRvD..94b2001A}
}

@ARTICLE{troxel18,
       author = {{Troxel}, M.~A. and {MacCrann}, N. and {Zuntz}, J. and {Eifler}, T.~F. and {Krause}, E. and others},
        title = "{Dark Energy Survey Year 1 results: Cosmological constraints from cosmic shear}",
      journal = {\prd},
         year = 2018,
        month = aug,
       volume = {98},
       number = {4},
          eid = {043528},
        pages = {043528},
          doi = {10.1103/PhysRevD.98.043528},
archivePrefix = {arXiv},
       eprint = {1708.01538},
 primaryClass = {astro-ph.CO},
       adsurl = {https://ui.adsabs.harvard.edu/abs/2018PhRvD..98d3528T}
}

@article{Raveri:2021wfz,
    author = "Raveri, Marco and Doux, Cyrille",
    title = "{Non-Gaussian estimates of tensions in cosmological parameters}",
    eprint = "2105.03324",
    archivePrefix = "arXiv",
    primaryClass = "astro-ph.CO",
    doi = "10.1103/PhysRevD.104.043504",
    journal = "Phys. Rev. D",
    volume = "104",
    number = "4",
    pages = "043504",
    year = "2021"
}

@ARTICLE{Pandya2025,
       author = {{Pandya}, Sneh and {Yang}, Yuanyuan and {Van Alfen}, Nicholas and {Blazek}, Jonathan and {Walters}, Robin},
        title = "{IAEmu: Learning Galaxy Intrinsic Alignment Correlations}",
      journal = {The Open Journal of Astrophysics},
         year = 2025,
        month = dec,
       volume = {8},
        pages = {51749},
          doi = {10.33232/001c.151749},
archivePrefix = {arXiv},
       eprint = {2504.05235},
 primaryClass = {astro-ph.CO},
       adsurl = {https://ui.adsabs.harvard.edu/abs/2025OJAp....851749P}
}

@article{Raveri:2024dph,
    author = "Raveri, Marco and Doux, Cyrille and Pandey, Shivam",
    title = "{Understanding posterior projection effects with normalizing flows}",
    journal = "arXiv",
    eprint = "2409.09101",
    archivePrefix = "arXiv",
    primaryClass = "astro-ph.IM",
    month = "9",
    year = "2024"
}

@ARTICLE{Reid_SDSS_LRG,
       author = {{Reid}, Beth A. and {Percival}, Will J. and {Eisenstein}, Daniel J. and {Verde}, Licia and {Spergel}, David N. and others},
        title = "{Cosmological constraints from the clustering of the Sloan Digital Sky Survey DR7 luminous red galaxies}",
      journal = {\mnras},
         year = 2010,
        month = may,
       volume = {404},
       number = {1},
        pages = {60-85},
          doi = {10.1111/j.1365-2966.2010.16276.x},
archivePrefix = {arXiv},
       eprint = {0907.1659},
 primaryClass = {astro-ph.CO},
       adsurl = {https://ui.adsabs.harvard.edu/abs/2010MNRAS.404...60R}
}

@ARTICLE{anbajagane25,
       author = {{Anbajagane}, D. and {Chang}, C. and {Zhang}, Z. and {Tan}, C.~Y. and {Adamow}, M. and others},
        title = "{The DECADE cosmic shear project I: A new weak lensing shape catalog of 107 million galaxies}",
      journal = {The Open Journal of Astrophysics},
         year = 2025,
        month = oct,
       volume = {8},
        pages = {46158},
          doi = {10.33232/001c.146158},
archivePrefix = {arXiv},
       eprint = {2502.17674},
 primaryClass = {astro-ph.CO},
       adsurl = {https://ui.adsabs.harvard.edu/abs/2025OJAp....846158A}
}

@ARTICLE{wright24,
       author = {{Wright}, Angus H. and {Kuijken}, Konrad and {Hildebrandt}, Hendrik and {Radovich}, Mario and {Bilicki}, Maciej and others},
        title = "{The fifth data release of the Kilo Degree Survey: Multi-epoch optical/NIR imaging covering wide and legacy-calibration fields}",
      journal = {\aap},
         year = 2024,
        month = jun,
       volume = {686},
          eid = {A170},
        pages = {A170},
          doi = {10.1051/0004-6361/202346730},
archivePrefix = {arXiv},
       eprint = {2503.19439},
 primaryClass = {astro-ph.GA},
       adsurl = {https://ui.adsabs.harvard.edu/abs/2024A&A...686A.170W}
}

@ARTICLE{kids-legacy,
       author = {{Wright}, Angus H. and {St{\"o}lzner}, Benjamin and {Asgari}, Marika and {Bilicki}, Maciej and {Giblin}, Benjamin and others},
        title = "{KiDS-Legacy: Cosmological constraints from cosmic shear with the complete Kilo-Degree Survey}",
      journal = {\aap},
         year = 2025,
        month = nov,
       volume = {703},
          eid = {A158},
        pages = {A158},
          doi = {10.1051/0004-6361/202554908},
archivePrefix = {arXiv},
       eprint = {2503.19441},
 primaryClass = {astro-ph.CO},
       adsurl = {https://ui.adsabs.harvard.edu/abs/2025A&A...703A.158W}
}

@ARTICLE{decade,
       author = {{Anbajagane}, D. and {Chang}, C. and others},
        title = "{Dark Energy Camera All Data Everywhere cosmic shear project. V. Constraints on cosmology and astrophysics from 270 million galaxies across 13,000 deg2 of the sky}",
      journal = {\prd},
         year = 2026,
        month = aug,
       volume = {114},
       number = {4},
          eid = {043527},
        pages = {043527},
          doi = {10.1103/k4lq-2s73},
archivePrefix = {arXiv},
       eprint = {2509.03582},
 primaryClass = {astro-ph.CO},
       adsurl = {https://ui.adsabs.harvard.edu/abs/2026PhRvD.114d3527A}
}

@ARTICLE{paus-cosmos,
       author = {{Alarcon}, Alex and {Gaztanaga}, Enrique and {Eriksen}, Martin and {Baugh}, Carlton M. and {Cabayol}, Laura and others},
        title = "{The PAU Survey: an improved photo-z sample in the COSMOS field}",
      journal = {\mnras},
         year = 2021,
        month = mar,
       volume = {501},
       number = {4},
        pages = {6103-6122},
          doi = {10.1093/mnras/staa3659},
archivePrefix = {arXiv},
       eprint = {2007.11132},
 primaryClass = {astro-ph.GA},
       adsurl = {https://ui.adsabs.harvard.edu/abs/2021MNRAS.501.6103A}
}

@ARTICLE{paus,
       author = {{Eriksen}, M. and {Alarcon}, A. and {Gaztanaga}, E. and {Amara}, A. and {Cabayol}, L. and others},
        title = "{The PAU Survey: early demonstration of photometric redshift performance in the COSMOS field}",
      journal = {\mnras},
         year = 2019,
        month = apr,
       volume = {484},
       number = {3},
        pages = {4200-4215},
          doi = {10.1093/mnras/stz204},
archivePrefix = {arXiv},
       eprint = {1809.04375},
 primaryClass = {astro-ph.GA},
       adsurl = {https://ui.adsabs.harvard.edu/abs/2019MNRAS.484.4200E}
}

@ARTICLE{cosmos2020,
       author = {{Weaver}, J.~R. and {Kauffmann}, O.~B. and {Ilbert}, O. and {McCracken}, H.~J. and {Moneti}, A. and others},
        title = "{COSMOS2020: A Panchromatic View of the Universe to z{\ensuremath{\sim}}10 from Two Complementary Catalogs}",
      journal = {\apjs},
         year = 2022,
        month = jan,
       volume = {258},
       number = {1},
          eid = {11},
        pages = {11},
          doi = {10.3847/1538-4365/ac3078},
archivePrefix = {arXiv},
       eprint = {2110.13923},
 primaryClass = {astro-ph.GA},
       adsurl = {https://ui.adsabs.harvard.edu/abs/2022ApJS..258...11W}
}

@ARTICLE{boss_color,
       author = {{Dawson}, Kyle S. and {Schlegel}, David J. and {Ahn}, Christopher P. and {Anderson}, Scott F. and {Aubourg}, {\'E}ric and {Bailey}, Stephen and {Barkhouser}, Robert H. and {Bautista}, Julian E. and {Beifiori}, Alessandra and {Berlind}, Andreas A. and {Bhardwaj}, Vaishali and {Bizyaev}, Dmitry and {Blake}, Cullen H. and {Blanton}, Michael},
        title = "{The Baryon Oscillation Spectroscopic Survey of SDSS-III}",
      journal = {\aj},
         year = 2013,
        month = jan,
       volume = {145},
       number = {1},
          eid = {10},
        pages = {10},
          doi = {10.1088/0004-6256/145/1/10},
archivePrefix = {arXiv},
       eprint = {1208.0022},
 primaryClass = {astro-ph.CO},
       adsurl = {https://ui.adsabs.harvard.edu/abs/2013AJ....145...10D}
}

@ARTICLE{eboss_dawson,
       author = {{Dawson}, Kyle S. and {Kneib}, Jean-Paul and {Percival}, Will J. and {Alam}, Shadab and {Albareti}, Franco D. and {Anderson}, Scott F. and {Armengaud}, Eric and {Aubourg}, {\'E}ric and {Bailey}, Stephen and {Bautista}, Julian E. and {Berlind}, Andreas A. and {Bershady}, Matthew A. and {Beutler}, Florian and others},
    title = "{The SDSS-IV Extended Baryon Oscillation Spectroscopic Survey: Overview and Early Data}",
      journal = {\aj},
         year = 2016,
        month = feb,
       volume = {151},
       number = {2},
          eid = {44},
        pages = {44},
          doi = {10.3847/0004-6256/151/2/44},
archivePrefix = {arXiv},
       eprint = {1508.04473},
 primaryClass = {astro-ph.CO},
       adsurl = {https://ui.adsabs.harvard.edu/abs/2016AJ....151...44D}
}

@ARTICLE{Lewis00,
       author = {{Lewis}, Antony and {Challinor}, Anthony and {Lasenby}, Anthony},
        title = "{Efficient Computation of Cosmic Microwave Background Anisotropies in Closed Friedmann-Robertson-Walker Models}",
      journal = {\apj},
         year = 2000,
        month = aug,
       volume = {538},
       number = {2},
        pages = {473-476},
          doi = {10.1086/309179},
archivePrefix = {arXiv},
       eprint = {astro-ph/9911177},
 primaryClass = {astro-ph},
       adsurl = {https://ui.adsabs.harvard.edu/abs/2000ApJ...538..473L}
}

@article{Lewis02,
   title={Cosmological parameters from CMB and other data: A Monte Carlo approach},
   volume={66},
   ISSN={1089-4918},
   url={http://dx.doi.org/10.1103/PhysRevD.66.103511},
   DOI={10.1103/physrevd.66.103511},
   number={10},
   journal={Physical Review D},
   publisher={American Physical Society (APS)},
   author={Lewis, Antony and Bridle, Sarah},
   year={2002},
   month=nov }

@ARTICLE{mccarthy17,
       author = {{McCarthy}, Ian G. and {Schaye}, Joop and {Bird}, Simeon and {Le Brun}, Amandine M.~C.},
        title = "{The BAHAMAS project: calibrated hydrodynamical simulations for large-scale structure cosmology}",
      journal = {\mnras},
         year = 2017,
        month = mar,
       volume = {465},
       number = {3},
        pages = {2936-2965},
          doi = {10.1093/mnras/stw2792},
archivePrefix = {arXiv},
       eprint = {1603.02702},
 primaryClass = {astro-ph.CO},
       adsurl = {https://ui.adsabs.harvard.edu/abs/2017MNRAS.465.2936M}
}

@article{anacal_1,
       author = {{Li}, Xiangchong and {Mandelbaum}, Rachel and {The LSST Dark Energy Science Collaboration}},
        title = "{Analytical noise bias correction for precise weak lensing shear inference}",
      journal = {\mnras},
         year = 2025,
        month = feb,
       volume = {536},
       number = {4},
        pages = {3663-3676},
          doi = {10.1093/mnras/stae2764},
archivePrefix = {arXiv},
       eprint = {2408.06337},
 primaryClass = {astro-ph.CO},
       adsurl = {https://ui.adsabs.harvard.edu/abs/2025MNRAS.536.3663L}
}

@ARTICLE{preston24,
       author = {{Preston}, Calvin and {Amon}, Alexandra and {Efstathiou}, George},
        title = "{Reconstructing the matter power spectrum with future cosmic shear surveys}",
      journal = {\mnras},
         year = 2024,
        month = sep,
       volume = {533},
       number = {1},
        pages = {621-631},
          doi = {10.1093/mnras/stae1848},
archivePrefix = {arXiv},
       eprint = {2404.18240},
 primaryClass = {astro-ph.CO},
       adsurl = {https://ui.adsabs.harvard.edu/abs/2024MNRAS.533..621P}
}

@ARTICLE{amon22,
       author = {{Amon}, Alexandra and {Efstathiou}, George},
        title = "{A non-linear solution to the S$_{8}$ tension?}",
      journal = {\mnras},
         year = 2022,
        month = nov,
       volume = {516},
       number = {4},
        pages = {5355-5366},
          doi = {10.1093/mnras/stac2429},
archivePrefix = {arXiv},
       eprint = {2206.11794},
 primaryClass = {astro-ph.CO},
       adsurl = {https://ui.adsabs.harvard.edu/abs/2022MNRAS.516.5355A}
}

@ARTICLE{preston23,
       author = {{Preston}, Calvin and {Amon}, Alexandra and {Efstathiou}, George},
        title = "{A non-linear solution to the S$_{8}$ tension - II. Analysis of DES Year 3 cosmic shear}",
      journal = {\mnras},
         year = 2023,
        month = nov,
       volume = {525},
       number = {4},
        pages = {5554-5564},
          doi = {10.1093/mnras/stad2573},
archivePrefix = {arXiv},
       eprint = {2305.09827},
 primaryClass = {astro-ph.CO},
       adsurl = {https://ui.adsabs.harvard.edu/abs/2023MNRAS.525.5554P}
}

@article{euclid,
	adsurl = {https://ui.adsabs.harvard.edu/abs/2011arXiv1110.3193L},
	archiveprefix = {arXiv},
	author = {{Laureijs}, R. and {Amiaux}, J. and {Arduini}, S. and {Augu{\`e}res}, J. -L. and {Brinchmann}, J. and others},
	eid = {arXiv:1110.3193},
	eprint = {1110.3193},
	journal = {arXiv e-prints},
	month = {October},
	pages = {arXiv:1110.3193},
	primaryclass = {astro-ph.CO},
	title = {{Euclid Definition Study Report}},
	year = {2011}}

@article{lsst,
	adsurl = {https://ui.adsabs.harvard.edu/abs/2009arXiv0912.0201L},
	archiveprefix = {arXiv},
	author = {{LSST Science Collaboration} and others},
	eid = {arXiv:0912.0201},
	eprint = {0912.0201},
	journal = {arXiv e-prints},
	month = {December},
	pages = {arXiv:0912.0201},
	primaryclass = {astro-ph.IM},
	title = {{LSST Science Book, Version 2.0}},
	year = {2009}}

@article{ivezic_lsst,
	adsurl = {https://ui.adsabs.harvard.edu/abs/2019ApJ...873..111I},
	archiveprefix = {arXiv},
	author = {{Ivezi{\'c}}, {\v{Z}}eljko and {Kahn}, Steven M. and {Tyson}, J. Anthony and others},
	eid = {111},
	eprint = {0805.2366},
	journal = {\apj},
	month = {March},
	number = {2},
	pages = {111},
	primaryclass = {astro-ph},
	title = {{LSST: From Science Drivers to Reference Design and Anticipated Data Products}},
	volume = {873},
	year = {2019}}

@article{spergel_roman,
	adsurl = {https://ui.adsabs.harvard.edu/abs/2015arXiv150303757S},
	archiveprefix = {arXiv},
	author = {{Spergel}, D. and {Gehrels}, N. and {Baltay}, C. and {Bennett}, D. and {Breckinridge}, J. and others},
	eid = {arXiv:1503.03757},
	eprint = {1503.03757},
	journal = {arXiv e-prints},
	month = {March},
	pages = {arXiv:1503.03757},
	primaryclass = {astro-ph.IM},
	title = {{Wide-Field InfrarRed Survey Telescope-Astrophysics Focused Telescope Assets WFIRST-AFTA 2015 Report}},
	year = {2015}}

@ARTICLE{gatti2024,
       author = {{Gatti}, M. and {Campailla}, G. and {Jeffrey}, N. and {Whiteway}, L. and {Porredon}, A. and others},
        title = "{Dark Energy Survey Year 3 results: Simulation-based cosmological inference with wavelet harmonics, scattering transforms, and moments of weak lensing mass maps. II. cosmological results}",
      journal = {\prd},
         year = 2025,
        month = mar,
       volume = {111},
       number = {6},
          eid = {063504},
        pages = {063504},
          doi = {10.1103/PhysRevD.111.063504},
archivePrefix = {arXiv},
       eprint = {2405.10881},
 primaryClass = {astro-ph.CO},
       adsurl = {https://ui.adsabs.harvard.edu/abs/2025PhRvD.111f3504G}
}

@article{DES:2025bxy,
       author = {{DES Collaboration}},
        title = "{Dark Energy Survey: Implications for cosmological expansion models from the final DES baryon acoustic oscillation and supernova data}",
      journal = {\prd},
         year = 2026,
        month = mar,
       volume = {113},
       number = {6},
          eid = {063530},
        pages = {063530},
          doi = {10.1103/z2q4-qcdq},
archivePrefix = {arXiv},
       eprint = {2503.06712},
 primaryClass = {astro-ph.CO},
       adsurl = {https://ui.adsabs.harvard.edu/abs/2026PhRvD.113f3530A}
}

@article{RaveriHu,
     author = {{Raveri}, Marco and {Hu}, Wayne},
        title = "{Concordance and discordance in cosmology}",
      journal = {\prd},
         year = 2019,
        month = feb,
       volume = {99},
       number = {4},
          eid = {043506},
        pages = {043506},
          doi = {10.1103/PhysRevD.99.043506},
archivePrefix = {arXiv},
       eprint = {1806.04649},
 primaryClass = {astro-ph.CO},
       adsurl = {https://ui.adsabs.harvard.edu/abs/2019PhRvD..99d3506R}
}

@article{mcal1,
       author = {{Sheldon}, Erin S. and {Huff}, Eric M.},
        title = "{Practical Weak-lensing Shear Measurement with Metacalibration}",
      journal = {\apj},
         year = 2017,
        month = may,
       volume = {841},
       number = {1},
          eid = {24},
        pages = {24},
          doi = {10.3847/1538-4357/aa704b},
archivePrefix = {arXiv},
       eprint = {1702.02601},
 primaryClass = {astro-ph.CO},
       adsurl = {https://ui.adsabs.harvard.edu/abs/2017ApJ...841...24S}
}

@article{mcal2,
	adsurl = {https://ui.adsabs.harvard.edu/abs/2017arXiv170202600H},
	archiveprefix = {arXiv},
	author = {{Huff}, Eric and {Mandelbaum}, Rachel},
	eid = {arXiv:1702.02600},
	eprint = {1702.02600},
	journal = {arXiv e-prints},
	month = {February},
	pages = {arXiv:1702.02600},
	primaryclass = {astro-ph.CO},
	title = {{Metacalibration: Direct Self-Calibration of Biases in Shear Measurement}},
	year = {2017}}

@article{bernstein_bfd1,
	adsurl = {https://ui.adsabs.harvard.edu/abs/2014MNRAS.438.1880B},
	archiveprefix = {arXiv},
	author = {{Bernstein}, Gary M. and {Armstrong}, Robert},
	doi = {10.1093/mnras/stt2326},
	eprint = {1304.1843},
	journal = {\mnras},
	month = feb,
	number = {2},
	pages = {1880-1893},
	primaryclass = {astro-ph.CO},
	title = {{Bayesian lensing shear measurement}},
	volume = {438},
	year = 2014}

@article{bernstein_bfd2,
	adsurl = {https://ui.adsabs.harvard.edu/abs/2016MNRAS.459.4467B},
	archiveprefix = {arXiv},
	author = {{Bernstein}, Gary M. and {Armstrong}, Robert and {Krawiec}, Christina and {March}, Marisa C.},
	doi = {10.1093/mnras/stw879},
	eprint = {1508.05655},
	journal = {\mnras},
	month = jul,
	number = {4},
	pages = {4467-4484},
	primaryclass = {astro-ph.IM},
	title = {{An accurate and practical method for inference of weak gravitational lensing from galaxy images}},
	volume = {459},
	year = 2016}

@ARTICLE{Schaye2023,
       author = {{Schaye}, Joop and {Kugel}, Roi and {Schaller}, Matthieu and {Helly}, John C. and {Braspenning}, Joey and {Elbers}, Willem and {McCarthy}, Ian G. and {van Daalen}, Marcel P. and {Vandenbroucke}, Bert and {Frenk}, Carlos S. and {Kwan}, Juliana and {Salcido}, Jaime and {Bah{\'e}}, Yannick M. and {Borrow}, Josh and {Chaikin}, Evgenii and {Hahn}, Oliver and {Hu{\v{s}}ko}, Filip and {Jenkins}, Adrian and {Lacey}, Cedric G. and {Nobels}, Folkert S.~J.},
        title = "{The FLAMINGO project: cosmological hydrodynamical simulations for large-scale structure and galaxy cluster surveys}",
      journal = {\mnras},
         year = 2023,
        month = dec,
       volume = {526},
       number = {4},
        pages = {4978-5020},
          doi = {10.1093/mnras/stad2419},
archivePrefix = {arXiv},
       eprint = {2306.04024},
 primaryClass = {astro-ph.CO},
       adsurl = {https://ui.adsabs.harvard.edu/abs/2023MNRAS.526.4978S}
}

@ARTICLE{Dave2019,
       author = {{Dav{\'e}}, Romeel and {Angl{\'e}s-Alc{\'a}zar}, Daniel and {Narayanan}, Desika and {Li}, Qi and {Rafieferantsoa}, Mika H. and {Appleby}, Sarah},
        title = "{SIMBA: Cosmological simulations with black hole growth and feedback}",
      journal = {\mnras},
         year = 2019,
        month = jun,
       volume = {486},
       number = {2},
        pages = {2827-2849},
          doi = {10.1093/mnras/stz937},
       adsurl = {https://ui.adsabs.harvard.edu/abs/2019MNRAS.486.2827D}
}

@ARTICLE{Hernandez-Aguayo2022,
       author = {{Hern{\'a}ndez-Aguayo}, C{\'e}sar and {Springel}, Volker and {Pakmor}, R{\"u}diger and {Barrera}, Monica and {Ferlito}, Fulvio and {White}, Simon D.~M. and {Hernquist}, Lars and {Hadzhiyska}, Boryana and {Delgado}, Ana Maria and {Kannan}, Rahul and {Bose}, Sownak and {Frenk}, Carlos},
        title = "{The MillenniumTNG Project: high-precision predictions for matter clustering and halo statistics}",
      journal = {\mnras},
         year = 2023,
        month = sep,
       volume = {524},
       number = {2},
        pages = {2556-2578},
          doi = {10.1093/mnras/stad1657},
archivePrefix = {arXiv},
       eprint = {2210.10059},
 primaryClass = {astro-ph.CO},
       adsurl = {https://ui.adsabs.harvard.edu/abs/2023MNRAS.524.2556H}
}

@ARTICLE{Henden2019,
       author = {{Henden}, Nicholas A. and {Puchwein}, Ewald and {Sijacki}, Debora},
        title = "{The redshift evolution of X-ray and Sunyaev-Zel'dovich scaling relations in the FABLE simulations}",
      journal = {\mnras},
         year = 2019,
        month = oct,
       volume = {489},
       number = {2},
        pages = {2439-2470},
          doi = {10.1093/mnras/stz2301},
archivePrefix = {arXiv},
       eprint = {1905.00013},
 primaryClass = {astro-ph.CO},
       adsurl = {https://ui.adsabs.harvard.edu/abs/2019MNRAS.489.2439H}
}

@ARTICLE{Bigwood2025,
       author = {{Bigwood}, Leah and {Bourne}, Martin A. and {Ir{\v{s}}i{\v{c}}}, Vid and {Amon}, Alexandra and {Sijacki}, Debora},
        title = "{The case for large-scale AGN feedback in galaxy formation simulations: insights from XFABLE}",
      journal = {\mnras},
         year = 2025,
        month = sep,
       volume = {542},
       number = {4},
        pages = {3206-3230},
          doi = {10.1093/mnras/staf1435},
archivePrefix = {arXiv},
       eprint = {2501.16983},
 primaryClass = {astro-ph.CO},
       adsurl = {https://ui.adsabs.harvard.edu/abs/2025MNRAS.542.3206H}
}

@ARTICLE{Bigwood2024,
       author = {{Bigwood}, L. and {Amon}, A. and {Schneider}, A. and {Salcido}, J. and {McCarthy}, I.~G. and others},
        title = "{Weak lensing combined with the kinetic Sunyaev-Zel'dovich effect: a study of baryonic feedback}",
      journal = {\mnras},
         year = 2024,
        month = oct,
       volume = {534},
       number = {1},
        pages = {655-682},
          doi = {10.1093/mnras/stae2100},
archivePrefix = {arXiv},
       eprint = {2404.06098},
 primaryClass = {astro-ph.CO},
       adsurl = {https://ui.adsabs.harvard.edu/abs/2024MNRAS.534..655B}
}

@ARTICLE{Arico2023,
       author = {{Aric{\`o}}, Giovanni and {Angulo}, Raul E. and {Zennaro}, Matteo and {Contreras}, Sergio and {Chen}, Angela and {Hern{\'a}ndez-Monteagudo}, Carlos},
        title = "{DES Y3 cosmic shear down to small scales: Constraints on cosmology and baryons}",
      journal = {\aap},
         year = 2023,
        month = oct,
       volume = {678},
          eid = {A109},
        pages = {A109},
          doi = {10.1051/0004-6361/202346539},
archivePrefix = {arXiv},
       eprint = {2303.05537},
 primaryClass = {astro-ph.CO},
       adsurl = {https://ui.adsabs.harvard.edu/abs/2023A&A...678A.109A}
}

@ARTICLE{siegel25,
       author = {{Siegel}, Jared and {Amon}, Alexandra and {McCarthy}, Ian G. and {Bigwood}, Leah and {Yamamoto}, Masaya and {Bulbul}, Esra and {Greene}, Jenny E. and {McCullough}, Jamie and {Schaller}, Matthieu and {Schaye}, Joop},
        title = "{Joint X-ray, kinetic Sunyaev-Zeldovich, and weak lensing measurements: toward a consensus picture of efficient gas expulsion from groups and clusters}",
       journal = {\apj},
         year = 2026,
        month = jun,
       volume = {1003},
       number = {2},
          eid = {151},
        pages = {151},
          doi = {10.3847/1538-4357/ae5dc2},
archivePrefix = {arXiv},
       eprint = {2509.10455},
 primaryClass = {astro-ph.CO},
       adsurl = {https://ui.adsabs.harvard.edu/abs/2026ApJ..1003..151S}
}

@ARTICLE{Dalal2025,
       author = {{Dalal}, Nihar and {To}, Chun-Hao and {Hirata}, Chris and {Hyeon-Shin}, Tae and {Hilton}, Matt and {Pandey}, Shivam and {Bond}, J. Richard},
        title = "{Deciphering baryonic feedback from ACT tSZ galaxy clusters}",
      journal = {\jcap},
         year = 2026,
        month = mar,
       volume = {2026},
       number = {3},
          eid = {036},
        pages = {036},
          doi = {10.1088/1475-7516/2026/03/036},
archivePrefix = {arXiv},
       eprint = {2507.04476},
 primaryClass = {astro-ph.CO},
       adsurl = {https://ui.adsabs.harvard.edu/abs/2026JCAP...03..036D}
}

@ARTICLE{Pandey2025,
       author = {{Pandey}, S. and {Hill}, J.~C. and {Alarcon}, A. and {Alves}, O. and {Amon}, A. and others},
        title = "{Constraints on cosmology and baryonic feedback with joint analysis of Dark Energy Survey Year 3 lensing data and ACT DR6 thermal Sunyaev-Zel'dovich effect observations}",
      journal = {arXiv e-prints},
         year = 2025,
        month = jun,
          eid = {arXiv:2506.07432},
        pages = {arXiv:2506.07432},
          doi = {10.48550/arXiv.2506.07432},
archivePrefix = {arXiv},
       eprint = {2506.07432},
 primaryClass = {astro-ph.CO},
       adsurl = {https://ui.adsabs.harvard.edu/abs/2025arXiv250607432P}
}

@ARTICLE{LaPosta2025,
       author = {{La Posta}, Adrien and {Alonso}, David and {Chisari}, Nora Elisa and {Ferreira}, Tassia and {Garc{\'\i}a-Garc{\'\i}a}, Carlos},
        title = "{Insights on gas thermodynamics from the combination of x-ray and thermal Sunyaev-Zel'dovich data cross correlated with cosmic shear}",
      journal = {\prd},
         year = 2025,
        month = aug,
       volume = {112},
       number = {4},
          eid = {043525},
        pages = {043525},
          doi = {10.1103/m77z-w7pl},
archivePrefix = {arXiv},
       eprint = {2412.12081},
 primaryClass = {astro-ph.CO},
       adsurl = {https://ui.adsabs.harvard.edu/abs/2025PhRvD.112d3525L}
}

@ARTICLE{Schneider2022,
       author = {{Schneider}, Aurel and {Giri}, Sambit K. and {Amodeo}, Stefania and {Refregier}, Alexandre},
        title = "{Constraining baryonic feedback and cosmology with weak-lensing, X-ray, and kinematic Sunyaev-Zeldovich observations}",
      journal = {\mnras},
         year = 2022,
        month = aug,
       volume = {514},
       number = {3},
        pages = {3802-3814},
          doi = {10.1093/mnras/stac1493},
archivePrefix = {arXiv},
       eprint = {2110.02228},
 primaryClass = {astro-ph.CO},
       adsurl = {https://ui.adsabs.harvard.edu/abs/2022MNRAS.514.3802S}
}

@article{zuntz18,
	archiveprefix = {arXiv},
        author = "Zuntz, J. and Sheldon, E. and others",
	eprint = {1708.01533},
	journal = {\mnras},
	month = {November},
	number = {1},
	pages = {1149-1182},
	primaryclass = {astro-ph.CO},
	title = {{Dark Energy Survey Year 1 results: weak lensing shape catalogues}},
	volume = {481},
	year = {2018}}

@ARTICLE{Siegel_BC25,
       author = {{Siegel}, Jared and {Bigwood}, Leah and {Amon}, Alexandra and {McCullough}, Jamie and {Yamamoto}, Masaya and {McCarthy}, Ian G. and {Schaller}, Matthieu and {Schneider}, Aurel and {Schaye}, Joop},
        title = "{The suppression of the matter power spectrum: strong feedback from X-ray gas mass fractions, kSZ effect profiles, and galaxy-galaxy lensing}",
      journal = {arXiv e-prints},
         year = 2025,
        month = dec,
          eid = {arXiv:2512.02954},
        pages = {arXiv:2512.02954},
archivePrefix = {arXiv},
       eprint = {2512.02954},
 primaryClass = {astro-ph.CO},
       adsurl = {https://ui.adsabs.harvard.edu/abs/2025arXiv251202954S}
}

@ARTICLE{Johnston19,
       author = {{Johnston}, Harry and {Georgiou}, Christos and {Joachimi}, Benjamin and {Hoekstra}, Henk and {Chisari}, Nora Elisa and {Farrow}, Daniel and {Fortuna}, Maria Cristina and {Heymans}, Catherine and {Joudaki}, Shahab and {Kuijken}, Konrad and {Wright}, Angus},
        title = "{KiDS+GAMA: Intrinsic alignment model constraints for current and future weak lensing cosmology}",
      journal = {\aap},
         year = 2019,
        month = apr,
       volume = {624},
          eid = {A30},
        pages = {A30},
          doi = {10.1051/0004-6361/201834714},
archivePrefix = {arXiv},
       eprint = {1811.09598},
 primaryClass = {astro-ph.CO},
       adsurl = {https://ui.adsabs.harvard.edu/abs/2019A&A...624A..30J}
}

@ARTICLE{krause16,
       author = {{Krause}, Elisabeth and {Eifler}, Tim and {Blazek}, Jonathan},
        title = "{The impact of intrinsic alignment on current and future cosmic shear surveys}",
      journal = {\mnras},
         year = 2016,
        month = feb,
       volume = {456},
       number = {1},
        pages = {207-222},
          doi = {10.1093/mnras/stv2615},
archivePrefix = {arXiv},
       eprint = {1506.08730},
 primaryClass = {astro-ph.CO},
       adsurl = {https://ui.adsabs.harvard.edu/abs/2016MNRAS.456..207K}
}

@ARTICLE{polychord,
       author = {{Handley}, W.~J. and {Hobson}, M.~P. and {Lasenby}, A.~N.},
        title = "{POLYCHORD: next-generation nested sampling}",
      journal = {\mnras},
         year = 2015,
        month = nov,
       volume = {453},
       number = {4},
        pages = {4384-4398},
          doi = {10.1093/mnras/stv1911},
archivePrefix = {arXiv},
       eprint = {1506.00171},
 primaryClass = {astro-ph.IM},
       adsurl = {https://ui.adsabs.harvard.edu/abs/2015MNRAS.453.4384H}
}

@ARTICLE{Torrado2021,
       author = {{Torrado}, Jes{\'u}s and {Lewis}, Antony},
        title = "{Cobaya: code for Bayesian analysis of hierarchical physical models}",
      journal = {\jcap},
         year = 2021,
        month = may,
       volume = {2021},
       number = {5},
          eid = {057},
        pages = {057},
          doi = {10.1088/1475-7516/2021/05/057},
archivePrefix = {arXiv},
       eprint = {2005.05290},
 primaryClass = {astro-ph.IM},
       adsurl = {https://ui.adsabs.harvard.edu/abs/2021JCAP...05..057T}
}

@ARTICLE{Doux2022,
       author = {{Doux}, C. and {Jain}, B. and {Zeurcher}, D. and {Lee}, J. and {Fang}, X. and others},
        title = "{Dark energy survey year 3 results: cosmological constraints from the analysis of cosmic shear in harmonic space}",
      journal = {\mnras},
         year = 2022,
        month = sep,
       volume = {515},
       number = {2},
        pages = {1942-1972},
          doi = {10.1093/mnras/stac1826},
archivePrefix = {arXiv},
       eprint = {2203.07128},
 primaryClass = {astro-ph.CO},
       adsurl = {https://ui.adsabs.harvard.edu/abs/2022MNRAS.515.1942D}
}

@ARTICLE{Chen23,
       author = {{Chen}, A. and {Aric{\`o}}, G. and {Huterer}, D. and {Angulo}, R.~E. and {Weaverdyck}, N. and others},
        title = "{Constraining the baryonic feedback with cosmic shear using the DES Year-3 small-scale measurements}",
      journal = {\mnras},
         year = 2023,
        month = feb,
       volume = {518},
       number = {4},
        pages = {5340-5355},
          doi = {10.1093/mnras/stac3213},
archivePrefix = {arXiv},
       eprint = {2206.08591},
 primaryClass = {astro-ph.CO},
       adsurl = {https://ui.adsabs.harvard.edu/abs/2023MNRAS.518.5340C}
}

@ARTICLE{garcia24,
       author = {{Garc{\'\i}a-Garc{\'\i}a}, Carlos and {Zennaro}, Matteo and {Aric{\`o}}, Giovanni and {Alonso}, David and {Angulo}, Raul E.},
        title = "{Cosmic shear with small scales: DES-Y3, KiDS-1000 and HSC-DR1}",
      journal = {\jcap},
         year = 2024,
        month = aug,
       volume = {2024},
       number = {8},
          eid = {024},
        pages = {024},
          doi = {10.1088/1475-7516/2024/08/024},
archivePrefix = {arXiv},
       eprint = {2403.13794},
 primaryClass = {astro-ph.CO},
       adsurl = {https://ui.adsabs.harvard.edu/abs/2024JCAP...08..024G}
}

@ARTICLE{Bigwood25_blueshear,
       author = {{Bigwood}, Leah and {McCullough}, Jamie and {Siegel}, Jared and {Amon}, Alexandra and {Efstathiou}, George and others},
        title = "{Confronting cosmic shear astrophysical uncertainties: DES Year 3 revisited}",
      journal = {arXiv e-prints},
         year = 2025,
        month = dec,
          eid = {arXiv:2512.04209},
        pages = {arXiv:2512.04209},
          doi = {10.48550/arXiv.2512.04209},
archivePrefix = {arXiv},
       eprint = {2512.04209},
 primaryClass = {astro-ph.CO},
       adsurl = {https://ui.adsabs.harvard.edu/abs/2025arXiv251204209B}
}

@ARTICLE{Huang21,
       author = {{Huang}, Hung-Jin and {Eifler}, Tim and {Mandelbaum}, Rachel and {Bernstein}, Gary M. and {Chen}, Anqi and others},
        title = "{Dark energy survey year 1 results: Constraining baryonic physics in the Universe}",
      journal = {\mnras},
         year = 2021,
        month = apr,
       volume = {502},
       number = {4},
        pages = {6010-6031},
          doi = {10.1093/mnras/stab357},
archivePrefix = {arXiv},
       eprint = {2007.15026},
 primaryClass = {astro-ph.CO},
       adsurl = {https://ui.adsabs.harvard.edu/abs/2021MNRAS.502.6010H}
}

@article{paulin08,
	adsurl = {https://ui.adsabs.harvard.edu/abs/2008A&A...484...67P},
	archiveprefix = {arXiv},
	author = {{Paulin-Henriksson}, S. and {Amara}, A. and {Voigt}, L. and {Refregier}, A. and {Bridle}, S.~L.},
	eprint = {0711.4886},
	journal = {\aap},
	month = {June},
	number = {1},
	pages = {67-77},
	primaryclass = {astro-ph},
	title = {{Point spread function calibration requirements for dark energy from cosmic shear}},
	volume = {484},
	year = {2008}}

@article{rowe10,
	adsurl = {https://ui.adsabs.harvard.edu/abs/2010MNRAS.404..350R},
	archiveprefix = {arXiv},
	author = {{Rowe}, Barnaby},
	eprint = {0904.3056},
	journal = {\mnras},
	month = {May},
	number = {1},
	pages = {350-366},
	primaryclass = {astro-ph.CO},
	title = {{Improving PSF modelling for weak gravitational lensing using new methods in model selection}},
	volume = {404},
	year = {2010}}

@article{jarvis16,
	adsurl = {https://ui.adsabs.harvard.edu/abs/2016MNRAS.460.2245J},
	archiveprefix = {arXiv},
	author = {{Jarvis}, M. and others},
	eprint = {1507.05603},
	journal = {\mnras},
	month = {August},
	number = {2},
	pages = {2245-2281},
	primaryclass = {astro-ph.IM},
	title = {{The DES Science Verification weak lensing shear catalogues}},
	volume = {460},
	year = {2016}}

@article{zhang23,
       author = {{Zhang}, Tianqing and {Almoubayyed}, Husni and {Mandelbaum}, Rachel and {Meyers}, Joshua E. and {Jarvis}, Mike and {Kannawadi}, Arun and {Schmitz}, Morgan A. and {Guinot}, Axel and {LSST Dark Energy Science Collaboration}},
        title = "{Impact of point spread function higher moments error on weak gravitational lensing - II. A comprehensive study}",
      journal = {\mnras},
         year = 2023,
        month = apr,
       volume = {520},
       number = {2},
        pages = {2328-2350},
          doi = {10.1093/mnras/stac3350},
archivePrefix = {arXiv},
       eprint = {2205.07892},
 primaryClass = {astro-ph.CO},
       adsurl = {https://ui.adsabs.harvard.edu/abs/2023MNRAS.520.2328Z}
}

@ARTICLE{desy6_bao,
       author = {{DES Collaboration}},
        title = "{Dark Energy Survey: A 2.1\% measurement of the angular baryonic acoustic oscillation scale at redshift zeff=0.85 from the final dataset}",
      journal = {\prd},
         year = 2024,
        month = sep,
       volume = {110},
       number = {6},
          eid = {063515},
        pages = {063515},
          doi = {10.1103/PhysRevD.110.063515},
archivePrefix = {arXiv},
       eprint = {2402.10696},
 primaryClass = {astro-ph.CO},
       adsurl = {https://ui.adsabs.harvard.edu/abs/2024PhRvD.110f3515A}
}

@ARTICLE{pantheon_plus22,
       author = {{Brout}, Dillon and {Scolnic}, Dan and others},
        title = "{The Pantheon+ Analysis: Cosmological Constraints}",
      journal = {\apj},
         year = 2022,
        month = oct,
       volume = {938},
       number = {2},
          eid = {110},
        pages = {110},
          doi = {10.3847/1538-4357/ac8e04},
archivePrefix = {arXiv},
       eprint = {2202.04077},
 primaryClass = {astro-ph.CO},
       adsurl = {https://ui.adsabs.harvard.edu/abs/2022ApJ...938..110B}
}

@ARTICLE{Riess2025,
       author = {{Riess}, Adam G. and others},
        title = "{The Perfect Host: JWST Cepheid Observations in a Background-free Type Ia Supernova Host Confirm No Bias in Hubble-constant Measurements}",
      journal = {\apjl},
         year = 2025,
        month = oct,
       volume = {992},
       number = {2},
          eid = {L34},
        pages = {L34},
          doi = {10.3847/2041-8213/ae0ad6},
archivePrefix = {arXiv},
       eprint = {2509.01667},
 primaryClass = {astro-ph.CO},
       adsurl = {https://ui.adsabs.harvard.edu/abs/2025ApJ...992L..34R}
}

@ARTICLE{DESY5SN2024,
       author = {{DES Collaboration} and others},
        title = "{The Dark Energy Survey: Cosmology Results with {\ensuremath{\sim}}1500 New High-redshift Type Ia Supernovae Using the Full 5 yr Data Set}",
      journal = {\apjl},
         year = 2024,
        month = sep,
       volume = {973},
       number = {1},
          eid = {L14},
        pages = {L14},
          doi = {10.3847/2041-8213/ad6f9f},
archivePrefix = {arXiv},
       eprint = {2401.02929},
 primaryClass = {astro-ph.CO},
       adsurl = {https://ui.adsabs.harvard.edu/abs/2024ApJ...973L..14D}
}

@ARTICLE{Popovic2025b,
       author = {{Popovic}, B. and others},
        title = "{The Dark Energy Survey supernova program: a reanalysis of cosmology results and evidence for evolving dark energy with an updated Type Ia supernova calibration}",
      journal = {\mnras},
         year = 2026,
        month = jun,
       volume = {548},
       number = {4},
          eid = {stag632},
        pages = {stag632},
          doi = {10.1093/mnras/stag632},
archivePrefix = {arXiv},
       eprint = {2511.07517},
 primaryClass = {astro-ph.CO},
       adsurl = {https://ui.adsabs.harvard.edu/abs/2026MNRAS.548ag632P}
}

@ARTICLE{miller13,
       author = {{Miller}, L. and {Heymans}, C. and {Kitching}, T.~D. and {van Waerbeke}, L. and {Erben}, T. and {Hildebrandt}, H. and {Hoekstra}, H. and {Mellier}, Y. and {Rowe}, B.~T.~P. and {Coupon}, J. and {Dietrich}, J.~P. and {Fu}, L. and {Harnois-D{\'e}raps}, J. and {Hudson}, M.~J. and {Kilbinger}, M. and {Kuijken}, K. and {Schrabback}, T. and {Semboloni}, E. and {Vafaei}, S. and {Velander}, M.},
        title = "{Bayesian galaxy shape measurement for weak lensing surveys - III. Application to the Canada-France-Hawaii Telescope Lensing Survey}",
      journal = {\mnras},
         year = 2013,
        month = mar,
       volume = {429},
       number = {4},
        pages = {2858-2880},
          doi = {10.1093/mnras/sts454},
archivePrefix = {arXiv},
       eprint = {1210.8201},
 primaryClass = {astro-ph.CO},
       adsurl = {https://ui.adsabs.harvard.edu/abs/2013MNRAS.429.2858M}
}

@ARTICLE{li23_kids,
       author = {{Li}, Shun-Sheng and {Kuijken}, Konrad and {Hoekstra}, Henk and {Miller}, Lance and {Heymans}, Catherine and {Hildebrandt}, Hendrik and {van den Busch}, Jan Luca and {Wright}, Angus H. and {Yoon}, Mijin and {Bilicki}, Maciej and {Bravo}, Mat{\'\i}as and {Lagos}, Claudia del P.},
        title = "{KiDS-Legacy calibration: Unifying shear and redshift calibration with the SKiLLS multi-band image simulations}",
      journal = {\aap},
         year = 2023,
        month = feb,
       volume = {670},
          eid = {A100},
        pages = {A100},
          doi = {10.1051/0004-6361/202245210},
archivePrefix = {arXiv},
       eprint = {2210.07163},
 primaryClass = {astro-ph.CO},
       adsurl = {https://ui.adsabs.harvard.edu/abs/2023A&A...670A.100L}
}

@ARTICLE{sailer25,
       author = {{Sailer}, Noah and {Kim}, Joshua and {Ferraro}, Simone and others},
        title = "{Cosmological constraints from the cross-correlation of DESI Luminous Red Galaxies with CMB lensing from Planck PR4 and ACT DR6}",
      journal = {\jcap},
         year = 2025,
        month = jun,
       volume = {2025},
       number = {6},
          eid = {008},
        pages = {008},
          doi = {10.1088/1475-7516/2025/06/008},
archivePrefix = {arXiv},
       eprint = {2407.04607},
 primaryClass = {astro-ph.CO},
       adsurl = {https://ui.adsabs.harvard.edu/abs/2025JCAP...06..008S}
}

@ARTICLE{farren24,
       author = {{Farren}, Gerrit S. and {Krolewski}, Alex and {MacCrann}, Niall and {Ferraro}, Simone and {Abril-Cabezas}, Irene and others},
        title = "{The Atacama Cosmology Telescope: Cosmology from Cross-correlations of unWISE Galaxies and ACT DR6 CMB Lensing}",
      journal = {\apj},
         year = 2024,
        month = may,
       volume = {966},
       number = {2},
          eid = {157},
        pages = {157},
          doi = {10.3847/1538-4357/ad31a5},
archivePrefix = {arXiv},
       eprint = {2309.05659},
 primaryClass = {astro-ph.CO},
       adsurl = {https://ui.adsabs.harvard.edu/abs/2024ApJ...966..157F}
}

@ARTICLE{desi_fs,
       author = {{Adame}, A.~G. and others},
        title = "{DESI 2024 VII: cosmological constraints from the full-shape modeling of clustering measurements}",
      journal = {\jcap},
         year = 2025,
        month = jul,
       volume = {2025},
       number = {7},
          eid = {028},
        pages = {028},
          doi = {10.1088/1475-7516/2025/07/028},
archivePrefix = {arXiv},
       eprint = {2411.12022},
 primaryClass = {astro-ph.CO},
       adsurl = {https://ui.adsabs.harvard.edu/abs/2025JCAP...07..028A}
}

@ARTICLE{desi,
       author = {{DESI Collaboration}},
        title = "{Overview of the Instrumentation for the Dark Energy Spectroscopic Instrument}",
      journal = {\aj},
         year = 2022,
        month = nov,
       volume = {164},
       number = {5},
          eid = {207},
        pages = {207},
          doi = {10.3847/1538-3881/ac882b},
archivePrefix = {arXiv},
       eprint = {2205.10939},
 primaryClass = {astro-ph.IM},
       adsurl = {https://ui.adsabs.harvard.edu/abs/2022AJ....164..207D}
}

@ARTICLE{desi_bao_dr1,
       author = {{DESI Collaboration}},
        title = "{DESI 2024 VI: cosmological constraints from the measurements of baryon acoustic oscillations}",
      journal = {\jcap},
         year = 2025,
        month = feb,
       volume = {2025},
       number = {2},
          eid = {021},
        pages = {021},
          doi = {10.1088/1475-7516/2025/02/021},
archivePrefix = {arXiv},
       eprint = {2404.03002},
 primaryClass = {astro-ph.CO},
       adsurl = {https://ui.adsabs.harvard.edu/abs/2025JCAP...02..021A}
}

@ARTICLE{armstrong24,
       author = {{Armstrong}, Robert and {Sheldon}, Erin and {Huff}, Eric and {Bosch}, Jim and {Rykoff}, Eli and {Mandelbaum}, Rachel and {Kannawadi}, Arun and {Melchior}, Peter and {Lupton}, Robert and {Becker}, Matthew R. and {Al-Sayyed}, Yusra and {The LSST Dark Energy Science Collaboration}},
        title = "{The little coadd that could: Estimating shear from coadded images}",
      journal = {arXiv e-prints},
         year = 2024,
        month = jul,
          eid = {arXiv:2407.01771},
        pages = {arXiv:2407.01771},
          doi = {10.48550/arXiv.2407.01771},
archivePrefix = {arXiv},
       eprint = {2407.01771},
 primaryClass = {astro-ph.CO},
       adsurl = {https://ui.adsabs.harvard.edu/abs/2024arXiv240701771A}
}

@ARTICLE{Ferreira23,
       author = {{Ferreira}, Tassia and {Alonso}, David and {Garcia-Garcia}, Carlos and {Chisari}, Nora Elisa},
        title = "{X-Ray-Cosmic-Shear Cross-Correlations: First Detection and Constraints on Baryonic Effects}",
      journal = {\prl},
         year = 2024,
        month = aug,
       volume = {133},
       number = {5},
          eid = {051001},
        pages = {051001},
          doi = {10.1103/PhysRevLett.133.051001},
archivePrefix = {arXiv},
       eprint = {2309.11129},
 primaryClass = {astro-ph.CO},
       adsurl = {https://ui.adsabs.harvard.edu/abs/2024PhRvL.133e1001F}
}

@article{y3-deepfields,
    author = "Hartley, W. G. and Choi, A. and others",
    collaboration = "DES",
    title = "{Dark Energy Survey Year 3 Results: Deep Field Optical + Near-Infrared Images and Catalogue}",
    eprint = "2012.12824",
    archivePrefix = "arXiv",
    primaryClass = "astro-ph.CO",
    reportNumber = "FERMILAB-PUB-20-670-AE",
    journal = {\mnras},
	year = 2022,
	month = jan,
	volume = {509},
	number = {3},
	pages = {3547-3579},
	doi = {10.1093/mnras/stab3055}
}

@article{y3-shapecatalog,
    author = "Gatti, M. and Sheldon, E. and others",
    collaboration = "DES",
    title = "{Dark Energy Survey Year 3 Results: Weak Lensing Shape Catalogue}",
    eprint = "2011.03408",
    archivePrefix = "arXiv",
    primaryClass = "astro-ph.CO",
    reportNumber = "FERMILAB-PUB-20-545-AE, DES-2015-0048",
    doi = "10.1093/mnras/stab918",
    journal = "Mon. Not. Roy. Astron. Soc.",
    volume = "504",
    number = "3",
    pages = "4312-4336",
    year = "2021"
}

@ARTICLE{y3-imagesims,
       author = {{MacCrann}, N. and {Becker}, M.~R. and others},
        title = "{Dark Energy Survey Y3 results: blending shear and redshift biases in image simulations}",
      journal = {\mnras},
         year = 2022,
        month = jan,
       volume = {509},
       number = {3},
        pages = {3371-3394},
          doi = {10.1093/mnras/stab2870},
archivePrefix = {arXiv},
       eprint = {2012.08567},
 primaryClass = {astro-ph.CO},
       adsurl = {https://ui.adsabs.harvard.edu/abs/2022MNRAS.509.3371M}
}

@ARTICLE{y3-sompz,
       author = {{Myles}, J. and {Alarcon}, A. and others},
        title = "{Dark Energy Survey Year 3 results: redshift calibration of the weak lensing source galaxies}",
      journal = {\mnras},
         year = 2021,
        month = aug,
       volume = {505},
       number = {3},
        pages = {4249-4277},
          doi = {10.1093/mnras/stab1515},
archivePrefix = {arXiv},
       eprint = {2012.08566},
 primaryClass = {astro-ph.CO},
       adsurl = {https://ui.adsabs.harvard.edu/abs/2021MNRAS.505.4249M}
}

@article{y3-sourcewz,
    author = "Gatti, M. and Giannini, G. and others",
    collaboration = "DES",
    title = "{Dark Energy Survey Year 3 Results: Clustering Redshifts -- Calibration of the Weak Lensing Source Redshift Distributions with redMaGiC and BOSS/eBOSS}",
    eprint = "2012.08569",
    archivePrefix = "arXiv",
    primaryClass = "astro-ph.CO",
    reportNumber = "FERMILAB-PUB-20-655-AE",
     journal = {\mnras},
	year = 2022,
	month = feb,
	volume = {510},
	number = {1},
	pages = {1223-1247},
	doi = {10.1093/mnras/stab3311}
}

@article{y3-shearratio,
        author = {{S\'anchez}, C. and {Prat}, J. and others},
        title = "{Dark Energy Survey Year 3 Results: Exploiting small-scale information with lensing ratios}",
    eprint = "2105.13542",
    archivePrefix = "arXiv",
    primaryClass = "astro-ph.CO",
    reportNumber = "DES-2015-0059, FERMILAB-PUB-15-309-AE",
        year = "2022",
        month = apr,
        volume = {105},
        number = {8},
        eid = {083529},
        pages = {083529},
        doi = {10.1103/PhysRevD.105.083529},
        journal = {\prd}
        }

@article{y3-samplers,
       author = {{Lemos}, P. and {Weaverdyck}, N. and others},
        title = "{Robust sampling for weak lensing and clustering analyses with the Dark Energy Survey}",
      journal = {\mnras},
         year = 2023,
        month = may,
       volume = {521},
       number = {1},
        pages = {1184-1199},
          doi = {10.1093/mnras/stac2786},
archivePrefix = {arXiv},
       eprint = {2202.08233},
 primaryClass = {astro-ph.CO},
       adsurl = {https://ui.adsabs.harvard.edu/abs/2023MNRAS.521.1184L}
}

@article{y3-tensions,
    author = "Lemos, P. Raveri, M. and others",
    collaboration = "DES",
    title = "{Assessing tension metrics with Dark Energy Survey and Planck data}",
    eprint = "2012.09554",
    archivePrefix = "arXiv",
    primaryClass = "astro-ph.CO",
    reportNumber = "FERMILAB-PUB-20-662-AE",
 journal = {\mnras},
year = 2021,
month = aug,
volume = {505},
number = {4},
pages = {6179-6194},
doi = {10.1093/mnras/stab1670}
}

@article{y3-covariances,
    author = "Friedrich, O. and others",
    collaboration = "DES",
    title = "{Dark Energy Survey Year 3 Results: Covariance Modelling and its Impact on Parameter Estimation and Quality of Fit}",
    eprint = "2012.08568",
    archivePrefix = "arXiv",
    primaryClass = "astro-ph.CO",
    reportNumber = "FERMILAB-PUB-20-663-E-SCD-V, DES-2019-0466",
    journal = {\mnras},
year = 2021,
month = dec,
volume = {508},
number = {3},
pages = {3125-3165},
doi = {10.1093/mnras/stab2384}
}

@article{y3-generalmethods,
       author = {{Krause}, E. and {Fang}, X. and {Pandey}, S. and {Secco}, L.~F. and others},
        title = "{Dark Energy Survey Year 3 Results: Multi-Probe Modeling Strategy and Validation}",
      journal = {arXiv e-prints},
         year = 2021,
        month = may,
          eid = {arXiv:2105.13548},
        pages = {arXiv:2105.13548},
          doi = {10.48550/arXiv.2105.13548},
archivePrefix = {arXiv},
       eprint = {2105.13548},
 primaryClass = {astro-ph.CO},
       adsurl = {https://ui.adsabs.harvard.edu/abs/2021arXiv210513548K}
}

@article{y3-cosmicshear1,
        author = {{Amon}, A. and others},
       title = "{Dark Energy Survey Year 3 results: Cosmology from cosmic shear and robustness to data calibration}",
  journal = {\prd},
  year = 2022,
  month = jan,
  volume = {105},
  number = {2},
  eid = {023514},
  pages = {023514},
  doi = {10.1103/PhysRevD.105.023514},
  archivePrefix = {arXiv},
  eprint = {2105.13543},
  primaryClass = {astro-ph.CO},
  adsurl = {https://ui.adsabs.harvard.edu/abs/2022PhRvD.105b3514A}
        }

@article{y3-cosmicshear2,
        author = {{Secco}, L.~F. and {Samuroff}, S. and others},
         title = "{Dark Energy Survey Year 3 results: Cosmology from cosmic shear and robustness to modeling uncertainty}",
 journal = {\prd},
 year = 2022,
 month = jan,
 volume = {105},
 number = {2},
 eid = {023515},
 pages = {023515},
 doi = {10.1103/PhysRevD.105.023515},
 archivePrefix = {arXiv},
 eprint = {2105.13544},
 primaryClass = {astro-ph.CO},
 adsurl = {https://ui.adsabs.harvard.edu/abs/2022PhRvD.105b3515S}
        }

@article{y6-das,
    author = {{DES Collaboration}},
    title = "{DES Year 6 Cosmology Data Release (Y6KP), DES Data Management}",
    eprint = {},
    url = {{https://des.ncsa.illinois.edu/releases/y6a2}},
    journal = {\url{https://des.ncsa.illinois.edu/releases/y6a2}},
    archivePrefix = {arXiv},
    primaryClass = {astro-ph.CO},
    reportNumber = "",
    year = 2026,
}

@article{y6-gold,
       author = {{Bechtol}, K. and others},
        title = "{Dark Energy Survey Year 6 Results: Photometric Dataset for Cosmology}",
      journal = {\apjs},
         year = 2026,
        month = feb,
       volume = {282},
       number = {2},
          eid = {62},
        pages = {62},
          doi = {10.3847/1538-4365/ae18d3},
archivePrefix = {arXiv},
       eprint = {2501.05739},
 primaryClass = {astro-ph.CO},
       adsurl = {https://ui.adsabs.harvard.edu/abs/2026ApJS..282...62B}
}

@article{y6-balrog,
       author = {{Anbajagane}, Dhayaa and {Tabbutt}, M. and others},
        title = "{Dark Energy Survey Year 6 Results: Synthetic-source Injection Across the Full Survey Using Balrog}",
      journal = {The Open Journal of Astrophysics},
         year = 2025,
        month = may,
       volume = {8},
          eid = {65},
        pages = {65},
          doi = {10.33232/001c.138627},
archivePrefix = {arXiv},
       eprint = {2501.05683},
 primaryClass = {astro-ph.CO},
       adsurl = {https://ui.adsabs.harvard.edu/abs/2025OJAp....8E..65A}
}

@article{y6-metadetect,
       author = {{Yamamoto}, M. and {Becker}, M.~R. and others},
       collaboration = {DES},
        title = "{Dark Energy Survey Year 6 results: cell-based coadds and METADETECTION weak lensing shape catalogue}",
      journal = {\mnras},
         year = 2025,
        month = nov,
       volume = {543},
       number = {4},
        pages = {4156-4186},
          doi = {10.1093/mnras/staf1661},
archivePrefix = {arXiv},
       eprint = {2501.05665},
 primaryClass = {astro-ph.CO},
       adsurl = {https://ui.adsabs.harvard.edu/abs/2025MNRAS.543.4156Y}
}

@article{y6-piff,
       author = {{Schutt}, T. and others},
        title = "{Dark Energy Survey Year 6 Results: Point-Spread Function Modeling}",
      journal = {The Open Journal of Astrophysics},
         year = 2025,
        month = mar,
       volume = {8},
          eid = {26},
        pages = {26},
          doi = {10.33232/001c.132299},
archivePrefix = {arXiv},
       eprint = {2501.05781},
 primaryClass = {astro-ph.CO},
       adsurl = {https://ui.adsabs.harvard.edu/abs/2025OJAp....8E..26S}
}

@article{y6-nzmodes,
author = {{Bernstein}, G.~M. and others},
        title = "{Dimensional Reduction for Sampled Priors and Application to Photometric Redshift Distributions}",
      journal = {\apj},
         year = 2026,
        month = may,
       volume = {1002},
       number = {1},
          eid = {95},
        pages = {95},
          doi = {10.3847/1538-4357/ae5baa},
archivePrefix = {arXiv},
       eprint = {2506.00758},
 primaryClass = {astro-ph.IM},
       adsurl = {https://ui.adsabs.harvard.edu/abs/2026ApJ..1002...95B}
}

@article{y6-mask,
       author = {{Rodr{\'\i}guez-Monroy}, M. and {Weaverdyck}, N. and others},
        title = "{Dark Energy Survey Year 6 Results: improved mitigation of spatially varying observational systematics with masking}",
      journal = {arXiv e-prints},
         year = 2025,
        month = sep,
          eid = {arXiv:2509.07943},
        pages = {arXiv:2509.07943},
          doi = {10.48550/arXiv.2509.07943},
archivePrefix = {arXiv},
       eprint = {2509.07943},
 primaryClass = {astro-ph.CO},
       adsurl = {https://ui.adsabs.harvard.edu/abs/2025arXiv250907943R}
}

@article{y6-lenspz,
       author = {{Giannini}, G. and others},
        title = "{Dark Energy Survey Year 6 Results: Redshift Calibration of the MagLim++ Lens Sample}",
      journal = {arXiv e-prints},
         year = 2025,
        month = sep,
          eid = {arXiv:2509.07964},
        pages = {arXiv:2509.07964},
          doi = {10.48550/arXiv.2509.07964},
archivePrefix = {arXiv},
       eprint = {2509.07964},
 primaryClass = {astro-ph.CO},
       adsurl = {https://ui.adsabs.harvard.edu/abs/2025arXiv250907964G}
}

@article{y6-cardinal,
       author = {{To}, Chun-Hao and others},
        title = "{Dark energy survey: Modeling strategy for multiprobe cluster cosmology and validation for the full six-year dataset}",
      journal = {\prd},
         year = 2025,
        month = sep,
       volume = {112},
       number = {6},
          eid = {063537},
        pages = {063537},
          doi = {10.1103/ynqj-6hsb},
archivePrefix = {arXiv},
       eprint = {2503.13631},
 primaryClass = {astro-ph.CO},
       adsurl = {https://ui.adsabs.harvard.edu/abs/2025PhRvD.112f3537T}
}

@article{y6-sourcepz,
       author = {{Yin}, B. and {Amon}, A. and {Campos}, A. and others},
        title = "{Dark Energy Survey Year 6 Results: Redshift Calibration of the Weak Lensing Source Galaxies}",
      journal = {arXiv e-prints},
         year = 2025,
        month = oct,
          eid = {arXiv:2510.23566},
        pages = {arXiv:2510.23566},
          doi = {10.48550/arXiv.2510.23566},
archivePrefix = {arXiv},
       eprint = {2510.23566},
 primaryClass = {astro-ph.CO},
       adsurl = {https://ui.adsabs.harvard.edu/abs/2025arXiv251023566Y}
}

@article{y6-maglim,
      author = {{Weaverdyck}, N. and {Rodr{\'\i}guez-Monroy}, M. and others},
      collaboration = {DES},
    title = "{Dark Energy Survey Year 6 Results: MagLim++ Lens Sample Selection and Measurements of Galaxy Clustering}",
      journal = {arXiv e-prints},
         year = 2026,
        month = jan,
          eid = {arXiv:2601.14484},
        pages = {arXiv:2601.14484},
          doi = {10.48550/arXiv.2601.14484},
archivePrefix = {arXiv},
       eprint = {2601.14484},
 primaryClass = {astro-ph.CO},
       adsurl = {https://ui.adsabs.harvard.edu/abs/2026arXiv260114484W}
}

@article{y6-methods,
       author = {{Sanchez-Cid}, D. and others},
        title = "{Dark Energy Survey Year 6 results: Weak lensing and galaxy clustering cosmological analysis framework}",
      journal = {\prd},
         year = 2026,
        month = sep,
       volume = {114},
       number = {6},
          eid = {063506},
        pages = {063506},
          doi = {10.1103/8n76-f6ln},
archivePrefix = {arXiv},
       eprint = {2601.14859},
 primaryClass = {astro-ph.CO},
       adsurl = {https://ui.adsabs.harvard.edu/abs/2026PhRvD.114f3506S}
}

@article{y6-ppd,
    author = {{Doux}, C. and {Muir}, J. and others},
    collaboration = {DES},
    title = "{Dark Energy Survey Year 6 Results: fast and interpretable posterior predictive checks for correlated cosmic probes}",
    journal = {\emph{in prep}},
    eprint = {},
    archivePrefix = {arXiv},
    primaryClass = {astro-ph.CO},
    reportNumber = "",
    year = 2025,
}

@article{y6-wz,
       author = {{d'Assignies}, W. and {Bernstein}, G.~M. and others},
        title = "{Dark Energy Survey Year 6 results: Clustering redshifts and importance sampling of self-organized-maps n(z) realizations for 3{\texttimes}2pt samples}",
      journal = {\prd},
         year = 2026,
        month = jun,
       volume = {113},
       number = {12},
          eid = {123549},
        pages = {123549},
          doi = {10.1103/6lfl-h6dg},
archivePrefix = {arXiv},
       eprint = {2510.23565},
 primaryClass = {astro-ph.CO},
       adsurl = {https://ui.adsabs.harvard.edu/abs/2026PhRvD.113l3549D}
}

@article{y6-imagesims,
    author = {{Mau}, S. and {Becker}, M.~R. and others},
    collaboration = {DES},
    title = "{Dark Energy Survey Year 6 Results: Image Simulations for Weak Lensing Shear and Photometric Redshift Calibration}",
    journal = {\emph{in prep}},
    eprint = {},
    archivePrefix = {arXiv},
    primaryClass = {astro-ph.CO},
    reportNumber = "",
    year = 2026,
}

@article{y6-magnification,
    author = {{Legnani}, E. and others},
    collaboration = {DES},
    title = "{Dark Energy Survey Year 6 Results: Magnification modeling and its impact on galaxy clustering and galaxy-galaxy lensing cosmology}",
      journal = {arXiv e-prints},
         year = 2026,
        month = jan,
          eid = {arXiv:2601.14833},
        pages = {arXiv:2601.14833},
          doi = {10.48550/arXiv.2601.14833},
archivePrefix = {arXiv},
       eprint = {2601.14833},
 primaryClass = {astro-ph.CO},
       adsurl = {https://ui.adsabs.harvard.edu/abs/2026arXiv260114833L}
}

@article{y6-gglens,
       author = {{Giannini}, G. and {Camacho-Ciurana}, G. and {Whyley}, A. and others},
        title = "{Dark Energy Survey Year 6 results: Galaxy-galaxy lensing}",
      journal = {\prd},
         year = 2026,
        month = may,
       volume = {113},
       number = {10},
          eid = {103503},
        pages = {103503},
          doi = {10.1103/c87p-j9nt},
archivePrefix = {arXiv},
       eprint = {2601.15175},
 primaryClass = {astro-ph.CO},
       adsurl = {https://ui.adsabs.harvard.edu/abs/2026PhRvD.113j3503G}
}

@misc{bechtol_DM,
       author = {{Bechtol}, Keith and {Birrer}, Simon and {Cyr-Racine}, Francis-Yan and others},
        title = "{Snowmass2021 Cosmic Frontier White Paper: Dark Matter Physics from Halo Measurements}",
      journal = {arXiv e-prints},
         year = 2022,
        month = mar,
          eid = {arXiv:2203.07354},
        pages = {arXiv:2203.07354},
          doi = {10.48550/arXiv.2203.07354},
archivePrefix = {arXiv},
       eprint = {2203.07354},
 primaryClass = {hep-ph},
       adsurl = {https://ui.adsabs.harvard.edu/abs/2022arXiv220307354B}
}

@article{y6-1x2pt,
    author = {{DES Collaboration}},
    collaboration = {DES},
    title = "{Dark Energy Survey Year 6 Results: Cosmological Constraints from Cosmic Shear}",
    journal = {\emph{in prep}},
    eprint = {},
    archivePrefix = {arXiv},
    primaryClass = {astro-ph.CO},
    reportNumber = "",
    year = 2025,
}

@article{y6-2x2pt,
    author = {{DES Collaboration}},
    collaboration = {DES},
    title = "{Dark Energy Survey Year 6 Results: Cosmological Constraints from Galaxy Clustering and Galaxy-Galaxy Lensing}",
    journal = {\emph{in prep}},
    eprint = {},
    archivePrefix = {arXiv},
    primaryClass = {astro-ph.CO},
    reportNumber = "",
    year = 2026,
}

@article{y6-3x2pt,
   author = {{DES Collaboration}},
    collaboration = {DES},
    title = "{Dark Energy Survey Year 6 Results: Cosmological Constraints from Galaxy Clustering and Weak Lensing}",
      journal = {arXiv e-prints},
         year = 2026,
        month = jan,
          eid = {arXiv:2601.14559},
        pages = {arXiv:2601.14559},
          doi = {10.48550/arXiv.2601.14559},
archivePrefix = {arXiv},
       eprint = {2601.14559},
 primaryClass = {astro-ph.CO},
       adsurl = {https://ui.adsabs.harvard.edu/abs/2026arXiv260114559D}
}

@article{y6-extensions,
   author = {{DES Collaboration}},
    collaboration = {DES},
    title = "{Dark Energy Survey Year 6 Results: Constraints on extensions to $\Lambda$CDM from Galaxy Clustering and Weak Lensing}",
    journal = {\emph{in prep}},
    eprint = {},
    archivePrefix = {arXiv},
    primaryClass = {astro-ph.CO},
    reportNumber = "",
    year = 2026,
}

@ARTICLE{DESDR2,
        author = {{DES Collaboration}},
        title = "{The Dark Energy Survey Data Release 2}",
      journal = {\apjs},
         year = 2021,
        month = aug,
       volume = {255},
       number = {2},
          eid = {20},
        pages = {20},
          doi = {10.3847/1538-4365/ac00b3},
archivePrefix = {arXiv},
       eprint = {2101.05765},
 primaryClass = {astro-ph.IM},
       adsurl = {https://ui.adsabs.harvard.edu/abs/2021ApJS..255...20A}
}

\appendix
\section{2pt measurement validation} \label{app:data_tests}

Prior to carrying out the cosmological parameter inference, our cosmic shear observable ($\xi_{\pm}$) was blinded according to \cite{muir_blinding}, which computes an additive factor based on the change in $\xi_{\pm}$ to shifts in reference cosmological parameters. We unblinded our observable once appropriate validations showed no significant systematics in our blinded data vectors. The tests are conducted to evaluate the shift in cosmological parameters, where we set the requirement to unblind is that the shift must be within 0.3$\sigma$ shift in 2D ($\Omega_{\rm m}$, $S_8$) and 1D $S_8$ posterior distributions. 
The tests we performed are
\begin{itemize}
    \item The impact of PSF modeling errors (Appendix~\ref{subsec:psf_sys})
    \item A null B-mode signal (Appendix~\ref{subsec:bmode_sys})
    \item The impact of additive bias correction (Appendix~\ref{subsec:additive_sys})
\end{itemize}

\begin{figure}
    \begin{center}
    \includegraphics[width=\columnwidth]{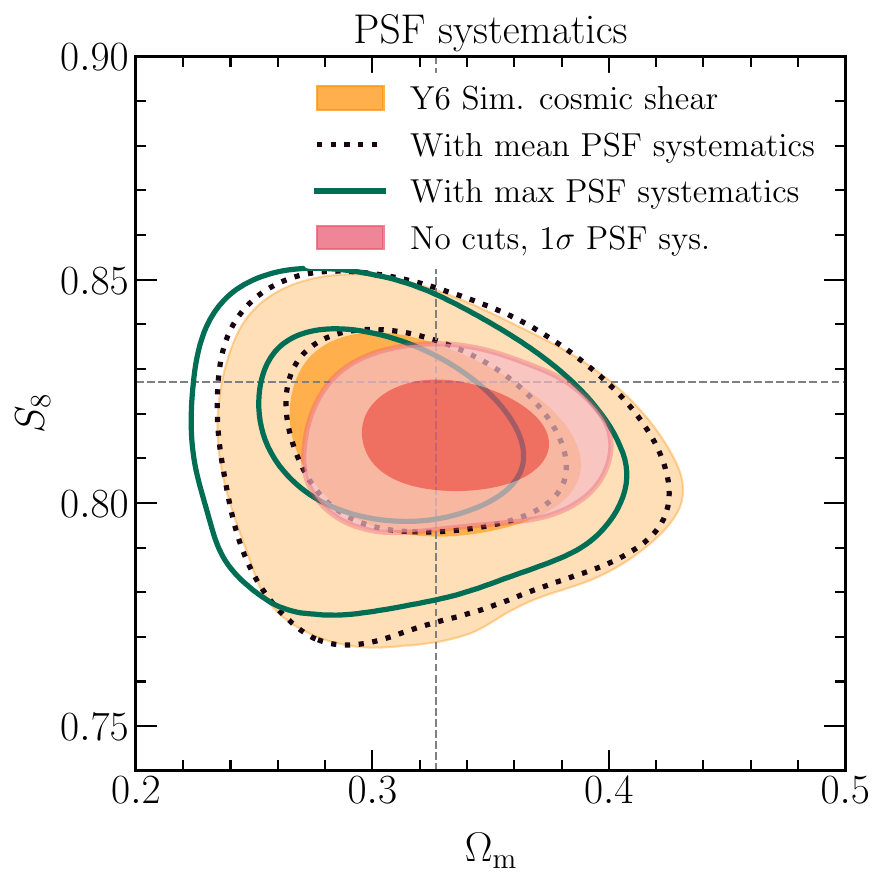}
    \end{center}
    \vspace{-0.75cm}
    \caption{Impact of PSF modeling errors on the cosmological constraints. We test the recovery of known cosmological parameters on simulated data vectors (produced with $\Omega_{\rm m}=0.327, S_8=0.827$) where they are contaminated with PSF modeling errors we measure from the data. The contaminated data vectors are produced with mean PSF systematics model (black dotted) and maximum PSF systematics model (green solid). We also test the contamination model at 1$\sigma$ uncertainty without imposing scale cuts on the simulated data vectors. All analyses result in the shift in both 2D ($S_8, \Omega_{\rm m}$) and 1D $S_8$ from the truth within $<0.3\sigma$. }
    \label{fig:PSF}
\end{figure}

\subsection{The impact of PSF modeling errors} \label{subsec:psf_sys}
The misuse of PSF models during shape measurement step (PSF leakage errors) and any PSF misestimation (PSF modeling errors) can cause additive biases. We consider the impact of PSF leakage and modeling errors on galaxy ellipticity as an additive signal to the observed shear $e^{\rm obs}$,
\begin{equation}\label{eqn:psfcont_xi}
    e^{\rm obs} = e^{\rm int} + \gamma + \delta e^{\rm sys}_{\rm PSF} + \delta e^{\rm noise}.
\end{equation}
Following \cite{paulin08, rowe10, jarvis16, zhang23, y6-metadetect}, we then model/parametrize $\delta e^{\rm sys}_{\rm PSF}$ by 
\begin{eqnarray}\label{eqn:model}
    \delta e^{\rm sys}_{\rm PSF} = && \alpha^{(2)} e_{\rm PSF} + \beta^{(2)} \Delta e_{\rm PSF} + \eta^{(2)}(e_{*}\Delta T_{\rm PSF}/T_{\rm PSF}) \\
    && + \alpha^{(4)} e^{(4)}_{\rm PSF} + \beta^{(4)}\Delta e^{(4)}_{\rm PSF} + \eta^{(4)}(e^{(4)}_{*}\Delta T^{(4)}_{\rm PSF}/T^{(4)}_{\rm PSF}) \nonumber \\
    && + \eta^{(24)}(e_{*}\Delta T^{(4)}_{\rm PSF}/T^{(4)}_{\rm PSF}) 
    + \eta^{(42)}(e^{(4)}_{*}\Delta T_{\rm PSF}/T_{\rm PSF}) \nonumber,
\end{eqnarray}
where $\Delta e_{\rm PSF} = e_{*} - e_{\rm PSF}$, $e_{\rm PSF}$ is the ellipticity of the PSF model, $e_{*}$ is the PSF ellipticity measured directly measured from stars, $\Delta T_{\rm PSF}=T_{*} - T_{\rm PSF}$, $T_{\rm PSF}$ is the PSF model size and $T_{*}$ is the PSF size measured from the star. The superscript $(2)$ on the ellipticity and size indicates a second-order moment and $(4)$ indicates a moment with a fourth-order radial term.

\begin{table}\label{tab:psf_sys}
    \centering
    \renewcommand{\arraystretch}{1.2}
    \caption{PSF systematics model parameters in each tomographic bin. }
    \resizebox{\columnwidth}{!}{
    \begin{tabular}{ccccc}
\hline\hline
      parameter & bin1 & bin2 & bin3 & bin4\\ \hline
    $\alpha$ & $0.028_{-0.013}^{+0.013}$ & $0.003_{-0.014}^{+0.014}$ & $0.000_{-0.015}^{+0.015}$ & $0.016_{-0.017}^{+0.017}$ \\ $\beta$ & $2.200_{-1.497}^{+1.421}$ & $-0.067_{-1.596}^{+1.609}$ & $2.500_{-1.613}^{+1.427}$ & $-1.719_{-1.695}^{+1.812}$ \\ $\eta$ & $8.999_{-11.582}^{+7.669}$ & $2.901_{-12.281}^{+10.784}$ & $-6.331_{-9.389}^{+12.772}$ & $9.994_{-11.918}^{+7.172}$ \\ $\alpha^{(4)}$ & $0.025_{-0.040}^{+0.041}$ & $0.089_{-0.042}^{+0.041}$ & $0.024_{-0.050}^{+0.050}$ & $0.069_{-0.055}^{+0.053}$ \\ $\beta^{(4)}$ & $3.054_{-0.544}^{+0.547}$ & $2.888_{-0.576}^{+0.570}$ & $2.696_{-0.573}^{+0.568}$ & $2.591_{-0.717}^{+0.722}$ \\ $\eta^{(4)}$ & $1.933_{-5.263}^{+5.223}$ & $-1.111_{-5.985}^{+6.065}$ & $-8.415_{-5.928}^{+5.946}$ & $-2.202_{-7.505}^{+7.485}$ \\ $\eta^{(24)}$ & $-4.995_{-2.863}^{+2.897}$ & $1.038_{-3.166}^{+3.171}$ & $0.687_{-3.163}^{+3.139}$ & $1.272_{-3.576}^{+3.595}$ \\ $\eta^{(42)}$ & $-50.385_{-31.102}^{+38.482}$ & $65.923_{-34.509}^{+23.465}$ & $23.089_{-43.278}^{+40.742}$ & $50.059_{-43.907}^{+32.953}$ \\ $\chi^{2}$ & $94.2835$ & $86.3405$ & $99.3924$ & $85.2875$ \\ $\chi^{2}_{\rm reduced}$ & $0.6203$ & $0.5680$ & $0.6539$ & $0.5611$ \\ $p$-value & $0.9999$ & $1.0000$ & $0.9997$ & $1.0000$ \\
\hline\hline
    \end{tabular}
    }
\end{table}

We first measure the eight parameters from the 2pt auto-correlation functions of PSF shapes/sizes and the cross-correlations of galaxy shapes in each tomographic bin and PSF quantities. Readers may refer Section 5.2 in \citepalias{y6-metadetect} for full details of how the model parameters are constrained. Table~\ref{tab:psf_sys} presents the constrained parameters from our best-fit model in each tomographic bin. Once our best-fit $\delta e^{\rm sys}_{\rm PSF}$ is measured, we can test the impact of PSF leakage and modeling errors on $\xi_{\pm}$ as $\xi^{\rm obs}_{\pm} = \xi^{\rm true}_{\pm} + \delta\xi^{\rm PSF}_{\pm}$, where $\delta\xi^{\rm PSF}_{\pm} = \langle \delta e^{\rm sys}_{\rm PSF} \delta e^{\rm sys}_{\rm PSF} \rangle$. We can, therefore, estimate the impact of PSF contamination on cosmological parameters. 

We explored two cases of contamination. First, we added $\delta\xi^{\rm PSF}_{\pm}$ computed from the best-fit model parameters to the synthetic data vector (described in \citepalias{y6-methods}) and created a contaminated data vector (i.e., mean contamination). We created another data vector where we added the 3$\sigma$ errors on the mean to the best-fit model parameters as a maximum PSF contamination (i.e., maximum contamination). We then ran cosmological parameter inference with these contaminated data vectors at the fiducial model and compared the $S_8$ inferred with the uncontaminated data vector. The Figure~\ref{fig:PSF} shows the results that both mean and maximum contamination scenarios do not lead to a significant shift in $S_8$, and both cases satisfy the requirement of the shift being smaller than 0.3$\sigma$. Furthermore, we test the impact of 1$\sigma$ contamination using the full-scale ($2.5<\theta<250$arcmin) cosmic shear measurement, without imposing scale cuts (pink). The resulting bias is $< 0.3\sigma$ -- the cosmic shear data vector is robust to a mis-modelled PSF to 1$\sigma$.

\subsection{A null B-mode signal} \label{subsec:bmode_sys}

In this section, we measure the B-mode signals in our data. To first order, we expect that gravitational lensing produces E-mode shear fields. At the current sensitivity of data, the detection of a spurious B-mode signal can point towards observational systematics in our data. Current literature suggests a few ways to estimate B-mode. These include COSEBIs \citep{schneider_cosebi} and Pseudo-$C_{\ell}$ \cite{hikage_pseudocell, alonso_namaster}, which require producing a shear map and/or interpolating $\xi_{\pm}$ to compute $C_{\ell}$. However, these methods are sensitive to the E- and B-mode mixing from masking, binning/interpolation, and pixelization (e.g., \cite{becker_bmode2013}). In order to construct an optimal estimator that avoids E/B-mode mixing, we utilize the Fourier Band-Power E/B-mode estimator\footnote{\url{https://github.com/beckermr/hybrideb}} \cite{becker_bandpower2016}, which enables us to estimate B-mode from the linear combinations of our real-space correlation function. This method takes the advantage of directly utilizing the binned-statistics without interpolation while accounting for survey mask in real space, showing that it can separate E/B-mode competitively or more compared with Pseudo-$C_{\ell}$ method.

We utilize estimators with Gaussian window functions with 20 band-powers. We computed the auto and cross two-point correlation functions of galaxy shapes in each tomographic bin (\citepalias{y6-metadetect}, \cite{y6-sourcepz}) with the angular scale of 0.25 to 250 arcmin in 250 bins. The measurement is then directly separated into E/B-mode using the estimators. Figure~\ref{fig:bmode} shows our B-mode measurement in each bin combination along with the p-value as the significance of non-null detection. The covariance of the measurement is computed by repeating the same measurement in mock catalogs used in \citep{y6-metadetect}. The result suggests that the B-mode in each tomographic bin combination shows a null detection within 3$\sigma$. 

\begin{figure}
    \begin{center}
    \includegraphics[width=0.45\textwidth]{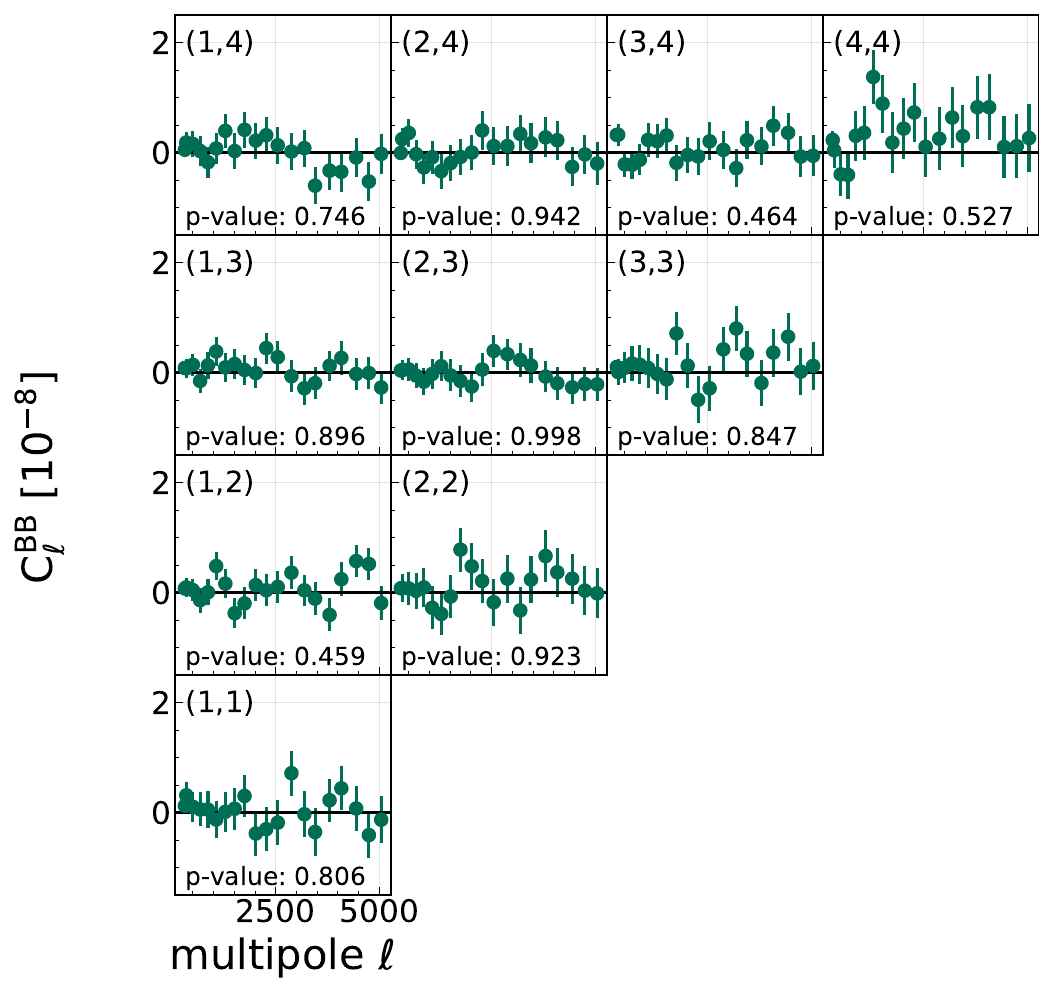}
    \end{center}
    \caption{B-mode estimated from $\xi_{\pm}$ computed from all the samples in Y6 shear catalog.}
    \label{fig:bmode}
\end{figure}

\subsection{Additive bias correction} \label{subsec:additive_sys}
In our shear catalog, we find the mean residual additive shear that is beyond cosmic variance and above what was measured in image simulations whose origins are unexplained \citep*{y6-metadetect, y6-imagesims}. As is done in Y3 by \cite*{y3-cosmicshear1, y3-cosmicshear2}, we treat the residual additive bias by subtracting it from each object's ellipticity in the catalog before the two-point correlation functions are measured. One must, however, check how significant the impact of subtracting mean residual shear from the catalog is before we infer cosmology. 

Table~\ref{tab:shape_stats} presents the mean residual shear in each tomographic bin. We can estimate the impact by contaminating the synthetic cosmic shear data vector $\xi^{ij}_{\pm}$ by adding $\langle e^{i}_1 \rangle \langle e^{j}_1 \rangle \pm \langle e^{i}_2 \rangle \langle e^{j}_2 \rangle$, and compare the inferred cosmological parameter shift in uncontaminated and contaminated data vectors. Assuming the same covariance matrix, we computed the difference to be $\Delta\chi_2 = 0.14$. We found that the parameter shift in 2D ($\Omega_{\rm m}$, $S_8$) and 1D ($S_8$) cases is within 0.3$\sigma$ on the simulated datavector and satisfies our criteria.

\section{Parameter constraints from all supporting tests}
\label{app:res}

This appendix summarizes the cosmological constraints obtained across the analyses performed in this paper. Table~\ref{tab:giant_results_table} shows the marginalized $S_8$, $\Omega_{\rm m}$ and $\sigma_8$ constraints. The second block in the table contains the results on various IA models which we discussed in Section~\ref{subsec:robust_ia} and ~\ref{subsec:insights_ia}. The third block presents the constraints discussed in Section~\ref{subsubsec:robust_baryons}, and the last row in the third block shows the constraints for fixed neutrino mass. The fourth block shows the results of our robustness tests in the data calibration we studied in Section~\ref{subsec:obssys_variants}. The last block presents the constraints from the subset of our data discussed in Section~\ref{sec:internal_consistency}.

\begin{table*}
\centering
\caption{Summary of marginalized parameter constraints $S_8$, $\Omega_{\rm m}$ and $\sigma_8$ in \lcdm. The mean and 68\% CL are provided for each cosmological parameter. The FoM for the $S_8-\Omega_{\rm m}$ plane are also shown. We distinguish variations on the fiducial model that are not required to give consistent
results (for example, by neglecting observational systematics) by an asterisk and an open symbol. 
}
\begin{tabular}{lcccccccc}
\hline
\hline
& \multicolumn{4}{c}{NLA} & \multicolumn{4}{c}{TATT} \\
\cmidrule(lr){2-5}\cmidrule(lr){6-9}
& $S_8$ & $\Omega_{\rm m}$ & $\sigma_8$ & FoM$_{S_8,\Omega_{\rm m}}$
& $S_8$ & $\Omega_{\rm m}$ & $\sigma_8$ & FoM$_{S_8,\Omega_{\rm m}}$ \\ 
\hline \\[0.2cm]
DES Y6 fiducial & $ 0.798^{+0.014}_{-0.015} $  & $ 0.332^{+0.035}_{-0.044} $  & $ 0.763^{+0.050}_{-0.057} $  & 2080  & $ 0.783^{+0.019}_{-0.015} $  & $ 0.321^{+0.036}_{-0.047} $  & $ 0.763^{+0.053}_{-0.062} $  & 1532  \\ [0.2cm] 
\hline
NLA-1 & $ 0.799^{+0.014}_{-0.015} $  & $ 0.336^{+0.036}_{-0.044} $  & $ 0.759^{+0.050}_{-0.055} $  & 2125  & --- & --- & --- & ---  \\ [0.2cm] 
NLA (flat $\eta_1$) & $ 0.798^{+0.014}_{-0.014} $  & $ 0.331^{+0.034}_{-0.044} $  & $ 0.765^{+0.051}_{-0.057} $  & 2113  & --- & --- & --- & ---  \\ [0.2cm] 
IA linear $z$ & $ 0.795^{+0.016}_{-0.014} $  & $ 0.328^{+0.033}_{-0.044} $  & $ 0.765^{+0.050}_{-0.055} $  & 1771  & $ 0.772^{+0.024}_{-0.019} $  & $ 0.322^{+0.040}_{-0.045} $  & $ 0.751^{+0.056}_{-0.063} $  & 1148  \\ [0.2cm] 
NLA per bin & $ 0.790^{+0.015}_{-0.017} $  & $ 0.325^{+0.034}_{-0.046} $  & $ 0.763^{+0.055}_{-0.056} $  & 1757  & --- & --- & --- & ---  \\ [0.2cm] 
TA ($b_{\rm TA}$=1) & --- & --- & --- & ---  & $ 0.802^{+0.014}_{-0.014} $  & $ 0.302^{+0.029}_{-0.040} $  & $ 0.803^{+0.051}_{-0.055} $  & 2325 \\ [0.2cm] 
TATT no $z$ & --- & --- & --- & ---  & $ 0.792^{+0.014}_{-0.015} $  & $ 0.326^{+0.038}_{-0.045} $  & $ 0.764^{+0.052}_{-0.065} $  & 1982 \\ [0.2cm] 
TATT ($A_1>0$) & --- & --- & --- & ---  & $ 0.783^{+0.017}_{-0.016} $  & $ 0.331^{+0.034}_{-0.044} $  & $ 0.750^{+0.049}_{-0.057} $  & 1784 \\ [0.2cm] 
TATT+$b_{\rm TA}$ & --- & --- & --- & ---  & $ 0.781^{+0.017}_{-0.017} $  & $ 0.354^{+0.042}_{-0.048} $  & $ 0.724^{+0.047}_{-0.063} $  & 1517 \\ [0.2cm] 
TATT (flat $\eta_1$, $\eta_2$) & --- & --- & --- & ---  & $ 0.782^{+0.020}_{-0.015} $  & $ 0.316^{+0.035}_{-0.046} $  & $ 0.768^{+0.056}_{-0.060} $  & 1410 \\ [0.2cm] 
TATT (NLA cuts) & --- & --- & --- & ---  & $ 0.786^{+0.019}_{-0.019} $  & $ 0.319^{+0.037}_{-0.046} $  & $ 0.768^{+0.054}_{-0.068} $  & 1540 \\ [0.2cm] 
IA $b_{\rm TA}$ free & $ 0.782^{+0.013}_{-0.015} $  & $ 0.373^{+0.041}_{-0.048} $  & $ 0.706^{+0.047}_{-0.057} $  & 1852  & $ 0.782^{+0.014}_{-0.016} $  & $ 0.365^{+0.039}_{-0.049} $  & $ 0.713^{+0.047}_{-0.059} $  & 1715  \\ [0.2cm] 
\hline
Free $\Theta_{\rm AGN} \in [7.2, 8.5]$ & $ 0.803^{+0.016}_{-0.018} $  & $ 0.326^{+0.035}_{-0.045} $  & $ 0.775^{+0.054}_{-0.062} $  & 1821  & $ 0.789^{+0.019}_{-0.020} $  & $ 0.312^{+0.040}_{-0.050} $  & $ 0.780^{+0.060}_{-0.075} $  & 1329  \\ [0.2cm] 
Free $\Theta_{\rm AGN} \in [7.2, 9.0]$ & $ 0.807^{+0.017}_{-0.019} $  & $ 0.323^{+0.035}_{-0.045} $  & $ 0.782^{+0.054}_{-0.064} $  & 1699  & $ 0.793^{+0.020}_{-0.021} $  & $ 0.308^{+0.039}_{-0.053} $  & $ 0.790^{+0.062}_{-0.080} $  & 1238  \\ [0.2cm] 
Free $A_{\rm mod}$ & $ 0.806^{+0.021}_{-0.025} $  & $ 0.327^{+0.037}_{-0.047} $  & $ 0.777^{+0.057}_{-0.073} $  & 1302  & $ 0.795^{+0.022}_{-0.025} $  & $ 0.316^{+0.039}_{-0.049} $  & $ 0.781^{+0.060}_{-0.078} $  & 1192  \\ [0.2cm]
Fixed $\sum m_{\nu}$ & $ 0.803^{+0.014}_{-0.014} $  & $ 0.318^{+0.032}_{-0.041} $  & $ 0.784^{+0.049}_{-0.056} $  & 2291  & $ 0.789^{+0.020}_{-0.014} $  & $ 0.305^{+0.032}_{-0.043} $  & $ 0.788^{+0.057}_{-0.061} $  & 1628  \\ [0.2cm] 
\hline
$m + \Delta z$ & $ 0.797^{+0.013}_{-0.013} $  & $ 0.336^{+0.034}_{-0.044} $  & $ 0.758^{+0.049}_{-0.055} $  & 2194  & $ 0.783^{+0.017}_{-0.014} $  & $ 0.324^{+0.038}_{-0.045} $  & $ 0.757^{+0.050}_{-0.063} $  & 1676  \\ [0.2cm] 
No WZ & $ 0.803^{+0.014}_{-0.015} $  & $ 0.336^{+0.034}_{-0.044} $  & $ 0.764^{+0.049}_{-0.056} $  & 1824  & $ 0.787^{+0.020}_{-0.015} $  & $ 0.325^{+0.034}_{-0.048} $  & $ 0.761^{+0.054}_{-0.058} $  & 1315  \\ [0.2cm] 
No blending correction & $ 0.797^{+0.013}_{-0.014} $  & $ 0.332^{+0.033}_{-0.045} $  & $ 0.763^{+0.049}_{-0.057} $  & 2116  & $ 0.782^{+0.018}_{-0.017} $  & $ 0.325^{+0.036}_{-0.047} $  & $ 0.757^{+0.054}_{-0.064} $  & 1551  \\ [0.2cm] 
With SR & $ 0.796^{+0.014}_{-0.014} $  & $ 0.342^{+0.035}_{-0.047} $  & $ 0.750^{+0.048}_{-0.056} $  & 2036  & $ 0.781^{+0.017}_{-0.016} $  & $ 0.333^{+0.036}_{-0.047} $  & $ 0.746^{+0.053}_{-0.061} $  & 1673  \\ [0.2cm] 
\hline
No bin1 & $ 0.787^{+0.015}_{-0.015} $  & $ 0.350^{+0.041}_{-0.049} $  & $ 0.734^{+0.050}_{-0.058} $  & 1516  & $ 0.781^{+0.015}_{-0.015} $  & $ 0.357^{+0.039}_{-0.048} $  & $ 0.720^{+0.047}_{-0.055} $  & 1517  \\ [0.2cm] 
No bin2 & $ 0.797^{+0.016}_{-0.015} $  & $ 0.349^{+0.039}_{-0.051} $  & $ 0.744^{+0.052}_{-0.060} $  & 1735  & $ 0.791^{+0.019}_{-0.014} $  & $ 0.315^{+0.033}_{-0.051} $  & $ 0.778^{+0.063}_{-0.062} $  & 1502  \\ [0.2cm] 
No bin3 & $ 0.807^{+0.018}_{-0.019} $  & $ 0.317^{+0.037}_{-0.049} $  & $ 0.792^{+0.058}_{-0.065} $  & 1346  & $ 0.788^{+0.023}_{-0.016} $  & $ 0.296^{+0.035}_{-0.046} $  & $ 0.800^{+0.060}_{-0.068} $  & 1028  \\ [0.2cm] 
No bin4 & $ 0.796^{+0.018}_{-0.019} $  & $ 0.311^{+0.037}_{-0.056} $  & $ 0.789^{+0.072}_{-0.075} $  & 1436  & $ 0.774^{+0.022}_{-0.021} $  & $ 0.314^{+0.040}_{-0.060} $  & $ 0.765^{+0.071}_{-0.082} $  & 1045  \\ [0.2cm] 
Low-z only & $ 0.752^{+0.040}_{-0.042} $  & $ 0.362^{+0.081}_{-0.102} $  & $ 0.702^{+0.082}_{-0.148} $  & 373  & $ 0.705^{+0.043}_{-0.042} $  & $ 0.354^{+0.080}_{-0.099} $  & $ 0.665^{+0.075}_{-0.136} $  & 292  \\ [0.2cm] 
High-z only & $ 0.759^{+0.033}_{-0.020} $  & $ 0.373^{+0.045}_{-0.057} $  & $ 0.685^{+0.051}_{-0.062} $  & 712  & $ 0.758^{+0.026}_{-0.020} $  & $ 0.373^{+0.045}_{-0.054} $  & $ 0.685^{+0.050}_{-0.062} $  & 773  \\ [0.2cm] 
$\xi_{+}$ only & $ 0.811^{+0.014}_{-0.013} $  & $ 0.326^{+0.033}_{-0.044} $  & $ 0.782^{+0.052}_{-0.057} $  & 1971  & $ 0.798^{+0.017}_{-0.015} $  & $ 0.317^{+0.035}_{-0.047} $  & $ 0.782^{+0.058}_{-0.062} $  & 1768  \\ [0.2cm] 
$\xi_{-}$ only & $ 0.797^{+0.019}_{-0.019} $  & $ 0.288^{+0.034}_{-0.062} $  & $ 0.824^{+0.088}_{-0.082} $  & 1189  & $ 0.787^{+0.025}_{-0.021} $  & $ 0.281^{+0.032}_{-0.060} $  & $ 0.823^{+0.083}_{-0.087} $  & 976  \\ [0.2cm]
\hline
\end{tabular}

\label{tab:giant_results_table}
\end{table*}

\end{document}